\documentclass[12pt,preprint]{aastex}

\shorttitle{Dust and Star Formation in NGC4449}
\shortauthors{Calzetti et al.}

\begin{document}

\title{Spatially Resolved Dust, Gas, and Star Formation in the Dwarf Magellanic Irregular NGC\,4449.\altaffilmark{1}}

\author{D. Calzetti\altaffilmark{2}, G.W. Wilson\altaffilmark{2}, B.T. Draine\altaffilmark{3}, H. Roussel\altaffilmark{4},
K.E. Johnson\altaffilmark{5},  M.H. Heyer\altaffilmark{2}, W. F. Wall\altaffilmark{6}, K. Grasha\altaffilmark{2}, A. Battisti\altaffilmark{2}, J.E. Andrews\altaffilmark{7}, A. Kirkpatrick\altaffilmark{8}, D. Rosa Gonz\'alez\altaffilmark{6}, O. Vega\altaffilmark{6}, J. Puschnig\altaffilmark{9}, M. Yun\altaffilmark{2}, G. \"Ostlin\altaffilmark{9},  A.S. Evans\altaffilmark{5,10},  Y. Tang\altaffilmark{2},  J. Lowenthal\altaffilmark{11}, D. S\'anchez-Arguelles\altaffilmark{6}}
  
\altaffiltext{1}{Based on observations obtained with the Large Millimeter Telescope Alfonso Serrano - a bi--national collaboration between 
INAOE (Mexico) and The University of Massachusetts - Amherst (USA). } 
\altaffiltext{2}{Dept. of Astronomy, University of Massachusetts -- Amherst, Amherst, MA 01003, USA;\\ calzetti@astro.umass.edu}
\altaffiltext{3}{Dept. of Astrophysical Science, Princeton University, Princeton, NJ, USA}
\altaffiltext{4}{Institut d'Astrophysique de Paris, Paris, France}
\altaffiltext{5}{Dept. of Astronomy, University of Virginia, Charlottesville, VA, USA}
\altaffiltext{6}{Instituto Nacional de Astrof\'\i sica, Optica, y Electr\'onica (INAOE), Tonantzintla, Puebla, Mexico}
\altaffiltext{7}{Dept. of Astronomy, University of Arizona, Tucson, AZ, USA}
\altaffiltext{8}{Dept. of Astronomy, Yale University, New Haven, CT, USA}
\altaffiltext{9}{Dept. of Astronomy, The Oskar Klein Centre, Stockholm University, Stockholm, Sweden }
\altaffiltext{10}{National Radio Astronomy Observatory, Charlottesville, VA, USA}
\altaffiltext{11}{Dept. of Astronomy, Smith College, Northampton, MA, USA}

\begin{abstract}
We investigate the relation between gas and star formation in sub--galactic regions, $\sim$360~pc to $\sim$1.5~kpc  in size, within the nearby starburst dwarf NGC\,4449, in order to  separate the underlying relation  from the effects of sampling at varying spatial scales. Dust and gas mass surface densities are derived by combining new observations at 1.1~mm, obtained with the AzTEC instrument on the Large Millimeter Telescope, with archival infrared images in the  range 8~$\mu$m--500~$\mu$m from the Spitzer Space Telescope and the Herschel Space Observatory. We extend the dynamic range of our mm (and dust) maps at the faint end, using a correlation between  the  far--infrared/millimeter colors F(70)/F(1100) [and F(160)/F(1100)] and  the mid--infrared color F(8)/F(24) that we establish for the first time for this and other galaxies. Supplementing our data with maps of the extinction--corrected star formation rate (SFR) surface density, we measure both the SFR--molecular~gas and the SFR--total~gas relations in NGC\, 4449. We find that the SFR--molecular~gas relation is described by a power law with exponent that decreases from $\sim$1.5 to $\sim$1.2 for increasing region size, while the exponent of the SFR--total~gas relation remains constant with value $\sim$1.5 independent of region size. We attribute the molecular law behavior to the increasingly better sampling of the molecular cloud mass function at larger region sizes; conversely, the total gas law behavior likely results from the balance between the atomic and molecular gas phases achieved in regions of active star formation. Our results indicate a  non-linear relation between SFR and gas surface density in NGC\,4449, similar to what is observed for galaxy samples. 
 \end{abstract}


\section{Introduction}
The evolution of the baryonic component of galaxies across cosmic times is inextricably linked to the relation between star formation and the gas reservoir. Large observational and theoretical efforts have therefore been expended on both measuring this relation and understanding its physical underpinning. 

Star formation and gas reservoir are related by a tight scaling \citep[the Schmidt-Kennicutt Law, or SK Law,][]{Kennicutt1998}, which  holds at both low  and high redshifts,  when these quantities are averaged over entire galaxies \citep[e.g.,][]{KennicuttEvans2012, Daddi2010, Genzel2010, Genzel2015}. The same relation breaks down when zooming into smaller regions, and recently--formed stars appear spatially uncorrelated with molecular clouds over scales smaller than $\sim$100--200 pc \citep[e.g.,][]{Onodera2010, Momose2010, Schruba2010}. The break--down may be due to the onset of two effects: (1) the increasing scatter in tracers of both star formation rate (SFR) and gas clouds, due to small number statistics in the random sampling \citep{Calzetti2012, Kruijssen2014}; and (2) the finite timescale of the physical association between young stars and their natal cloud \citep[e.g.,][]{Kawamura2009, Whitmore2014}. Within molecular clouds, the dust--enshrouded star formation is most closely associated with the densest gas components \citep[e.g.,][]{Gao2004, Lada2010, Lada2012, Evans2014}, while it appears  only indirectly  related to the molecular gas reservoir \citep[e.g.,][]{Heiderman2010, Gutermuth2011, Faesi2014,  Hony2015}. At the kpc-- or larger scale, where multiple clouds are averaged together, the relation becomes better defined, but its exact functional form, which carries physical information, remains controversial  \citep{Kennicutt2007, Bigiel2008, Rahman2012, Leroy2013, Lada2013, Shetty2014}. The ill--quantified CO-dark gas in galaxies \citep{Pineda2013} and the presence of non--star forming, diffuse molecular gas \citep{Shetty2014, Mogotsi2016} contribute to complicate an already complex picture. 

The interlocking of these still-disjointed pieces of evidence requires the investigation of the relation between star formation and gas reservoir across a range of spatial scales, probing a variety of galactic environments and star formation activity, from the molecular cloud size regions all the way to whole galaxies. \citet{Liu2011} derive the SK Law as a function of spatial scale for two late--type star--forming spirals: NGC\,5194 (M\,51) and NGC\,3521, covering the range $\sim$250~pc--1.3~kpc. When expressing the SK Law as: $\Sigma_{SFR}\propto \Sigma_{H2}^{\gamma_{H2}}$, with $\Sigma$ indicating the surface density and H$_2$ indicating that only the molecular phase is included, Liu et al. find that the exponent $\gamma_{H2}$ is a decreasing function of increasing spatial scale, converging towards unity beyond $\sim$1--1.3~kpc. They also find that the scatter of the data about the best--fitting relation decreases for increasing spatial scale. \citet{Wall2016}  measure $\gamma_{H2}$ in both NGC\, 5194 and NGC\, 5236, and find values that are in agreement with those of \citet{Liu2011}, once the differences in the fitting routines are factored in \citep[see][for a quantification of the impact of fitting routine choices]{Calzetti2012}. 

\citet{Calzetti2012} model these results as the effect of stochastic sampling of the molecular clouds mass function and of star formation within galaxy regions. At regions sizes larger than $\sim$1~kpc, those authors find that the cloud mass function becomes fully sampled, and the relation between $\Sigma_{SFR}$ and $\Sigma_{H2}$ converges to an exponent $\gamma_{H2}$=1 irrespective of the underlying, intrinsic relation between SFR and gas reservoir. The exact size of the galactic region where the convergence to $\gamma_{H2}$=1 occurs depends on the maximum size of the molecular clouds formed in the galaxy, with region sizes in the range $\sim$1--2~kpc for M$_{cloud}$(max)$\sim$10$^6$--10$^{7.5}$~M$_{\odot}$. Similar results, at least qualitatively, are obtained by \citet{Kruijssen2014}, where they model their galaxy regions by centering on either a star forming region or a molecular cloud; they are, however, quantitatively different from those of \citet{Calzetti2012}, who sample their regions randomly within galaxies. Both studies agree that once the full sampling of the cloud mass function is achieved, the relation between $\Sigma_{SFR}$ and $\Sigma_{H2}$ is only `counting clouds'; the physical information is contained in the trends of $\gamma_{H2}$ and the scatter of the data as a function of the spatial scale of galaxy regions. 

The literature on the analysis of the observed SFR--gas relation as a function of region size is currently sparse; the relevant studies have so far concentrated on large, star--forming spiral galaxies (see above), and, albeit not in a systematic manner, on the Magellanic Clouds \citep{Jameson2016,Hony2015}. In order to expand on those studies, we have recently mapped the nearby starburst dwarf NGC\,4449 with the Large Millimeter Telescope, using the AzTEC instrument at 1.1~mm, which we combine with imaging data at shorter wavelengths to investigate the relation between SFR and gas densities at sub--kpc and kpc scales. 

NGC\,4449 is a well-studied local  Magellanic irregular dwarf galaxy hosting a central starburst. Its basic parameters are listed in Table~\ref{tab1}. The specific SFR [SFR/(Stellar Mass)] places NGC\, 4449 solidly, about a factor 2.5--9, above the Main Sequence for local star forming galaxies, with the exact offset depending on the Main Sequence relation used \citep{Whitaker2012, Cook2014}. The starburst has been possibly triggered by a minor merger \citep{Lelli2014} or by the interaction with another galaxy \citep{Hunter1998}. It has sub--solar metallicity with a modest gradient  \citep[Table~\ref{tab1},][]{Pilyugin2015}. The central oxygen abundance value  determined by \citet{Pilyugin2015} is in agreement with that of  \citet{Berg2012}, 12+Log(O/H)=8.26$\pm$0.09. 
Our analysis concentrates on the inner $\sim$5~kpc, for which, when necessary, we adopt a mean value of the oxygen abundance 12+Log(O/H)$\sim$8.2 or 30\% the solar value\footnote{We adopt a solar oxygen abundance of 12+Log(O/H)=8.69, \citet{Asplund2009}.}. 

The low metallicity of NGC\,4449 is consistent with its relatively low dust content. The IR/UV ratio indicates that only about 40\% of the light from young stars is absorbed by dust in this galaxy \citep{Hao2011, Grasha2013}. \citet{Reines2008} find that, with one exception, even embedded sources have A$_V<$1.5~mag. This characteristic enables us to effectively decouple tracers of SFR from those of gas content. In dusty galaxies, the use of the blue and red sides of the IR spectral energy distribution (SED) to derive SFR and dust(gas) masses, respectively, can potentially introduce  a covariance between the two physical quantities \citep{Bolivar2016}. The low extinction of NGC\,4449 ensures that most of the SFR will be traced at UV--optical (e.g., H$\alpha$) wavelengths; this decouples the SFR measurements from those of the dust mass, the latter used to derive gas masses. 
Despite being relatively metal deficient, NGC\,4449 is still sufficiently metal--rich that measurements of dust masses yield reliable measurements of gas masses \citep{Draine2007,  RemyRuyer2014, Accurso2017}. 

The properties listed above, together with the relative proximity of the galaxy (10$^{\prime\prime}$ subtends a linear scale of 204~pc), make NGC\,4449 an ideal target for investigating the relation between SFR and gas surface densities as a function of spatial scale. 

\section{Observations and Archival Data}

\subsection{LMT Data}

\subsubsection{Observations}
Observations of NGC\,4449 (RA(2000)=12$^h$28$^m$13$^s$.60, DEC(2000)=$+$44$^o$05$^{\prime}$35$^{\prime\prime}$.14) were carried out with the AzTEC instrument on the 32--m Large Millimeter Telescope near the city of Puebla (Mexico) between February 25 and March
9, 2015, in excellent weather conditions with a median zenith opacity, $\tau_{\rm 225GHz}$, of 0.055. AzTEC is a 144 element bolometer array configured to observe in the 1.1~mm atmospheric window \citep[225 GHz,][]{Wilson2008}, with a half-power beam width of 8.5$^{\prime\prime}$ (corresponding to a beam area of 164~arcsec$^2$, Table~\ref{tab2}). The total observing time was 9.03 hours, divided into  23 individual maps. Each map  consisted of a combination of a fast motion of the telescope boresight in a lissajous pattern superimposed on a slow raster scan of the center of the lissajous pattern over the field.  Both motions were performed in the azimuth-elevation coordinate system.  This observational technique, known colloquially as a ``rastajous scan,'' allows for deep imaging of moderate sized fields with high scan speeds and good cross-linking of
pixels but without the inefficiencies of lost observing time due to rapid turn-arounds of the telescope during fast raster scan.  

Corrections to the LMT's native pointing model were made by interspersing the maps with pointed observations of the quasar 1203$+$480, which has an unresolved flux of 190~mJy. Pointing corrections were typically $\pm$3$^{\prime\prime}$ in azimuth and $-$4$^{\prime\prime}$ to 11$^{\prime\prime}$ in elevation.  These corrections are typical for AzTEC observations due to a known boresight mismatch between the AzTEC and the LMT's pointing model. Reconstructed pointing errors following these corrections are typically of-order 1$^{\prime\prime}$ which is much smaller than the beam size (8.5$^{\prime\prime}$).

Data were calibrated based on beammap observations of the asteroids Ceres and Pallas taken during each evening of observations along with observations of the pre-planetary nebula CRL618.  All calibrations are
self-consistent and we estimate the calibration error to be approximately 10\%.

\subsubsection{Data Processing and Final Maps}
The data were processed using the recently developed AzTEC C++ pipeline which is now the standard pipeline for AzTEC data on the LMT.  This new pipeline offers the same reduction
processes reported in previous AzTEC studies \citep[e.g.,][]{Perera2008,Scott2008,Austermann2010,Liu2010,Scott2012}, including removal of atmospheric contributions (the dominant noise source
 in the AzTEC maps), correction of pointing errors,  data calibration, and coaddition of the ensemble of maps, but with improvements to achieve better performance and speed.  
The updated pipeline also implements a new approach to estimate the contribution of the atmosphere into the observed raw time streams, aiming at improving the recovery of extended
emission in the AzTEC data. 

The new approach is based on the Cottingham Method \citep{Cottingham1987},  where an unbiased estimator of the atmosphere signal is constructed using a set of B-spline basis vectors and the pointing information from individual detectors.  \cite{Hincks2010} demonstrate that this
method is a maximum--likelihood estimator for both the atmosphere signal and the astronomical surface brightness distribution. Our implementation has been adapted to maximize the compatibility with the
existing pipeline code and other atmosphere subtraction techniques; therefore, it only estimates and subtracts the atmosphere component using Equation 7 from \citet{Hincks2010}.  Details on the technique are 
given in the Appendix. 

In addition to a surface brightness map, a noise map is constructed from  jack--knifed noise realizations of each AzTEC map in which 
the time stream is randomly multiplied by $\pm$1.  The effect of the jack--knife step is to remove sources while retaining 
noise properties.  Each jack--knifed time--stream is converted into a map and coadded in weighted quadrature  to produce a final image of rms noise values, $\sigma$.    

The primary data products from the AzTEC pipeline are an image of surface brightness at 1.1~mm, in units of Jy/beam, and a 
corresponding weight image (1/$\sigma^2$)  that reflects both instrumental and atmospheric contributions to the noise budget. The final images 
are constructed in equatorial coordinates (J2000). As in all such maps, the outer edges  have larger errors owing to 
fewer number of collected samples at these positions, and therefore, less accumulated time, relative to the central regions. The map  has a diameter of about 600$^{\prime\prime}$ ($\sim$12~kpc at the distance of NGC\,4449), with a lower noise region of about 450$^{\prime\prime}$ diameter (Figure~\ref{fig1}). Our galaxy fits well within this central region: NGC\,4449 has a 500~$\mu$m diameter $<$450$^{\prime\prime}$, as measured from the Herschel SPIRE map (see next section), and optical major axis $\sim$370$^{\prime\prime}$\footnote{From NED, the NASA Extragalactic Database.}. The central 450$^{\prime\prime}$ diameter region has median sensitivity of 1~mJy/beam, with 1$\sigma$ variations of less than 10\%.

The measured fluxes for NGC\,4449  at 1.1~mm (=1100~$\mu$m) are listed in Table~\ref{tab2}. The AzTEC bandpass excludes the CO(2--1) line, so no contamination is expected from this transition. Estimates of the free--free and synchrotron emission by extrapolating the  measurements  at cm wavelengths for this galaxy\footnote{Photometry at cm wavelengths is from NED, the NASA Extragalactic Database. See references therein.} yield a $\lesssim$8\% contribution of these processes to the light in the AzTEC band.  Our estimate uses spectral indices of $-$0.5 for the synchrotron emission and $-$0.1 for the free-free emission, from \citet{Srivastava2014}. Thus, the measured 1100~$\mu$m flux can be attributed almost entirely to dust emission.

\subsection{Archival Images}

For our analysis we make use of several images and maps from archives, covering a wide range of wavelengths: ground--based H$\alpha$ and HI, Spitzer Space Telescope (SST) 3.6~$\mu$m to 24~$\mu$m, and Herschel Space Observatory (HSO) 70~$\mu$m to 500~$\mu$m. 

The ground--based H$\alpha$ image was obtained at the Bok Telescope as part of the LVL (Local Volume Legacy) project, a Spitzer Legacy program that has observed 
almost 300 galaxies in the local 11~Mpc \citep{Dale2009, Lee2011}. The image is available through NED, already continuum--subtracted and calibrated. We correct the fluxes for the effects of foreground Milky Way extinction (E(B-V)=0.017, corresponding to A$_{H\alpha}$=0.043~mag) and of [NII] contamination ([NII]/H$\alpha$=0.23), \citep{Kennicutt2008}. The final image has a Point Spread Function (PSF) FWHM$\sim$1.5$^{\prime\prime}$.

The  SST IRAC 3.6~$\mu$m, 4.5~$\mu$m, 5.8~$\mu$m, and  8.0~$\mu$m and MIPS 24~$\mu$m, 70~$\mu$m, and 160~$\mu$m images were also obtained as part of the LVL project \citep{Dale2009}, and are available already fully processed and calibrated through IRSA and NED. All Spitzer images are in units of MJy/sr. For our pixel--resolved analysis, we make use of the highest angular resolution images  from Spitzer, i.e., up to and including the 24~$\mu$m image. For the purpose of fitting the dust SED, we  remove the stellar contribution from both the 8~$\mu$m and 24~$\mu$m images, using the formulae of \citet{Helou2004} and \citet[][see also \citet{Calapa2014}]{Calzetti2007}: f$_{\nu, D}$(8) = f$_{\nu}$(8) - 0.25 f$_{\nu}$(3.6), and f$_{\nu, D}$(24)=f$_{\nu}$(24) - 0.035 f$_{\nu}$(3.6), where the flux densities are in units of Jy, and the subscript `D' indicates the dust--only emission component.  Using these formulae, the typical stellar continuum contribution to the fluxes at 8~$\mu$m and 24~$\mu$m of individual pixels is about 6\%--7\% and $<$1\%, with maximum values of about 
20\% and 2.4\% at 8 and 24~$\mu$m, respectively. The contamination of the 3.6~$\mu$m image by the 3.3~$\mu$m Polycyclic Aromatic Hydrocarbons (PAH) emission feature is minimal \citep[$\sim$5\%--15\%,][]{Meidt2012}, and the IRAC~3.6 can be used as a reasonable stellar continuum tracer for the purpose of subtraction from the 8~$\mu$m and 24~$\mu$m fluxes. When required by the analysis (see later sections), we will instead use the uncontaminated 4.5~$\mu$m IRAC image as a reliable tracer of the stellar continuum distribution. 

The HSO PACS 70~$\mu$m, 100~ $\mu$m, and 160~$\mu$m, and SPIRE 250~$\mu$m, 350~$\mu$m, and 500~$\mu$m images were retrieved from 
the Herschel archive, and reprocessed by us using Scanamophos v24.0 \citep{Roussel2013}, which dramatically improves the final maps by removing the 
low frequency noise associated with them. The PACS images are in units of Jy/pixel, and the SPIRE images are in units of Jy/beam. For photometry, we use the following SPIRE beam areas: 469~arcsec$^2$,  831~arcsec$^2$, and 1804~arcsec$^2$, for 250~$\mu$m, 350~$\mu$m, and 500~$\mu$m, respectively\footnote{see: http://herschel.esac.esa.int/twiki/bin/view/Public/SpirePhotometerBeamProfile2}.

Color corrections to the SST and HSO photometry for the typical SEDs of star forming galaxies are at the level of 10\%--12\% for the HSO/PACS70 band, and less than 5\% for all other HSO and SST bands. We apply color corrections only to the HSO/PACS70 photometry, as the corrections for the other bands are small relative to the measurement uncertainties. In addition, we implement aperture corrections of about 8\% to the SST/IRAC 8~$\mu$m, as appropriate for partially--extended sources, to compensate for the scattered light into this band\footnote{The sources in our Field--of--View have typical extent of 4$^{\prime\prime}$--8$^{\prime\prime}$, and we apply the aperture corrections described in section 4.11 of the IRAC Instrument Handbook (http://irsa.ipac.caltech.edu/data/SPITZER/docs/irac/iracinstrumenthandbook/1/).}. 

HI maps of NGC\.4449 were obtained as part of the THINGS \citep[The HI Nearby Galaxy Survey,][]{Walter2008} project, which observed about three dozen 
nearby galaxies with the NRAO VLA\footnote{Fully-processed maps are publicly available at http://www.mpia.de/THINGS/Overview.html}. The highest angular resolution velocity--integrated HI maps for NGC\,4449, obtained with the `robust weighted data' \citep{Walter2008}, have a beam of 13.74$^{\prime\prime}\times$12.50$^{\prime\prime}$, and are calibrated in units of Jy~beam$^{-1}$~m~s$^{-1}$. 

\subsection{Additional Processing and Photometry}

After verifying the alignment  in sky coordinates, we convolve all the images to common resolution, using the kernels of \citet{Aniano2011}.  We then resample all images to common pixel sizes that we refer to as `spaxels', since a crude spectrum made of multi--wavelength photometry is associated with each resampled pixel. The spaxels are used for the spatially--resolved analysis. We select two resolutions for our photometry: that of the SPIRE 500~$\mu$m images for large--scale analysis; and that  of the 
PACS 160~$\mu$m images for the spatially--resolved analysis (Table~\ref{tab2}). The latter choice is dictated by a compromise between exploiting the excellent angular resolution of the AzTEC images, while preserving the largest possible number of photometric data points for obtaining reliable fits of the dust SEDs. 

The two resolutions dictate the choice of photometric measurements included in each dust SED fit: for the large--scale fits we include all dust measurements between 8~$\mu$m and 1100~$\mu$m, while for the spaxel--based fits, we include only measurements at 8, 24 (from the SST), 70, 100, 160~$\mu$m (from the HSO), and 1100~$\mu$m (from the LMT). 
Table~\ref{tab2} lists, for each of the dust emission bands used in this work, the central wavelength, the facility/instrument combination that has provided the measurement, the PSF's FWHM,  the whole--galaxy photometry\footnote{For the whole galaxy photometry at mm wavelengths, we measure the total flux within an aperture of diameter 360$^{\prime\prime}$ on the AzTEC map; this is the diameter that encompasses all pixels 
detected with signal--to--noise S/N$\ge$3 in the HSO/SPIRE500 map of NGC\, 4449. For the other images, which are significantly larger than the extent of the galaxy, we use the curve--of--growth technique. In all cases, the image/sky   background is subtracted.} and the photometry of a central rectangular region with projected size 4.7$\times$4.7~kpc$^2$, and de--projected size 4.7$\times$12.5~kpc$^2$ for a 68$^o$ inclination \citep[][Table~\ref{tab1}]{Hunter2002, Hunter2005}.  

The central rectangular region is chosen to include all AzTEC pixels with signal--to--noise S/N$>$3.5 that also have PACS 160~$\mu$m values with S/N$>$10, when the data are binned in spaxels of 11$^{\prime\prime}$ squares. Not 
coincidentally, this is also very close to the region imaged by \citet{Kohle1998} and \citet{Kohle1999} in CO(1--0) \citep[see Figure~1 in][]{Bottner2003}. In section~6.2, we use this CO measurement to derive a M(H$_2$)--L(CO) conversion factor for the central region. 

We compare our photometry in the AzTEC band with similar measurements available either from archives or from the literature.  For the whole galaxy, we use the Planck photometry at the position of NGC\, 4449 \citep{Planck2016}, as retrieved from the Planck Legacy Archive. We employ the archival photometry extracted using an aperture with Gaussian weights and with axes of about 5$^{\prime}$,  close to the FWHM of the effective beam in the four bands at wavelengths between 350~$\mu$m and 1,380~$\mu$m. The Planck photometry is listed in Table~\ref{tab2}. Interpolation between the two Planck bands at 850~$\mu$m and 1,380~$\mu$m predicts a flux of 0.53~Jy at 1,100~$\mu$m for this galaxy, implying that our recovered flux of $\sim$0.4~Jy (Table~\ref{tab2}) is a little over 70\% of total. Although the Planck data may suffer from modest flux increase due to potential foreground and background source contamination \citep{Dale2017},  our AzTEC observations are below the Planck interpolated flux, although not significantly when uncertainties are included.  Some of the discrepancy may be due to the fact that we do not recover all the extended mm emission from the galaxy.  
This is expected, due to the large spatial scale filtering applied to the mm data in order to remove contributions from the atmosphere (see section~2.1); this process also filters out extended, low--surface--brightness emission from the galaxy, and is a common outcome of all ground-based millimeter and submillimeter camera systems. The size of the AzTEC map, which is only slightly larger than the galaxy itself, may also contribute to underestimating the total flux. We further verify this hypothesis in section~5, where we model the low S/N spaxels in the AzTEC image. For the central region, we compare our AzTEC photometry, which yields a flux density of 0.25$\pm$0.09~Jy (Table~\ref{tab2}),  with the value of 0.26$\pm$0.04~Jy obtained by \citet{Kohle1999} at 1.2~mm for roughly the same region, after removal of the CO(2--1) line contamination \citep{Bottner2003}. The two values are identical within the uncertainties.

We note that while the HSO/PACS100 and HSO/PACS160 whole galaxy photometry are virtually identical to the IRAS100 \citep{Hunter1989} 
and SST/MIPS160 photometry, respectively, the HSO/PACS70 is $\sim$30\% higher than the SST/MIPS70 measurement. The discrepancy between MIPS70 and PACS70 has already been reported for other galaxies \citep{Aniano2012, Draine2014}, with striking variations observed in M31, where the ratio between the two bands covers the range 0.5--2 across the body of the galaxy, with no obvious correlation with any specific property \citep{Draine2014}. Figure~\ref{fig3} shows the ratio between the HSO/PACS70 and SST/MIPS70 images for NGC4449, in spaxel  20$^{\prime\prime}$ in size \citep[comparable to the FHWM of the MIPS70 PSF,][]{Aniano2011}. We observe spaxel--to--spaxel  variations in the PACS70/MIPS70 ratio between $\sim$0.9 and 1.8, about half as large as those observed in M31. We use the PACS70 data as our default in what follows, so as to leverage the higher angular resolution of the HSO, but bearing in mind this potential difficulty. In section~6.1, we also assess the impact on the inferred dust masses of using the SST/MIPS70 photometry instead of the HSO/PACS70 one.

For the spaxel--based analysis, we make use of the HI map at the original resolution, since the FWHM of the HI beam is close to the FWHM of the Herschel  PACS 160~$\mu$m PSF. 

We divide the central region into 21$\times$21 square spaxels with size 11$^{\prime\prime}$ each. These are the primary resolution elements used in our spatially resolved analysis. Before binning the central region, we convolve all images to the HSO/P160 PSF \citep{Aniano2011}, with the exclusion of the SPIRE images and Planck data which are not included in the spatially resolved analysis.   Out of 441 total spaxels, 429 have S/N$\ge$10 at 8, 24, 70, 100, and 160~$\mu$m. The flux distribution of the 429 spaxels in the AzTEC image is 
consistent with a gaussian, as expected; 42 spaxels ($\sim$10\%) have S/N$\ge$3.5, 131 have S/N$\ge$2, and 91 have negative or null 1100~$\mu$m flux. We also analyze larger spaxels with sides of 22$^{\arcsec}$, 33$^{\arcsec}$, and 44$^{\arcsec}$ (2, 3, and 4 times the size of the original spaxels). This binning provides information as a 
function of increasing spatial scale, in the range $\sim$360~pc--1.5~kpc. At the largest binning size, there are 25 spaxels in the central region. 
We do not attempt to go beyond 1.5~kpc, as the number of spaxels becomes too small for statistical purposes.

In section~6, we use the photometry of both the total galaxy and of central region (see above and Figures~\ref{fig1} and \ref{fig2}) from 8~$\mu$m to 1100~$\mu$m  to perform SED fits and derive physical parameters, including dust masses.  These are used  to validate the procedure we utilize later to derive spaxel--based dust mass surface densities. In what follows, photometric values are expressed as F($\lambda$)=$\nu$F$_{\nu}$($\lambda$), in units of erg~s$^{-1}$~cm$^{-2}$, unless otherwise noted.

\section{Star Formation Rates}

We derive dust-corrected, spaxel--based SFRs using the hybrid indicator that combines H$\alpha$ emission with 24~$\mu$m emission. We use the calibration of \citet{Calzetti2007}, which is appropriate for HII regions and sub--kpc star forming regions:
\begin{equation}
SFR(M_{\odot} yr^{-1}) = 5.4\times 10^{-42} [L(H\alpha) + 0.031 L(24)],
\end{equation}
where both luminosities are in units of erg~s$^{-1}$, 
and the SFR is calculated in each spaxel. The calibration factor in this equation assumes a \citet{Kroupa2001} stellar Initial Mass Function (IMF) in the stellar mass range 0.1--100~M$_{\odot}$. The SFRs are then converted to SFR surface densities, $\Sigma_{SFR}$, by dividing the SFR by the de--projected area of each spaxel. We use the inclination of 68$^o$ from Table~\ref{tab1}. The advantage of using ionized gas and dust emission for deriving the SFR is that both ionized and neutral gas have the same inclination relative to the plane of the sky for this galaxy \citep{Hunter2002}. Stars have a different inclination in NGC\, 4449, and could be viewed face--on, if the galaxy is dominated by a strong bar \citep{Hunter2005}. This characteristic makes the use of hybrid SFR indicators that combine direct stellar light (e.g., UV) with dust emission (e.g., 24~$\mu$m emission) more complicated to use, and we avoid this.

At each position, the H$\alpha$ light and 24~$\mu$m light may receive contribution from populations other than those associated with the  local current star formation, and which may form a diffuse background to the actual light from star formation \citep[e.g.,][]{Calzetti2005}. In the case of H$\alpha$, ionizing photon leakage from massive stars at other locations (up to $\gtrsim$1~kpc) can contribute to the observed ionized gas emission (commonly known as diffuse ionized gas, or DIG). Part of the 24~$\mu$m emission can be contributed by non--equilibrium heating of the dust by pre--existing stellar populations, older than the typical age of the current event of star formation \citep{Calzetti2005,Draine2007,Calapa2014}. The impact of these spurious contributions to $\Sigma_{SFR}$ is quantified in \citet{Calzetti2012}. An estimate of the contribution of older stellar populations to the heating at 24~$\mu$m  is obtained from the best fit solution to the dust SED of the central region, presented in section~6.2:  about 30\% of the emission at this wavelength is provided by the general, diffuse starlight (section~4), which is unrelated to the current star formation.

We use GALFIT \citep{Peng2002, Peng2010} to produce a smooth map of the galaxy from  the 4.5~$\mu$m image; this step removes the low--level, small--scale contribution of recent star formation to the 4.5~$\mu$m emission, and produces an image  that traces the old, distributed stellar population \citep{Kendall2008}. The GALFIT best fit parameters include an exponential disk (Sersic index=1.2), with effective radius $\sim$1.4~kpc and position angle of 45$^o$. The effective radius (corresponding to a scale length of $\sim$0.8~kpc) and the position angle we recover are in agreement with the parameters of the stellar disk obtained by \citet{Hunter1999} and \citet{Hunter2002}. The smooth stellar disk map produced by GALFIT is then subtracted from both the 24~$\mu$m and H$\alpha$ images, by removing 30\% and 10\% of the total emission from the central region at 24~$\mu$m and H$\alpha$, respectively  \citep{Calzetti2005,Draine2007,Liu2011,Calzetti2012}.  The 30\% contaminating fraction to the 24~$\mu$m emission is discussed in the previous paragraph.  The 10\% fraction of diffuse H$\alpha$ emission is consistent with the fraction of DIG present in starburst galaxies  \citep{Oey2007}. With these fractions, the number of oversubtracted spaxels, i.e.,  containing zero or negative SFR, is less than 5\%. The negative/null spaxels are distributed at the edges of the central region, in agreement with most of the star formation being clustered towards the center of the galaxy. After subtraction of the smooth background emission, we combine the H$\alpha$ and 24~$\mu$m emission to obtain dust--free $\Sigma_{SFR}$ estimates at each spaxel. We verify that in NGC\, 4449 most of the emission from star formation emerges directly, instead of being reprocessed by dust. Indeed, for 85\% of the 11$^{\arcsec}\times$11$^{\arcsec}$ spaxels, 60\%--to--100\% of the star formation is recovered through the H$\alpha$ emission and less than 40\% through the 24~$\mu$m emission. For the remaining 15\% pixels, the light from star formation emerges through each of the H$\alpha$ and 24~$\mu$m emission in roughly equal fractions.

\section{Models and Fits of the Infrared Spectral Energy Distribution}

For our infrared SED fits, we employ the models of \citet{DraineLi2007}, as implemented in \citet{Draine2007}\footnote{The models are publicly available, and can be retrieved from:
https://www.astro.princeton.edu/~draine/dust/irem.html}. The models consist of mixtures of dust grains that combine 
carbonaceous grains (including PAHs) and amorphous silicates, with size distributions that aim at reproducing the Milky Way (MW),  
Large Magellanic Cloud (LMC), and Small Magellanic Cloud (SMC) extinction curves, respectively. For each grain distribution (extinction curve), a range of PAH 
dust mass fractions is considered, between q$_{PAH}$=0.01\% and 4.6\%, in several discrete values. The highest q$_{PAH}$ value is consistent with Milky Way--type dust, while the SMC and LMC--type dust have lower q$_{PAH}$ values.  In the models, the size distribution of PAHs and the ratio of neutral to ionized species are held constant, consistent with results that show little variation in the neutral/ionized PAH ratio in galaxies that do not host AGNs \citep{Smith2007}. Even in low metallicity galaxies like the Small Magellanic Cloud, which has a factor $\sim$1.5 times lower metal content than NGC\, 4449, the PAH estimates based on the 8~$\mu$m photometry are in good agreement with those based on Spitzer/IRS spectroscopy \citep{Sandstrom2010}.

The dust mixture is heated by the combination of two starlight intensity 
components:  the diffuse starlight that permeates the interstellar medium, described by the energy density parameter {\em U$_{min}$}, and a range of regions with a power law 
distribution of intensities, $dM_{dust}/dU \propto U^{-\alpha}$, between  {\em U$_{min}$}, and {\em U$_{max}$}  and slope $\alpha$=$2$. The two starlight intensity components are added together in proportion to ($1-\gamma$) and $\gamma$, where 0$\le\gamma\le$1. The parameter $\gamma$ is related to the fraction of starlight intensity due to 
current star formation or other activity. A more intuitive parameter is f$_{PDR}$, the fraction of total dust luminosity radiated from regions  with U$>$10$^2$, as defined in \citet[][eq. 29]{DraineLi2007}; f$_{PDR}$ is a measure of the dust emission from the photodissociation regions (PDRs) of star--forming regions, and is a combination of 
$\gamma$, U$_{min}$, and U$_{max}$. We use f$_{PDR}$ instead of $\gamma$ in the analysis that follows. The emission in each band will also be proportional to the total amount of 
dust M$_{dust}$. Thus, for each given extinction curve, there are a total of five parameters for the models: q$_{PAH}$, U$_{min}$, U$_{max}$, $\gamma$ (or f$_{PDR}$), 
and M$_{dust}$.  The parameters increase to six, if we include the extinction curve as one of them.

For the large--scale SED fits (section~6), we employ photometric data in nine bands, implying three degrees of freedom (or two, if the extinction curve is included among the parameters to be fit). Thus, we have enough data points to avoid under--constrained fits. We implement $\chi^2$ minimization for the best fits \citep{DraineLi2007}, and derive the uncertainties on our best--fit quantities from the shape of the reduced $\chi^2$ probability distribution. For the $\chi^2$ minimization, we apply the standard technique of weighting each data point by the inverse of the square of the measurement uncertainty. For the spaxel--based SED fits (section~7), we employ photometric data from six bands, implying that the fits are under-constrained (zero degrees of freedom). Thus, for the spaxels we fix several free parameters using the large-scale results for guidance. 

\citet{Karczewski2013} and \citet{RemyRuyer2015}  derive q$_{PAH}\sim$2\% and 3\%, respectively, from the global photometry of NGC\, 4449. These authors include SST spectroscopy in their fits, although they do not have a mm point. Our SED fits for the global photometry (section~6.1) yield a best extinction curve/q$_{PAH}$ combination of MW/3.2\%.  Fits that use MW/2.5\% and LMC/2.4\% also yield acceptable results, with reduced $\chi^2$ values that are within a factor 1.5--2 of those of the MW/3.2\% parameter combination. All other combinations produce significantly larger (by factors 3 or more) reduced $\chi^2$ values, and we consider 
these unacceptable. On account of the sparse sampling in both extinction curve and q$_{PAH}$ in the models, it is difficult to assign an uncertainty to these parameters.  Our best fit values of q$_{PAH}\sim$2.4--3.2\% are in excellent agreement with those of \citet{Karczewski2013} and \citet{RemyRuyer2015}. The q$_{PAH}$ values from the best fits of \citet{Karczewski2013} and \citet{RemyRuyer2015} are consistent with NGC\, 4449 being a low--metallicity galaxy, where q$_{PAH}$ is expected to be lower than the MW q$_{PAH}\sim$4.6\%; however, NGC\, 4449 is still above the threshold of 12+Log(O/H)$\sim$8.1, below which the PAH emission is observed to be significantly suppressed in galaxies \citep{Smith2007, Draine2007}. In summary, for the modeling of the infrared SEDs, we adopt the MW extinction curve with a fraction of 3.2\% of PAHs. 

\section{Mid/Far-Infrared Color Trends}

\subsection{Color Trends Within NGC\, 4449}

The dynamic range covered by the AzTEC image for the spaxels with S/N$\ge$3.5 is relatively small, about a factor 2 (Figure~\ref{fig2}, top--right panel); furthermore, 90\%  of the spaxels in the central region have insufficient S/N to be considered reliable, i.e., they are below S/N=3.5. In order to increase the number of spaxels used in our analysis, we leverage properties of the infrared--mm SEDs of regions in this galaxy. Figure~\ref{fig4} shows that the flux ratios  F(70)/F(1100) and F(160)/F(1100) are anti--correlated with F(8)/F(24) for all the high S/N$_{1100~\mu m}$ spaxels. This can be understood if regions of high star formation rate correspond to increased values of F(24), F(70), and F(160) relative to F(8) and F(1100). F(8) and F(1100) have appreciable contributions from dust heated by the diffuse starlight, with F(8) being produced by single--photon heating, and F(1100) coming from the larger grains with nearly
steady temperatures maintained by the diffuse starlight. This is shown in Figure~{\ref{fig5}, where we derive the best SED fits for the two extreme, i.e., the lowest F(8)/F(24) and the lowest F(70)/F(1100), spaxels with S/N$_{1100~\mu m}\ge$3.5. The spaxel with the lowest F(8)/F(24) ratio corresponds to a region of high star formation rate surface density, which is contributing significantly to the heating of the dust (f$_{PDR}$=0.57). 
The spaxel with the lowest F(70)/F(1100) ratio corresponds to a region of very low star formation rate density, where the dust is entirely (within our uncertainties) heated by the general interstellar radiation field, i.e., f$_{PDR}$=0. Figure~{\ref{fig5} also identifies the location of both spaxels within the central region, confirming that they correspond to high and low local star formation, respectively. In summary, the trends marked by the high S/N$_{1100~\mu m}$ ratio spaxels in Figure~\ref{fig4} identify a smooth progression from regions of high star formation, with low F(8)/F(24) and high F(70)/F(1100) [or F(160)/F(1100)] ratios, to regions of low--to--negligible star formation, with high F(8)/F(24) and low F(70)/F(1100) [or F(160)/F(F(1100)] ratios. 

Comparison of the observed color--color distributions with the expectations from the \citet{DraineLi2007} models provides further support to the above picture (bottom two panels of Figure~\ref{fig4}). Spaxels with low F(8)/F(24) and high F(70)/F(1100) [or F(160)/F(1100)] correspond to models with high U$_{min}$, high U$_{max}$, and high $\gamma$ values, 
i.e., with a high contribution to the dust heating by current star formation. Conversely, spaxels with high F(8)/F(24) and low F(70)/F(1100) [or F(160)/F(1100)] correspond to models with relatively low U$_{min}$, low U$_{max}$, and low $\gamma$ values, as expected from a larger contribution to the dust heating by the diffuse starlight. The location of the data relative to the models is also consistent with the fact that the spaxels in the central region of NGC\, 4449 typically require values of U$_{min}>$2, as appropriate for an actively star--forming galaxy \citep[see section~7;][]{Draine2007}. The DL07 models do not produce values log[F(8)/F(24)]$>$0.5, because of the assumed properties for the PAH population in the models, with single--photon heating contributing  F(24)$\sim$ (1/3)F(8) (see Fig. 15 of DL07). This assumption does not correspond to a physical limitation, and, indeed, several of the spaxels in NGC\, 4449 have log[F(8)/F(24)]$>$0.5. 

When plotting the observed 8~$\mu$m to 24~$\mu$m luminosity ratio as a function of both the H$\alpha$ luminosity surface density and the H$\alpha$ to 4.5~$\mu$m luminosity ratio for all the 11$^{\prime\prime}\times$11$^{\prime\prime}$ spaxels in the central region, irrespective of the S/N$_{1100 \mu m}$ value, we find a general anti--correlation between the two quantities (Figure~\ref{fig6}).  The H$\alpha$ luminosity surface density is a proxy for the SFR surface density, while the ratio H$\alpha$/4.5~$\mu$m is a proxy for the spaxel--based specific SFR, since  the 4.5~$\mu$m emission traces the old, diffuse stellar population; both quantities measure the local strength of the star formation.

The strength of the correlations is evaluated by applying the non--parametric Kendall $\tau$ test to the data. In this work, we prefer Kendall $\tau$ over Spearman $\rho$, as it is less sensitive to outliers \citep{FeigelsonJogesh2012, Ivezic2014}; thus Kendall $\tau$ works well also in case of small datasets (e.g., section~7). Figure~\ref{fig6} lists, in each panel, the probability p$_{\tau}$ that the data are uncorrelated: small values of p$_{\tau}$ correspond to high probability of correlation, with p$_{\tau}$=0.05 indicating a 2~$\sigma$ significance level for a two--tailed test. In our case, the F(8)/F(24) luminosity ratio is highly anti--correlated with both $\Sigma$(H$\alpha$) and the F(H$\alpha$)/F(4.5) luminosity ratio, with a 6.4$\times$10$^{-18}$ and 4.4$\times$10$^{-25}$, respectively, likelihood that the pairs of data are drawn from a random distribution. Thus, more strongly star forming regions have lower F(8)/F(24) ratio, in agreement with the results of Figures~\ref{fig4} and \ref{fig5}.  As the plots show, this result holds for {\em all} spaxels in the central region, irrespective of their S/N$_{1100~\mu m}$. Regions of low 1100~$\mu$m flux, and, in first approximation, low surface dust density, are still well described by the behavior of the 8/24~$\mu$m color ratio observed for higher S/N$_{1100~\mu m}$ spaxels.

The smaller scatter of the F(70)/F(1100)--vs.--F(8)/F(24) correlation relative to the F(160)/F(1100)--vs.--F(8)/F(24) one in Figure~\ref{fig4} is also in line with the findings above, since the F(160) receives a  smaller contribution than F(70) from dust heated by recent star formation \citep[e.g.,][]{Calzetti2010, Li2013}.  A minimum $\chi^2$ fit through the highest (S/N$_{1100~\mu m}\ge$3.5) significance data of Figure~\ref{fig4} gives:
\begin{equation}
Log [F(70)/F(1100)] = 4.19 - \Bigl\{(0.65\pm0.05) Log[F(8)/F(24)] +(0.75\pm0.04)\Bigr\}^5,
\end{equation}
and
\begin{equation}
Log [F(160)/F(1100)] = 3.67 - \Bigl\{(0.72\pm0.05) Log[F(8)/F(24)] +(0.62\pm0.04)\Bigr\}^4.
\end{equation}
When including also the spaxels with S/N$_{1100~\mu m}\ge$2.0 in the 1100~$\mu$m emission, the fits still agree within the 1~$\sigma$ uncertainty. Thus, we consider these results sufficiently 
robust to be used to extrapolate F(1100), and therefore  dust masses, also to spaxels with insufficient S/N (S/N$_{1100~\mu m}<$3.5) from the AzTEC observations. 

\subsection{Comparison with Other Galaxies}

We compare the IR--mm relations of equations~4 and 5 with the trends of an independent galaxy sample. We concentrate on 
the relation between Log[F(8)/F(24)]  and Log [F(70)/F(1100)], which has the lowest dispersion of the two. \citet{Dale2017} collected and 
homogenized the UV, optical, infrared, sub-mm, and radio measurements for the galaxies in the KINGFISH sample \citep{Kennicutt2011}, augmented by 
the galaxies in the SINGS sample \citep{Kennicutt2003} that are not already in KINGFISH. The KINGFISH sample, and its parent SINGS sample, are representative of star forming 
galaxies within the local 30~Mpc \citep{Kennicutt2003}, and are, therefore, reasonable templates against which compare our spatially--resolved results.
 Measurements at 850~$\mu$m obtained with SCUBA are available for 27 of the 79 galaxies in \citet{Dale2017}; an additional 14 galaxies 
have Planck photometry.  Of these 41 galaxies, 38 were observed with both the SST and HSO, and the remaining three (MRK\,33, NGC\,5033, and NGC\,7552) with SST only. 

For the galaxies with Planck photometry, we simply interpolate between the two closest frequency bands to the AzTEC frequency.  For the galaxies with 
SCUBA data, we extrapolate the SCUBA 850~$\mu$m photometry to 1100~$\mu$m using the DL07 models appropriate for each galaxy\footnote{Assumptions, found sometimes in the literature, that the dust emission follows a Rayleigh-Jeans tail approximation beyond 500~$\mu$m should be taken with caution. The deviation of the HSO SPIRE 500~$\mu$m emission from the R-J tail approximation ranges from a factor 2.2 at 20~K to 68\% at 30~K.}, as described in \citet{Draine2007} and \citet{Dale2009}. 
For the 70~$\mu$m emission of the galaxies, we use the 
HSO  measurements when available; for the three galaxies with SST--only observations, we increase the MIPS70 values by 15\% (0.06 in log scale), to account for 
the slight systematic discrepancy in the photometry of the two instruments \citep[see photometric values listed in][]{Dale2017}\footnote{Our results do not change 
significantly if this small shift in flux is included or not for all three galaxies: MRK\,33, NGC\,5033, and NGC\,7552.}.  Photometry is corrected 
as discussed in section~2.2.

Figure~\ref{fig7} shows the results for the 41 KINGFISH$+$SINGS galaxies, in comparison with the S/N$_{1100 \mu m}\ge$2 spaxels from NGC\,4449.  The general locus 
occupied by the galaxies is consistent with that of the spaxels in NGC\,4449, supporting the use of equations~4 and 5 as tools to obtain dust mass measurements below the 
detection limits of the AzTEC observations. 

The KINGFISH$+$SINGS galaxies, however, span a smaller range of F(8)/F(24) colors than the spaxels in NGC\,4449,  with the maximum value for the galaxies being Log[F(8)/F(24)]$\sim$0.53. This is consistent with the picture that the KINGFISH$+$SINGS galaxies are star forming, while several spaxels in NGC\, 4449 are areas of little current 
star formation (and F(8)/F(24)$>$0.53, see, also, Figure~\ref{fig5}). The galaxy data also justify the assumption in the DL07 models to only include PAH populations that yield Log[F(8)/F(24)]$\le$0.5. In order  to apply the DL07 
models in a regime where they are valid, in the following sections we only use spaxels  with Log[F(8)/F(24)]$\le$0.55\footnote{We include in our fits data with F(8)/F(24) ratios about  10\% larger than the DL07 model limit of Log[F(8)/F(24)]=0.5, to make allowance for measurement uncertainties.}. The right panel of Figure~\ref{fig7} shows 
the distribution of F(8)/F(24) values for all the spaxels in the central region of NGC\,4449, irrespective of their S/N$_{1100 \mu m}$, and 91\% of them (total of 393) 
are below Log[F(8)/F(24)]$=$0.55, suggesting that our upper limit to the F(8)/F(24) ratio  is not expected to impact the generality of our results.

\subsection{Extending the Millimeter Flux to Low Significance Regions}

Figure~\ref{fig8} provides additional sanity checks for the model--derived 1100~$\mu$m flux. The left panel compares the original 1100~$\mu$m measurements, irrespective of their 
S/N$_{1100 \mu m}$ ratio, with the model--derived fluxes, from the application of equation~2. As expected, the spread between the two sets of data is large, but the two sets show a general positive trend, i.e., larger observed 1100~$\mu$m fluxes get assigned larger values also by the model, as also highlighted by the 1--to--1 relation drawn on the plot.  A Kendall $\tau$ test returns a probability of 2$\times$10$^{-30}$ that the two are drawn from a random distribution. This result, which indicates  a $\sim$14.5~$\sigma$ correlation, lends additional credibility to the model of equation~2. 

There are highly discrepant points from the general relation of Figure~\ref{fig8}, left, which we investigate further. Highly deviant spaxels, i.e. more than a factor 10 away from the 1--to--1 relation, have S/N$_{1100 \mu m}<$1.8, and they are  found at the edges of the starburst, in the NW and SE areas of the central region. Thus, the large deviations from our model flux predictions are attributable to the low significance of the measured fluxes. In all cases, the model of equation~2 provides a far more physical 1100~$\mu$m flux than the (low significance) measurements. Spaxels located in the bottom--right region of Figure~\ref{fig8}, left, have  unphysically high 1100~$\mu$m observed fluxes, i.e., fluxes that are larger than any of the shorter wavelength fluxes, while the spaxels located in the top--left region of Figure~\ref{fig8} have  too low 1100~$\mu$m observed fluxes, which would require unphysically low emissivity for the dust relative to that at shorter wavelengths. Based on all these considerations we consider the model of equation~2 a robust approach to derive 1100~$\mu$m flux in the low--significance spaxels of the central region of NGC\, 4449.

The right panel of Figure~\ref{fig8} compares the model 1100~$\mu$m fluxes with the 24~$\mu$m fluxes. This comparison is crucial for our analysis, since a linear 
correlation between the logarithm of the two quantities would undermine any conclusion relating the SFR to the gas density. The reason for this is because the derivation of the SFR includes 
use of the 24~$\mu$m flux and the surface density of gas is derived from the 1100~$\mu$m intensity (see below); however, F(24) and F(1100) are also linked by being measured from  the dust emission SED, and could, therefore, be covariant. The right panel of Figure~\ref{fig8} serves the purpose of verifying that the two quantities 
are not covariant in a simple way, i.e., through a linear correlation, although they are linked via a complex (polynomial) relation. From this, we conclude that {\em our results, based on the use of F(1100) for deriving dust (and gas) masses and of F(24) as the non--dominant contributor to the SFR, are not trivially covariant}. This is further helped by the fact that for most spaxels the SFR is mainly traced by H$\alpha$.

In section~7, we derive dust mass surface densities for all spaxels with Log[F(8)/F(24)]$\le$0.55, using directly the measured AzTEC 1100~$\mu$m flux for the 36 11$^{\prime\prime}\times$11$^{\prime\prime}$ spaxels with S/N$\ge$3.5 and equation~2 for the remaining 357 spaxels. Maps of the model 1100$\mu$m flux for the central region at 11$^{\prime\prime}$ and at 44$^{\prime\prime}$ resolution are shown in Figure~\ref{fig9}. The sum of the values in all the spaxels gives a flux density of 0.28~Jy, about 10\% higher than the value directly measured from the original AzTEC map, but well within the uncertainties of the central region 1100~$\mu$m flux. If we extend the model of equation~2 to all spaxels in the galaxy, we obtain a total 1100~$\mu$m  flux of $\sim$0.47~Jy. This is  a little over 20\% higher than the flux directly measured from the AzTEC image (Table~\ref{tab2}), although consistent within the uncertainties. We caution the reader that this result is obtained by applying equation~2 to spaxels that are outside the central region, with S/N$<$10 at 8, 24, and 70~$\mu$m, i.e. outside the regime where equation~2 has been derived. The trend for larger regions to yield larger discrepancies between the measured and modeled 1100~$\mu$m flux supports our earlier statement that some of the extended flux in the original image is lost due to spatial filtering. The flux of 0.47~Jy is about 10\% lower than value of 0.53~Jy from the interpolation of the Planck photometry. As already stated in section~2.3, this small discrepancy could be due to contamination from foreground and background sources in the Planck data \citep{Dale2017}.

\section{Global Dust Masses}

We derive both the total and central region (Figures~\ref{fig1} and \ref{fig2}) dust masses for NGC\,4449,  in order to constrain some of the free parameters in the DL07 models, to be used in the spaxel--based analysis. For the large--scale fits, we increase the number of constraints in the fits by using three additional photometric datapoints (the three HSO/SPIRE bands), that cannot be used for the spatially--resolved analysis due to resolution limitations.

\subsection{The Whole Galaxy}

For the whole galaxy photometry, the best fit values of $U_{min}$, U$_{max}$, and $f_{PDR}$ (Figure~\ref{fig10}, left, and Table~\ref{tab3}) are consistent with those derived by \citet{Draine2007} for the starburst galaxies in the SINGS sample \citep{Kennicutt2003}. Like for these authors, our results are not very sensitive to the specific value of U$_{max}$, although we continue to carry this as a free parameter. We do not include the Planck photometry in the fits, for uniformity of treatment of both the whole galaxy and central region. Figure~\ref{fig10}, left, shows the data points with the best fit from the models. Although not used for the fit, the Planck photometry is in general agreement with the model expectation: there is a small positive offset, by about 10\%, for the data longward of 350~$\mu$m, which has been discussed in the previous section. The best-fit  parameters for the entire galaxy remain virtually unchanged whether we include our mm data point or not, as the data from SPIRE provide sufficient constraints to the long--wavelength behavior of the SED. The total dust mass we derive (Table~\ref{tab3}) is consistent with the value of 3.5$\times$10$^6$~M$_{\odot}$ obtained by \citet{Karczewski2013}, once their value is rescaled to our preferred distance of 4.2~Mpc.  

If we perform the fits using the SST/MIPS70, rather than the HSO/PACS70, i.e., include a 30\% lower 70~$\mu$m photometric point, we obtain results that are consistent with those that use the PACS70 photometry, within the uncertainties. As an example, the best--fit dust mass now is M$_{dust}\sim$3.8$\times$10$^6$~M$_{\odot}$, about 12\% higher than the mass obtained when using the PACS70 point, but consistent with it within the 1~$\sigma$ error bar. 

Recent results from Planck indicate that the dust masses derived using the \citet{DraineLi2007} models need to be reduced by some factor, in order to reconcile the implied dust opacities with the measurements for the MW \citep{PlanckIntermediateXXIX2016, Fanciullo2015}. The reduction factor is a function of U$_{min}$, decreasing for increasing U$_{min}$ and with average value $\sim$2. The analysis in both papers only includes values of U$_{min}\lesssim$1, which are well below what we measure in NGC\,4449. A simple linear extrapolation of the observed trends for U$_{min}>$1 suggests that the reduction factor for our case is $\lesssim$1.5. We carry forward our analysis into the next section adopting two cases: a no--reduction case and a case in which dust masses need to be reduced by a factor $\sim$1.5 (Table~\ref{tab3}).

\subsection{The Central Region}

The best--fit model to the IR SED of the central region gives parameters that are similar to those of the whole galaxy, with the exception of the dust mass, which is about 2/3 in value (Figure~\ref{fig10}, right).  A result of the fit is also that at 24~$\mu$m, about 1/3 of the emission is contributed by dust heated by the diffuse starlight, which is not directly related to the current star formation event and needs to be removed when using the 24~$\mu$m emission as a SFR estimator. 

In order to further verify that our dust fitting parameters are reasonable, we derive a CO--to--H$_2$ conversion factor from the the dust mass of the central region, and compare it with other derivations in the literature.  The dust mass is related to the total (atomic$+$molecular) hydrogen mass via the relation:
\begin{equation}
M_{dust} = (D/H) M_H, 
\end{equation}
where $M_H=[M(HI)+M(H_2)]$ is the hydrogen mass, and D/H is the dust--to--hydrogen ratio, i.e., it is the dust--to--gas mass ratio uncorrected for the helium factor. We will  refer all our calculations in this section to the hydrogen mass, rather than the gas mass.  Taking into account the sub--solar metallicity of NGC\,4449, we expect:
\begin{equation}
D/H \sim 0.01 {(O/H)\over(O/H)_{MW}} \sim 0.003,
\end{equation}
\citep{Draine2007}, where $(O/H)/(O/H)_{MW}$ is the ratio of the oxygen abundance in NGC\,4449 to the Milky Way oxygen abundance, which we adopt to be 0.3 on average (Table~\ref{tab1}). Our assumed value for D/H is consistent with the preferred fit to the D/G--versus--metallicity data of \citet{RemyRuyer2014}, once we account for the factor 1.36  for helium and metals included in the models by those authors. The inferred hydrogen mass is listed in Table~\ref{tab3}; the uncertainties include the 20\% variation  within the central region due to the gradient in oxygen abundance. The HI mass (uncorrected for helium), as calculated from the THINGS map for the central region, represents about 40\% (or 60\% if we consider the Planck correction factor for the dust mass) of the hydrogen mass. Subtracting the HI mass from the total hydrogen mass yields the molecular hydrogen masses listed in Table~\ref{tab3}. The ratio M(H$_2$)/M(HI)$\sim$1.5 and 0.7, for the uncorrected and Planck--corrected dust mass cases, respectively, is consistent with the range of molecular--to--atomic hydrogen ratio found in nearby galaxies \citep{Leroy2005, Saintonge2011}. 

The relation between H$_2$ mass and CO luminosity is:
\begin{equation}
{M(H_2)\over M_{\odot}}=3.2 \Biggl({X_{CO}\over X_{CO,MW}}\Biggr) {L(CO) \over K~km~s^{-1}~pc^2},
\end{equation}
\citep[e.g.,][]{Bolatto2013}, where $X_{CO}$ is the `X'-factor, with the Milky Way value being 
$X_{CO,MW}$=2$\times$10$^{20}$~cm$^{-2}$~(K~km~s$^{-1}$)$^{-1}$; the factor 3.2 reflects our choice not to include helium corrections. The CO(1-0) luminosity of the central region is L(CO)= 1.02$\times$10$^7$~K~km~s$^{-1}$~pc$^2$ \citep{Bottner2003}, which yields the X--factors listed in Table~\ref{tab3}. 
Within the uncertainties, both values are consistent with the value $(X_{CO}/X_{CO,MW})\approx 11$\footnote{\citet{Bottner2003} do not provide an uncertainty for their $X_{CO}$ value, although they state that the dust mass from which it is derived has a large uncertainty, within a factor of 3.} determined by  \citet{Bottner2003} and with the range $5-25$ expected for galaxies with metallicity $\sim$1/3 solar 
\citep{Bolatto2013}. However, the model by \citet{Accurso2017} predicts $(X_{CO}/X_{CO,MW})\sim 7$ for a galaxy with the oxygen abundance and the Main Sequence offset of NGC4449; this value is marginally more consistent with the lower $X_{CO}/X_{CO,MW}$ derived with a 1.5 reduction factor on the dust mass (last column of Table~\ref{tab3}).  One additional consideration is that, based on the above, the reduction factor is unlikely to be as large as  the factor $\sim$2 appropriate for the diffuse ISM \citep{PlanckIntermediateXXIX2016},  which would yield a molecular gas mass consistent with zero within the uncertainties and inconsistent with the presence of strong star formation within this galaxy. 

Outside of the central region, the total hydrogen mass is between (3.6$\pm$1.5)$\times$10$^8$~M$_{\odot}$ and (2.5$\pm$1.1)$\times$10$^8$~M$_{\odot}$, the latter after inclusion of the Planck correction factor. In the same outer region, the HI mass is (1.9$\pm$0.1)$\times$10$^8$~M$_{\odot}$. Since the central region contains almost 90\%  of the star formation in the galaxy, we expect the external regions to contain little molecular gas, and the total hydrogen mass should be the same as the HI mass. In this case, the total hydrogen mass reduced by the 1.5 Planck factor is marginally more consistent with the HI mass than the unreduced case, leading to a  preference for using the former value over the latter. 

The results above, from the $(X_{CO}/X_{CO,MW})$ comparison within the central region and the M$_{dust}$/M(HI) comparison outside the central region, show better consistency when the dust masses derived from the \citet{DraineLi2007} models are decreased by a factor of 1.5. We adopt this correction in the rest of this work.

\section{Spatially--Resolved Dust and Gas Masses}

Spaxel--based dust masses are derived by fitting the luminosity densities at  8, 24, 70, 100, 160, and 1100~$\mu$m in each spaxel. Since we are only fitting six photometric data points for each spaxel, we are only allowed to have four free parameters in order not to overfit the data. Our results from section~6 enable us to constrain several model parameters, thus simplifying the fits. We fix the extinction curve/q$_{PAH}$ combination to be MW/3.2\%, like in the whole galaxy and central region best fits (section~6). As already mentioned above and in other papers \citep{Draine2007}, the derived dust masses are not sensitive to the choice of U$_{max}$, which we fix at a value of 10$^5$.  We find that most of the regions with S/N$_{1100 \mu m}\ge$3.5 are fit by U$_{min}$ in the range 2--8; we thus adopt U$_{min}=$5, as derived for the central region fit (section~6.2),  for all spaxels in the region; variations in the derived dust masses for different choices of U$_{min}$ in the range 2--8 are about 30\%, much less than our typical uncertainty in the dust mass value. In summary, we can fix four of the six free parameters in the models (extinction curve, q$_{PAH}$, U$_{min}$, and U$_{max}$) without losing generality; the two remaining parameters, f$_{PDR}$ and M$_{dust}$, are left as free parameters in the fits. Figure~\ref{fig11} shows, for the spaxels with S/N$_{1100 \mu m}\ge$3.5, the range of f$_{PDR}$ we fit as a function of the star formation rate surface density (section~3). Dust mass surface densities, $\Sigma_{dust}$, in units of M$_{\odot}$~pc$^{-2}$, are derived by correcting each area by the inclination of  68$^o$ (Table~\ref{tab1}). We further divide the dust mass surface densities by a factor 1.5, as discussed in section~6.  Figure~\ref{fig12} shows the maps of the fitted values of f$_{PDR}$ and $\Sigma_{dust}$ at 11$^{\prime\prime}$ and 44$^{\prime\prime}$ resolution, together with the maps of $\Sigma_{SFR}$. The sum of the dust masses derived individually for each spaxel is within 12\% of the dust mass derived from the global fit of the central region. This discrepancy is smaller than the typical uncertainties on the derived dust masses, including the global ones, supporting the robustness of our approach \citep[see discussion in][]{Galametz2012}.

The use of the 1100~$\mu$m data point, or other sub--mm/mm dust--dominated emission,  provides important constraints to the dust masses, as already found and discussed in \citet{Draine2007}. Figure~\ref{fig13} shows the effect of including or excluding the 1100~$\mu$m data in the SED fits of the spaxels with S/N$_{1100 \mu m}\ge$3.5. For this experiment, we have allowed U$_{min}$ to vary in the fitting. The fits yield higher dust surface densities when the 1100~$\mu$m data are excluded from the fits, with a median offset of about 60\%. The reason is that, in the absence of a sub--mm or mm data point, the best fits tend to yield lower U$_{min}$ values, and, therefore, larger dust masses, relative to the cases that include the long--wavelength data \citep{Draine2007}. If we exlcude the data at 1100~$\mu$m, the best fit U$_{min}$ is in the range 0.5--8 for the S/N$_{1100 \mu m}\ge$3.5 spaxels, as opposed to 2--8, as reported above. Figure~\ref{fig13} also shows a mild trend for larger offsets in dust surface densities for regions with larger $\Sigma_{SFR}$, which can directly impact any conclusion that relates SFRs to gas(dust) masses. Specifically, for $\Sigma_{SFR}>0.015$~M$_{\odot}$~yr$^{-1}$~kpc$^{-2}$, the dust surface densities are overestimated by about a factor 2, in the absence of the 1100~$\mu$m data point.

Gas mass surface densities, $\Sigma_{HI+H2}$, are derived from the dust mass surface densities using equation~4, after implementation of a metallicity--dependent gas--to--dust ratio (equation~5), using the metallicity gradient derived by \citet{Pilyugin2015}. Variations in the D/H ratio can be as large as a factor $\sim$3 within galaxies \citep{Roman-Duval2014, Roman-Duval2017}, which is larger than the systematic change from the metallicity gradient ($\Delta(O/H)/(O/H)<0.15$) across the Central Region;  however, they are smaller than the dynamic range of our gas mass data. We multiply all gas mass densities by 1.36 to include contribution from helium and metals, for easier comparison with results from the literature. The gas surface densities are then converted to H$_2$ surface densities, $\Sigma_{H2}$, by subtracting the HI (also corrected for He and metals) in each spaxel. We carry out our analysis on all positive  values of $\Sigma(H_2)$ that are larger than the combined uncertainty of total gas measurement and HI detection limit; we set this as our lower limit to $\Sigma(H_2)$, in order to avoid spurious values for the molecular gas surface density. Since  $\Sigma(H_2)$ results from subtracting $\Sigma_{HI}$ from $\Sigma_{HI+H2}$, spurious or unphysical values can arise when the gas mass surface density is close in value, within the uncertainties, to the HI surface density.  There are about 230 spaxels above this threshold (out of 393), or about 60\% of the available ones. This fraction, $\sim$60\%, of usable spaxels remains the same also at larger binning scales. Figure~\ref{fig14} shows the H$_2$/HI mass ratio as a function of the H$_2$ surface density; higher surface densities correspond to higher H$_2$/HI ratios, in agreement with the small dynamic range in HI mass surface densities  observed within galaxies \citep[e.g.,][]{Kennicutt2007, Bigiel2008}. The range of mass ratios  is similar for both the smaller (11$^{\arcsec}\times$11$^{\arcsec}$) and larger (44$^{\arcsec}\times$44$^{\arcsec}$) regions, and is in agreement with what is observed in most local galaxies \citep{Saintonge2011}. 

\section{Results and Discussion}

\subsection{The Relation between Star Formation and Gas}

Following previous authors \citep[e.g.,][]{Kennicutt1998, Kennicutt2007, Bigiel2008, Rahman2012, KennicuttEvans2012, Leroy2013}, we produce scatter plots of the surface density of the star formation rate as a function of both the surface density of molecular gas (H2) and of the total atomic$+$molecular gas (HI$+$H2). We fit linear relations to the scatter plots in logarithmic scale, using the Ordinary Least-Square (OLS) bi--sector linear fitting algorithm \citep{Isobe1990}. This algorithm is commonly employed in studies of the scaling relations of star formation, and we adopt it to facilitate comparisons with previous investigations\footnote{The value of the linear fit slope $\gamma$ depends on the algorithm used. \citet{Calzetti2012} show from simulations that the bi--linear regression fitting algorithm FITEXY (from the Numerical recipes) yields systematically larger values of $\gamma$ than the OLS bi--sector fitting algorithm, by $\Delta \gamma\sim$0.2--0.5 for an intrinsic linear slope $\gamma$=1.5, when reasonable uncertainties are included in the data. The differences are larger or smaller for higher and lower slope, respectively.}. The best fit lines through the data are expressed as:
\begin{equation}
Log(\Sigma_{SFR}) = \gamma_{H2} \times  Log(\Sigma_{H2}) + A_{H2}, 
\end{equation}
and 
\begin{equation}
Log(\Sigma_{SFR}) = \gamma_{HI+H2} \times Log(\Sigma_{HI+H2}) + A_{HI+H2}, 
\end{equation}
with $\gamma_{H2}$ and $ \gamma_{HI+H2}$ the best fit slopes, and A$_{H2}$ and A$_{HI+H2}$ the intercepts. The width of the distribution of the data about 
the best--fitting lines are also measured, in terms of standard deviations $\sigma_{H2}$ and $\sigma_{HI+H2}$. 

The scatter plots, together with the best fit lines, are shown in Figure~\ref{fig15} for the smallest and the largest spaxel sizes: 365~pc (11$^{\arcsec}$) and 1460~pc (44$^{\arcsec}$), and listed in 
Table~\ref{tab4} for all the four spaxel sizes considered here. The best fit parameters for the $\Sigma_{SFR}$--$\Sigma_{H2}$ relation are derived in two conditions:  (1) the fits are limited to the H2 surface densities that are above the threshold discussed in the previous section, i.e., those to the right of the hatched yellow (cyan) region in Figure~\ref{fig14}; these fits are listed in the first line for each region size in Table~\ref{tab4}; and (2)  the fits include data  down to 1/10th of the threshold (second line of Table~\ref{tab4} for each regions size, in parenthesis). The second sets of fits is performed to evaluate the stability of the fits with the more stringent limit.  

The Kendall $\tau$ test is applied to each set of $\Sigma_{SFR}$ and $\Sigma_{H2}$ (or $\Sigma_{HI+H2}$) pairs, to evaluate the strength of each correlation, and the resulting p$_{\tau}$ values are listed in Table~\ref{tab4}. For all spaxel sizes, the correlation between $\Sigma_{SFR}$ and $\Sigma_{H2}$ is generally weak, between 2~$\sigma$ and 3~$\sigma$. Conversely, the correlations between $\Sigma_{SFR}$ and $\Sigma_{HI+H2}$ are significant, with values between 4~$\sigma$ and 12~$\sigma$, for region sizes between 1.1~kpc and 360~pc, with only the largest regions at 1.5~kpc showing a weak (2.5~$\sigma$) correlation. The largest spaxels suffer from small number statistics which account for the low correlation significance.

The smallest region sizes we analyze, $\sim$360~pc in side, are sufficiently small to be affected by stochastic sampling of the stellar IMF, in the sense that the IMF may not be fully populated, especially at the lowest $\Sigma_{SFR}$ values \citep[see][and references therein]{Calzetti2013}. We evaluate the scatter introduced by stochastic IMF sampling by using the models for the ionizing 
photon rate calculated by {\citet{Cervino2002}. We convert these authors' values to our preferred \citet{Kroupa2001} IMF, and recall that only 60\% of the SFR is emerging directly in H$\alpha$. The remaining $\sim$40\% is captured by dust and measured through the 24~$\mu$m emission. This band receives most of its heating contribution from the non--ionizing UV radiation, which is about 3--4 times less sensitive to stochastic sampling than the ionizing radiation \citep[e.g.,][]{Andrews2013}. The combination of the two contributions of stochastic sampling to the ionizing and non--ionizing photons are shown in Figure~\ref{fig15} (top panels) as vertical bars: the effect of stochastic IMF sampling decreases for increasing $\Sigma_{SFR}$, and the scatter it produces  is generally smaller than the observed scatter in the data. Thus, stochastic sampling of the IMF has a small effect on our results, even at the smallest region sizes.

Our best fit lines are compared with similar results from the literature; where possible, we attempt to match the region sizes used by the authors we compare our results against. \citet{Kennicutt2007} measured the SFR-gas relation of $\sim$500~pc size star--forming regions in NGC\,5194, both for the molecular gas, obtaining $\gamma_{H2}$=1.37, and the total gas, obtaining $\gamma_{HI+H2}$=1.56. We compare these results with those of our 365~pc spaxels (top panels of Figure~\ref{fig12}). \citet{Bigiel2008} and \citet{Rahman2012} measured the SFR--molecular~gas relation of samples of nearby galaxies with spatial scale $\sim$0.75--1~kpc, obtaining $\gamma_{H2}$=0.95 and $\gamma_{H2}$=1.1, respectively. Similarly, \citet{Leroy2013} obtains a slope of 1$\pm$0.15, although \citet{Shetty2014b}, using a Bayesian approach to the fits, finds a sub--linear slope, $\gamma_{H2}\sim$0.7--0.8. 
We compare the Bigiel et al.'s and the Rahman et al's  results with our 1460~pc sized spaxels, although an equally good agreement would be obtained if we compared those authors' results with our 1~kpc spaxels. For the comparison with $\gamma_{HI+H2}$ at our largest spaxel size, we use the mean value derived for whole galaxies by \citet{Kennicutt1998}, $\gamma_{HI+H2}$=1.4. In contrast with this, we should note that \citet{Bigiel2008} find $\gamma_{HI+H2}$ to vary from galaxy to galaxy in the range 1.1--2.7, when analyzing regions $\lesssim$1~kpc in size. We briefly discuss these results in  section~8.3.

\subsection{Simulations}

We compare our results with the simulations of \citet{Calzetti2012}, for the relation $\Sigma_{SFR}$--$\Sigma_{H2}$. Many details are provided in that paper, and here we summarize only the elements relevant to the current analysis. In the simulations, realizations of galaxy regions with sizes in the range 0.2--5~kpc are randomly populated with molecular clouds distributed according to an exponentially decreasing filling factor (to favor low--filling factor regions, $<$15\%, as observed). The clouds are extracted from a  mass distribution with slope $\alpha$, in the range 500-M$_{max}$~M$_{\odot}$. We adopt here the same $\alpha=$2 as those authors, but we lower their value  of M$_{max}$ from 3$\times$10$^7$~M$_{\odot}$ (appropriate for large late--type spirals like NGC\,5194) down to $\sim$10$^6$~M$_{\odot}$. This value of M$_{max}$ is what is observed in the Large Magellanic Cloud \citep{Hughes2010}, and we consider it more appropriate for low--mass dwarfs as NGC\,4449. The simulated molecular clouds, furthermore, obey the Larson's Laws \citep{Larson1981}, which allow us to link their masses to sizes, and generate filling factors within spaxels; scatter in the parameters is added to the simulated data, as derived from observations. 

Star formation is related to the molecular cloud mass as:
\begin{equation}
SFR\propto M_{H2}^{\beta},
\end{equation}
with $\beta$ in the range [1,2]. The relation between SFR and M$_{H2}$ has a built-in scatter expressed as a gaussian in log--log space with a factor 2 standard deviation. Changing this standard deviation to a factor 4 or zero has minimal impact on the results \citep{Calzetti2012}. 
The choice of a direct relation between SFR and molecular gas mass, instead of a relation between the volume densities of these quantities, stems from a model limitation: Calzetti et al. simulate clouds, but not their internal structure, which is necessary to obtain a relation between $\rho_{SFR}$ and $\rho_{H2}$ \citep{Lada2010, Lada2012}. However, as discussed in that paper, a power law relation between volume densities translates into a power law relation between SFR and gas mass. At the most basic level, $\beta$=1 implies $\rho_{SFR}\propto\rho_{H2}$. The data on molecular clouds within $\sim$1~kpc of the Sun \citep{Heiderman2010, Evans2014} imply SFR$\propto$M$_{H2}^{1.3}$ and $\rho_{SFR}\propto\rho_{H2}^{1.7}$. 

The Schmidt--Kennicutt relation is then measured in the simulations following the same strategy as the observational approach, i.e., by deriving the parameters of equations~7 and 8 
 from the simulated data.
\citet{Calzetti2012} apply selection biases and detection `limits' which attempt to mimic as closely as possible typical observational conditions, including a dynamic range $\lesssim$1.5 in Log($\Sigma_{H2}$). This is consistent with the dynamic range we obtain for NGC\,4449, although, thanks to the IR/mm relations (equations~4 and 5) our range is about 0.5~dex wider than most previous studies.  

We compare the data for NGC\,4449 with simulations for  $\beta$=1 and $\beta$=1.5, which bracket the range of observed slopes. We also consider a third model, with $\beta$=1 and with a threshold for the star formation: the clouds in the lowest 18\% bin, by mass, of the mass function do not form stars. This third model aims at simulating the case in which our method for removing a  smooth background from the H$\alpha$ and 24~$\mu$m images (section~5), which averages to $\sim$18\% of the total emission from the two images, artificially removes some of the true star formation.

\subsection{Comparing Data with Simulations}

The slopes and scatter predicted by the simulations are compared with the observed values for NGC\,4449 as a function of increasing region size in Figure~\ref{fig16}. The same algorithms are used to measure $\gamma$ and $\sigma$ in the simulated and actual data, to ensure that they can be compared; we use the OLS bisector fitting routine to measure the slope $\gamma$ and we fit the distributions of perpendicular distances of the data from the best fit lines with Gaussians; we adopt the standard deviation $\sigma$ of the Gaussian as our measure of the spread of the data. Among the three models described above, only the case of $\beta$=1.5 consistently, albeit not perfectly, agrees with the observations. We do not attempt to improve the agreement between the $\beta$=1.5 simulation and the data because of parameter degeneracy. The trends of both $\gamma_{H2}$ and $\sigma_{H2}$ can be made steeper towards smaller spaxel sizes by either increasing (slightly) $\beta$ or by changing the slope of the cloud mass function, or by increasing the maximum value of the molecular cloud mass by a factor 2--3. The latter is within the regime of observed values for the LMC, which, as mentioned above, we take as being similar to NGC\,4449. 

The model comparisons in Figure~\ref{fig16} indicate that the observed trends for $\gamma_{H2}$ and $\sigma_{H2}$ are driven by the increasingly better sampling of the molecular cloud mass function at larger region sizes, as expected. They also suggest that the relation between SFR and cloud mass in NGC\,4449 is super--linear, which implies a super--linear relation between volume densities of the same parameters. A linear relation between the SFR and cloud mass (or between $\rho_{SFR}$ and $\rho_{H2}$) is excluded by the data at more than the 5~$\sigma$ level, from the combined trends of both $\gamma_{H2}$ and $\sigma_{H2}$.

From Figure~\ref{fig16}, the value of $\gamma_{H2}$ at 1,100~pc is lower than what would be expected for a smooth trend between $\gamma_{H2}$ and region size, although still consistent within the uncertainties. A close inspection of the data and fits does not reveal any unusual circumstance, beyond the known sensitivity of the OLS bi--sector fitting algorithm to outliers at the edges of the dynamic range of the data. Data on other galaxies will be required to fully understand this deviation.

Our observed trends for the relation between $\Sigma_{SFR}$ and $\Sigma_{H2}$ in NGC\, 4449 are in agreement with those found by \citet{Liu2011}. These authors analyze two star--forming galaxies, NGC\,3521 and NGC\,5194 (M\,51), and find that both the power law exponent and the scatter about the best fit lines of the molecular gas--SFR relation are a decreasing function of increasing region size, between 250~pc and 1.3~kpc.  We report the data of \citet{Liu2011} in Figure~\ref{fig16}. The data for M\,51 are systematically higher than those for NGC\,4449, while the data for the slope $\gamma_{H2}$ of NGC\,3521 are roughy consistent with those of our galaxy. While M51 is an almost face--on grand design spiral, NGC\,3521 is highly inclined. The methodology of  \citet{Liu2011} consists in identifying  the HII regions before removing the diffuse stellar light from older stellar populations, in order to measure the SFRs; this method is subject to line--of--sight confusion for highly inclined galaxies. We also report on Figure~\ref{fig16} the results of  \citet{Kennicutt2007}, \citet{Bigiel2008}, and \citet{Rahman2012}. These authors only measure one single region size, common to all galaxies, as specified in the caption of Figure~\ref{fig16}. \citet{Kennicutt2007} and \citet{Rahman2012} obtain results for $\gamma_{H2}$ (and, in the case of \citet{Kennicutt2007}, also for $\sigma_{H2}$) that are consistent with our results for NGC\, 4449, once we take into account the appropriate region sizes for the measurements. These authors use a methodology for removing the diffuse stellar emission from the SFR indicators that are the closest to our method. Conversely, the data of \citet{Bigiel2008} are the most discrepant from our results; we speculate that the reason for the discrepancy is that the fluxes used to derive the SFR indicators in that work include the contribution of diffuse emission from old stellar populations, which may yield artificially low values of the slopes and scatter. This effect is quantified in \citet{Calzetti2012}. The data from the spirals show both similarities and differences with our observed trends for $\gamma_{H2}$ and $\sigma_{H2}$. Some of the discrepancies could be intrinsic, but others could be due to differences in the measurement techniques. All results from the literature discussed here, for instance, use CO emission to measure the molecular gas mass, and do not include CO--dark molecular gas \citep{Pineda2013}. Conversely, our dust--based measurements include the CO--dark H$_2$. Meaningful comparisons between galaxies, thus, will first need to address the issue of getting homogeneous measurements.

Our main conclusion from the analysis of Figure~\ref{fig16} is that trends of $\gamma_{H2}$ and $\sigma_{H2}$ with region size are present both in the current and previous data, but had not been recognized before \citet{Liu2011}.

In contrast, the slope $\gamma_{HI+H2}$ between the surface densities of the SFR and total gas remains roughly constant as a function of region size (Figure~\ref{fig17}), and is in agreement with the slope observed for whole galaxies \citep{Kennicutt1998, KennicuttEvans2012} and for the star--forming regions in M\,51 \citep{Kennicutt2007}. The spread of the data about the best fit lines, $\sigma_{HI+H2}$ also remains roughly constant with region size (Table~\ref{tab4}), and comparable in value to the largest values observed for $\sigma_{H2}$; this suggests that the spread in HI$+$H2  is driven predominantly by the scatter in the HI component. The roughly constant values of both $\gamma_{HI+H2}$ and $\sigma_{HI+H2}$, when combined with the high significance of the  $\Sigma_{SFR}$--$\Sigma_{HI+H2}$ correlations, suggest that the onset of the super--linear trend between the SFR and the total gas surface densities  (the SK Law) occurs already at the sub--galactic scale, and that the atomic gas is an integral part in establishing the scaling between gas and SFR. In other words, the process that determines the balance between the atomic and molecular gas phases and between the gas and the SFR is established already at scales of a few hundred pc, much larger than the typical size of a single molecular cloud, but much smaller than the size of a galaxy. 

\citet{Bigiel2008} find the value of $\gamma_{HI+H2}$ to vary considerably from galaxy to galaxy, with a range 1.1--2.7, when performing their analysis using region sizes of 750~pc in each galaxy (Figure~\ref{fig17}). The authors' interpretation is that, since HI is not directly related to star formation, its influence on $\gamma_{HI+H2}$ is to make it a galaxy--dependent measure. This interpretation appears in contradiction with our result that $\gamma_{HI+H2}$ is independent of region's size in NGC\, 4449. In addition to the already--mentioned difference in the method to derive SFR indicators, the authors measure the molecular mass surface density directly from CO(2-1). We speculate that the large range of values they find may arise from the combination of region--dependent contribution of the old stellar population to the SFRs, spatially--variable CO(2-1)/CO(1-0) ratios \citep{Koda2012}, and uncertain CO--to--H2 conversion values. As already noted above, meaningful comparisons between different samples and galaxies will first require that homogeneous measurements are performed across all diagnostics. A larger sample of galaxies observed in the cold dust regime, thus capturing the entire gas component with one measurement technique, is  needed, in order to confirm or refute our speculations.

\citet{Hopkins2014} and \citet{Orr2017} perform high--resolution cosmological simulations of star--forming galaxies, finding that the Schmidt--Kennicutt relation arises naturally as an effect of feedback on local scales within the galaxies, irrespective of the specific star formation prescription used. The local stellar feedback, including supernovae, stellar winds, radiation pressure and photo heating, generates turbulence and acts as a self--regulating mechanism, which not only keeps star formation inefficient, but also imparts a scaling between gas (both total and molecular) and SFR. Our results tend to support this scenario: in NGC\,4449 we observe a balance between total gas and SFR, which occurs at all scales probed, from $\sim$350~pc to 1.5~kpc. A scale--free mechanism like turbulence would be able to account for this result, although there may be additional favorable conditions unique to the starburst environment of NGC\,4449, such a high filling factor for star formation. This has been shown to be a common feature of starbursts, as opposed to low--filling factors for normal star--forming disks \citep{Elmegreen2014}. The intense, localized star formation is likely to favor a scenario in which feedback--induced turbulence reaches equilibrium already at the small scales of a few hundred pc. Observations of larger, and more varied, samples of galaxies will enable testing this and other models.  

\section{Summary and Conclusions}

We have combined 1.1~mm (=1100~$\mu$m) maps of the nearby starburst galaxy NGC\,4449 from the LMT/AzTEC with images at infrared wavelengths from the Spitzer Space Telescope and the Herschel Space Observatory to investigate the distribution and properties of the dust in the central $\sim$4.7$\times$12.5~kpc$^2$ active region. Our main findings are:
\begin{enumerate}
\item the dust emission in this dwarf has similar characteristics to those of other nearby starburst dwarfs, including a relatively high value of U$_{min}\sim$4--5, indicating a relatively high mean dust temperature \citep{Draine2007}. We find that the dust masses derived from the \citet{DraineLi2007} models require a downward correction by a factor $\lesssim$1.5. This is in line with what has been found by the Planck Observatory for the dust in our Milky Way, for which correction factors around 1.8--2.0 have been determined \citep{PlanckIntermediateXXIX2016, Fanciullo2015}. The dust in NGC\,4449 appears to require a smaller correction factor than the Milky Way, likely a reflection of its higher mean starlight intensity. 

\item The mid--IR (F(8)/F(24)) color is correlated with the far--IR/mm (F(70)/F(1100) and F(160)/F(1100)) colors, for regions of sizes in the range $\sim$360~pc  (11$^{\arcsec}\times$11$^{\arcsec}$) to $\sim$1.5~kpc  (44$^{\arcsec}\times$44$^{\arcsec}$) in NGC\, 4449. This IR--mm correlation, which is recovered also for whole galaxies, enables us to increase the dynamic range of our dust mass determinations. We use high signal--to--noise data (mid--IR and far--IR data close to the peak of infrared emission) to derive the luminosity at 1100~$\mu$m for spaxels that are undetected or marginally detected by AzTEC. This extends our dust mass derivations down to one order of magnitude fainter limit in 1100~$\mu$m luminosity than would be possible by using only the mm data points. 

\item The total, HI$+$H$_2$, and molecular, H$_2$, mass surface densities, derived from the combination of dust and HI mass surface densities, are compared with the SFR surface density as a function of region size. The $\Sigma_{SFR}$--$\Sigma_{H2}$ scatter plots can be fit by power laws with exponents that decrease from $\sim$1.5 to $\sim$1.2 for increasing  region size from 360~pc to 1.5~kpc; the same decreasing trend is observed for the scatter of the data about the best fit lines. The $\Sigma_{SFR}$--$\Sigma_{H2}$ trend with region size is consistent with random sampling of both molecular clouds and star--forming regions within randomly--selected galactic regions \citep{Calzetti2012}. The presence of such trends had already been recognized by \citet{Liu2011}.

\item Conversely, the power laws describing the $\Sigma_{SFR}$--$\Sigma_{HI+H2}$ scatter plots show a constant exponent with value $\sim$1.5 and a constant scatter about the best fit lines, independent of region size. The dominant factor in the spread of the data is likely the scatter in the HI component. We also find that the $\Sigma_{SFR}$--$\Sigma_{HI+H2}$ correlation is significantly stronger than the  $\Sigma_{SFR}$--$\Sigma_{H2}$ correlation. We suggest that the constant slope  reflects the fact that both gas components play a role in determining the scaling with the SFR in this HI--dominated galaxy, and their balance is established at galactic scales that are larger than those of individual clouds, but considerably smaller than those of the whole galaxy. This is in agreement with a scenario in which the equilibrium between gas and SFR is set by feedback--induced turbulence \citep{Hopkins2014, Orr2017}.
\end{enumerate}

The results for NGC\,4449, albeit tantalizing, are still limited by having a single target. Analyses of larger samples of nearby galaxies, conducted with either the LMT or other millimeter facilities, will be required to place the results from this work on a more solid footing. The IR/mm color--color relation is of particular interest. If confirmed and calibrated for a significant number of star--forming galaxies and regions within galaxies, it will provide a venue for expanding the dynamic range of dust mass determinations. By using high--luminosity portions of the infrared spectral energy distributions of galaxies to infer their millimeter wavelength emission (which is intrinsically faint), this relation would increase the dynamic range of dust mass and surface density determinations by more than an order of magnitude relative to what possible with millimeter data alone. 

\acknowledgments

\section{Acknowledgments}

The authors would like to thank the anonymous referee for many constructive comments that have helped significantly improve the presentation of this paper.

Based on observations made with the Large Millimeter Telescope Alfonso Serrano, a binational project between Mexico and the United States of America, led by the Instituto Nacional de Astrof\'\i sica, Optica, y Electr\'onica (INAOE)  and by the University of Massachusetts at Amherst, respectively. 

This research has made use of the NASA/IPAC Extragalactic Database (NED) which is operated by the Jet
Propulsion Laboratory, California Institute of Technology, under contract with the National Aeronautics and Space
Administration.

This work is based in part on data downloaded from the Spitzer Heritage Archive. The Spitzer Space Telescope is operated by the Jet Propulsion Laboratory, California Institute of Technology under a contract with NASA. 

This work used data obtained from the ESA Herschel Science Archive. Herschel was an ESA space observatory with science instruments provided by European-led Principal 
Investigator consortia and with important participation from NASA. 

\facility{LMT(AzTEC)}

\appendix
\section{The Cottingham Method applied to AzTEC Data}
In this Appendix, we summarize the Cottingham \citep{Cottingham1987} method as applied to the AzTEC data. 

The Cottingham Method is a fitting approach that results in a maximum-likelihood estimate of parameters in a model of the temporally varying atmospheric contamination.  The approach begins with the despiked and calibrated time stream detector signals from each of AzTEC's detectors. The median value of each individual time stream is estimated and subtracted, an additional linear gain correction is applied to remove a long term time drift -- observed as variable tilt in the raw time streams -- and then the atmosphere coefficients, $\alpha$ for a B-spline basis matrix, $B$, (the atmosphere model) are found by solving the matrix equation:
\begin{equation}
AB\alpha = Ad^{\rm row},
\end{equation}
where the matrix $A$ is defined as
\begin{equation}
A = B^{\rm T}(1-P\Pi).
\end{equation}
$\Pi(P^TP)^{-1}P^T$ is known as the ``pointing matrix'' and $d^{\rm row} = (d_1, d_2, ..., d_{\rm nbolo})$ is a row vector holding the concatenation of the different detectors.  These equations are written assuming uncorrelated noise among the detectors, i.e., $N=I$ where $N$ is the noise covariance matrix.

Because of AzTEC's relatively small number of detectors, and that key detectors near the center of the array are inactive, the pointing matrix is not directly invertible for most of the data.  To solve this, we perform a first-order polynomial fit to the detector time stream residuals in azimuthal coordinates of the form
\begin{equation}
P^i_j(a,e) = A + ba^i_j + Ce^i_j
\end{equation}
where $a^i_j$ and $e^i_j$ are the azimuth and elevation coordinates of detector $i$ at sample $j$.  This is equivalent to fitting an instantaneous plane to the array sky brightness in azimuth and elevation to each set of detector samples.  Finally, the atmosphere cleaned timestreams are calculated by subtracting both the B-spline atmosphere template and the polynomial fit from the observed raw timestreams.  The reduction process then continues without modification from the standard AzTEC pipeline approach \citep[e.g.,][and references therein]{Scott2012}.

\subsection{Iterative Flux Recovery}
Although the Cottingham Method is an unbiased surface brightness estimator, some of the reduction steps, like the median estimate, the linear gain and
the polynomial fit, are susceptible to introduce some degree of bias.  In order to recover a map which better represents the astronomical surface brightness
distribution, the Flux Recovery Using Iterations Techniques (FRUIT) was used in a similar fashion as described in \citet{Liu2010}; the main difference relative to that paper is that we 
use the Cottingham Method instead of the standard PCA based atmospheric removal approach, together with some other minor modifications to the pixel selection criteria.

\clearpage

\begin{deluxetable}{rrr}
\tablecolumns{3}
\tabletypesize{\footnotesize}
\tablecaption{Adopted Parameters for NGC\,4449.\label{tab1}}
\tablewidth{0pt}
\tablehead{
\colhead{Parameter (Units)} &  \colhead{Value} & \colhead{Reference\tablenotemark{a}} 
\\
}
\startdata
\hline
Distance (Mpc)  & 4.2         & 1, 2         \\
Inclination (degrees) & 68 & 3, 4 \\
Stellar Mass (M$_{\odot}$)    & 1$\times$10$^9$    & 5       \\
HI Mass   (M$_{\odot}$)\tablenotemark{b} &  2$\times$10$^9$   &  6       \\
SFR (M$_{\odot}$~yr$^{-1}$)\tablenotemark{c}  & 0.5      &  7       \\
12$+$Log(O/H)\tablenotemark{d}  & 8.26 & 8 \\
Metall. Gradient (dex/kpc)\tablenotemark{d}   & $-$0.055          & 8    \\
\hline
\enddata

\tablenotetext{a}{1-- \citet[][from TRGB]{Karachentsev2003}; 2 -- \citet[][from TRGB]{Tully2013};  3 -- \citet{Hunter2002}; 4 -- \citet{Hunter2005}; 5 -- \citet{Calzetti2015}; 6 -- \citet{Huchtmeier1989}; 7 -- \citet{Lee2009};  8 -- \citet{Pilyugin2015}.}
\tablenotetext{b}{Atomic hydrogen mass associated with the galaxy.}
\tablenotetext{c}{Star formation rate from the extinction--corrected ultraviolet luminosity.}
\tablenotetext{d}{Central oxygen abundance and abundance gradient, respectively. The central oxygen abundance has an uncertainty of $\pm$0.01 \citep{Pilyugin2015}.}
 \end{deluxetable}

\clearpage

\begin{deluxetable}{rlrrr}
\tablecolumns{5}
\tabletypesize{\footnotesize}
\tablecaption{Infrared and mm flux densities for NGC\,4449.\label{tab2}}
\tablewidth{0pt}
\tablehead{
\colhead{Wavelength} &  \colhead{Facility/Instrument\tablenotemark{a}} & \colhead{PSF\tablenotemark{b}}  & \colhead{Galaxy \tablenotemark{c}} 
& \colhead{Central\tablenotemark{d}}  
\\
\colhead{} & \colhead{} & \colhead{}  & \colhead{Flux density} & \colhead{Flux Density} 
\\
\colhead{$\mu$m} &\colhead{} & \colhead{($^{\prime\prime}$)}  & \colhead{(Jy)}  & \colhead{(Jy)}    
\\
}
\startdata
\hline
8       & SST/IRAC               & 2.8      & 1.41$\pm$0.17  & 1.35$\pm$0.16      \\
24     & SST/MIPS               & 6.4      & 3.04$\pm$0.34  & 2.93$\pm$0.33  \\
70     & HSO/ PACS            &  5.7     &  51.2$\pm$ 6.8  &   47.3 $\pm$  6.2  \\
100   & HSO/ PACS           &  7.0      &  74.4$\pm$  9.1   &  68.5 $\pm$ 8.6 \\
160   & HSO/ PACS           & 11.2     &  74.2$\pm$  8.9   &  61.3$\pm$ 7.4 \\
 250  &  HSO/ SPIRE         &  18.2    &   34.3$\pm$  4.1 & 25.6$\pm$ 3.1  \\
 350  &  HSO/ SPIRE         &   24.9   &   15.5$\pm$ 2.0  &  11.3$\pm$  1.5 \\
 350  &  Planck                   &   278.   &    15.5$\pm$0.4 &  ... \\
 500  &  HSO/ SPIRE         &   36.1   &    6.0$\pm$   0.7  &  4.2$\pm$ 0.5 \\
 550  &  Planck                   &   290.   &    4.6$\pm$0.1 & ...\\
 850  &  Planck                   &  292.    &   1.24$\pm$0.09 & ...\\
 1100 & LMT/AzTEC\tablenotemark{e}          & 8.5       & 0.38$\pm$0.16   &   0.25$\pm$0.09  \\
 1380 & Planck                   & 301.     & 0.25$\pm$0.04 & ...\\
\hline
\enddata

\tablenotetext{a}{Facility/Instrument combination: SST= Spitzer Space Telescope, IRAC and MIPS Cameras; HSO=Herschel Space Observatory, PACS and SPIRE instruments; Planck=Planck Observatory, LMT=Large Millimeter Telescope with the AzTEC instrument.}
\tablenotetext{b}{Point Spread Function (PSF) FWHM. The SST and HSO PSFs are from \citet{Aniano2011}; the Planck PSFs (effective FWHMs) are from the Planck Collaboration, as listed at: https://wiki.cosmos.esa.int/planckpla/index.php/Effective\_Beams.}
\tablenotetext{c}{Flux density of the entire galaxy. Except for the Planck photometry, all measurements are performed on images matched in angular resolution to the HSO/SPIRE 500~$\mu$m PSF. The Planck photometry is retrieved from the Planck Legacy Archive at: https://www.cosmos.esa.int/web/planck/pla.}
\tablenotetext{d}{Flux density of a central, rectangular region with size $\sim$4.7$\times$4.7~kpc$^2$, corresponding to a de--projected size of $\sim$4.7$\times$12.5~kpc$^2$ (Figure~\ref{fig2}).  All measurements performed on images matched in angular resolution to the HSO/SPIRE 500~$\mu$m PSF.}
\tablenotetext{e}{The LMT/AzTEC flux density obtained by including only regions with S/N$\ge$3.5 in the map is 0.078$\pm$0.004~Jy. The sum area of these regions is $\sim$10\% of the area of the Central Region, or 1.46$\times$1.46~kpc$^2$ (corresponding to a de--projected size of 1.46$\times$3.90~kpc$^2$).}
 \end{deluxetable}

\clearpage

\begin{deluxetable}{lrrrr}
\tablecolumns{3}
\tabletypesize{\footnotesize}
\tablecaption{Derived Quantities for NGC\,4449.\label{tab3}}
\tablewidth{0pt}
\tablehead{
\colhead{Parameter (Units)} & \multicolumn{2}{c}{Galaxy \tablenotemark{a}}  & \multicolumn{2}{c}{Central Region\tablenotemark{a}}  
\\
}
\startdata
\hline
U$_{min}$\tablenotemark{b} & \multicolumn{2}{c}{4$\pm$1} & \multicolumn{2}{c}{5$\pm$1} \\
U$_{max}$\tablenotemark{b} &  \multicolumn{2}{c}{$\sim$10$^5$} & \multicolumn{2}{c}{$\sim$10$^5$} \\
f$_{PDR}$\tablenotemark{b} &  \multicolumn{2}{c}{0.12$\pm$0.04} & \multicolumn{2}{c}{0.15 $\pm$0.04}\\
M$_{HI}$ (M$_{\odot}$)\tablenotemark{c}  &  \multicolumn{2}{c}{(5.0$\pm$0.3)$\times$10$^8$} & \multicolumn{2}{c}{(3.1$\pm$0.1)$\times$10$^8$}\\
\hline
      &  D\&L07     &   D\&L07$+$Planck       &  D\&L07     &   D\&L07$+$Planck\\
\hline
M$_{dust}$ (M$_{\odot}$)\tablenotemark{d}  & (3.4$\pm$0.6)$\times$10$^6$ & (2.3$\pm$0.4)$\times$10$^6$ & (2.3$\pm$0.4)$\times$10$^6$ & (1.5$\pm$0.3)$\times$10$^6$\\
M$_{H}$ (M$_{\odot}$)\tablenotemark{e}  & (11.3$\pm$2.9)$\times$10$^8$ & (7.6$\pm$1.9)$\times$10$^8$ & (7.7$\pm$2.0)$\times$10$^8$\ & (5.1$\pm$1.3)$\times$10$^8$\\
M$_{H2}$ (M$_{\odot}$)\tablenotemark{f}  & (6.3$\pm$3.0)$\times$10$^8$ & (2.6$\pm$2.0)$\times$10$^8$  & (4.6$\pm$2.0)$\times$10$^8$ & (2.0$\pm$1.3)$\times$10$^8$\\
${X_{CO} / X_{CO,MW}}$\tablenotemark{g}  & \nodata & \nodata & 14.4$\pm$6.1 & 6.1$\pm$3.9\\
\hline
\enddata

\tablenotetext{a}{Derived quantities, from best--fit dust SED models \citep{DraineLi2007}, for the entire galaxy (second column) and for the central region as 
defined in the text and in Table~\ref{tab2} (third column).}
\tablenotetext{b}{The minimum and maximum, U$_{min}$ and U$_{max}$, energy density parameters and the fraction, f$_{PDR}$, of dust luminosity due to current star 
formation and other activity, as defined in \citep{DraineLi2007}. The SED fits are not strongly sensitive to the value of U$_{max}$.}
\tablenotetext{c}{HI mass, in solar masses and uncorrected for helium, derived from the THINGS maps of \citet{Walter2008}, using the relation 
M(HI)= 2.356$\times$10$^5$ D$^2$ $\Sigma_i$ S$_i\, \delta v$, 
where D is the galaxy's distance in Mpc and $\Sigma_i$ S$_i\, \delta v$ is the integral of the intensity along the line-of-sight velocity, in units of Jy~km~s$^{-1}$. 
The uncertainty in the HI mass estimate is $\sim$6\%, after including registration errors. For the `galaxy' measurement, the HI mass is determined in the same area used for the dust mass, and the resulting mass is about 1/4 of the total atomic hydrogen mass associated with the galaxy \citep{Huchtmeier1989}.}  
\tablenotetext{d}{The mass in dust for the entire galaxy and for the central region, in solar masses, as derived from the models of \citet[][D\&L07]{DraineLi2007} and from the same models reduced by a factor 1.5, to account for recent findings by the Planck Collaboration (D\L07$+$Planck, sections~6.1 and 6.2).}
\tablenotetext{e}{Total gas mass, uncorrected for helium, from the dust mass using equations~4 and 5. }
\tablenotetext{f}{The molecular hydrogen mass, uncorrected for helium, derived from subtracting M$_{HI}$ from M$_H$.}
\tablenotetext{g}{The ratio of the NGC\, 4449 CO--to--H$_2$ conversion factor, or X--factor $X_{CO}$, to the Milky Way's $X_{CO,MW}$, derived from equation~6, using the H$_2$ molecular hydrogen mass in this Table together with the CO luminosity measured  by \citet{Bottner2003}.  We adopt a Milky Way value $X_{CO,MW}$=2$\times$10$^{20}$~cm$^{-2}$~(K~km~s$^{-1}$)$^{-1}$.}
 \end{deluxetable}

\clearpage

\begin{deluxetable}{rrrrrrrrrrrr}
\tablecolumns{12}
\rotate
\tabletypesize{\footnotesize}
\tablecaption{Star Formation Rate and Gas Surface Density Correlations.\label{tab4}}
\tablewidth{0pt}
\tablehead{
\colhead{Region Size\tablenotemark{a}} &  \multicolumn{5}{c}{$\Sigma_{SFR}$ vs. $\Sigma_{H2}$\tablenotemark{b}} &  & \multicolumn{5}{c}{$\Sigma_{SFR}$ vs. $\Sigma_{HI+H2}$\tablenotemark{b}}  
\\
\\
  \cline{2-6}  \cline{8-12} 
\\
\colhead{(pc)} &\colhead{N\tablenotemark{c}} & \colhead{$\gamma_{H2}$\tablenotemark{c}} & \colhead{A$_{H2}$\tablenotemark{c}} & \colhead{$\sigma_{H2}$\tablenotemark{c}} & \colhead{p$_{\tau}$\tablenotemark{c}} & & \colhead{N\tablenotemark{d}}  & \colhead{$\gamma_{HI+H2}$\tablenotemark{d}} & \colhead{A$_{HI+H2}$\tablenotemark{d}} & \colhead{$\sigma_{HI+H2}$\tablenotemark{d}} & \colhead{p$_{\tau}$\tablenotemark{d}}  
\\
}
\startdata
\hline
360    &  229 & 1.49$\pm$0.13   & $-$3.81$\pm$0.12   & 0.41$\pm$0.05 &2.3E-03 &   & 390 & 1.51$\pm$0.05 & $-$4.18$\pm$0.06  & 0.37$\pm$0.03 &4.6E-34\\
           & (247) &  (1.34$\pm$0.09)   & ($-$3.61$\pm$0.08)   &                  &    &   &       &                              &                                      &                           & \\
730     &  55 & 1.35$\pm$0.16     & $-$3.68$\pm$0.15   & 0.33$\pm$0.06 &1.7E-02& & 90   & 1.57$\pm$0.12 & $-$4.25$\pm$0.14   & 0.36$\pm$0.04 & 1.5E-06\\
            & (59) & (1.17$\pm$0.14)    & ($-$3.41$\pm$0.15)   &                     &     & &       &                            &                                      &                            & \\
1100   & 26 & 1.06$\pm$0.14    &  $-$3.62$\pm$ 0.23  & 0.28$\pm$0.10  & 2.2E-02&   & 43 &1.42$\pm$0.21 & $-$4.17$\pm$0.23   & 0.30$\pm$0.07& 4.3E-05\\
            &  (27) &(1.02$\pm$0.10)  &  ($-$3.54$\pm$0.17) &                       &    &   &       &                            &                                     &                            &  \\
1460   & 13 &   1.18$\pm$0.21   &  $-$3.36$\pm$0. 15  & 0.20$\pm$0.10 &4.2E-02 & & 22 & 1.53$\pm$0.20 &  $-$4.06$\pm$0.20  & 0.37$\pm$0.06 &1.1E-02\\
            &  (13) & (1.18$\pm$0.21) &  ($-$3.36$\pm$0. 15) &                       &    &   &     &                             &                                     &                            & \\
\hline
\enddata

\tablenotetext{a}{Size of each spaxel in pc. The size is derived from the square root of the inclination--corrected area of the spaxel.}
\tablenotetext{b}{The star formation rate density $\Sigma_{SFR}$ is in units of M$_{\odot}$~yr$^{-1}$~kpc$^{-2}$. The gas surface density, both $\Sigma_{H2}$ and $\Sigma_{HI+H2}$, is in units of M$_{\odot}$~pc$^{-2}$. $\Sigma_{HI+H2}$ is derived directly from equation~4, and is directly proportional to the dust surface density.  
}
\tablenotetext{c}{The best--fit slope $\gamma_{H2}$ and intercept A$_{H2}$ obtained by applying  the OLS bi--sector linear fitting method to the N data in the $\Sigma_{SFR}$--$\Sigma_{H2}$ scatter plots; N is the total number of spaxels used in the fit; $\sigma_{H2}$ is the scatter of the data about the best fit line. The first line includes the best fits to the spaxels that have F(8)/F(24)$\le$0.55 and molecular gas values above a threshold calculated from the combined uncertainty of the total gas measurement and HI detection limit (section~7). The p$_{\tau}$ values indicate the probability that the data are uncorrelated. On the second line for each region, in parenthesis, we report also the number of spaxels used and the best fit values derived by including spaxels that have $\Sigma_{H2}$ values as low as 1/10th of the combined uncertainty.}
\tablenotetext{d}{The best--fit slope $\gamma_{HI+H2}$ and intercept A$_{HI+H2}$ obtained by applying  the OLS bi--sector linear fitting method to the N data in the $\Sigma_{SFR}$--$\Sigma_{HI+H2}$ scatter plots; N is the total number of spaxels used in the fit; $\sigma_{HI+H2}$ is the scatter of the data about the best fit line. The p$_{\tau}$ values indicate the probability that the data are uncorrelated. For the $\Sigma_{SFR}$--$\Sigma_{HI+H2}$ fits we use all spaxels with F(8)/F(24)$\le$0.55, since the total gas uncertainty is low enough that no spaxel is rejected.}
 \end{deluxetable}

\clearpage
\begin{figure}[h]
\figurenum{1}
\plotone{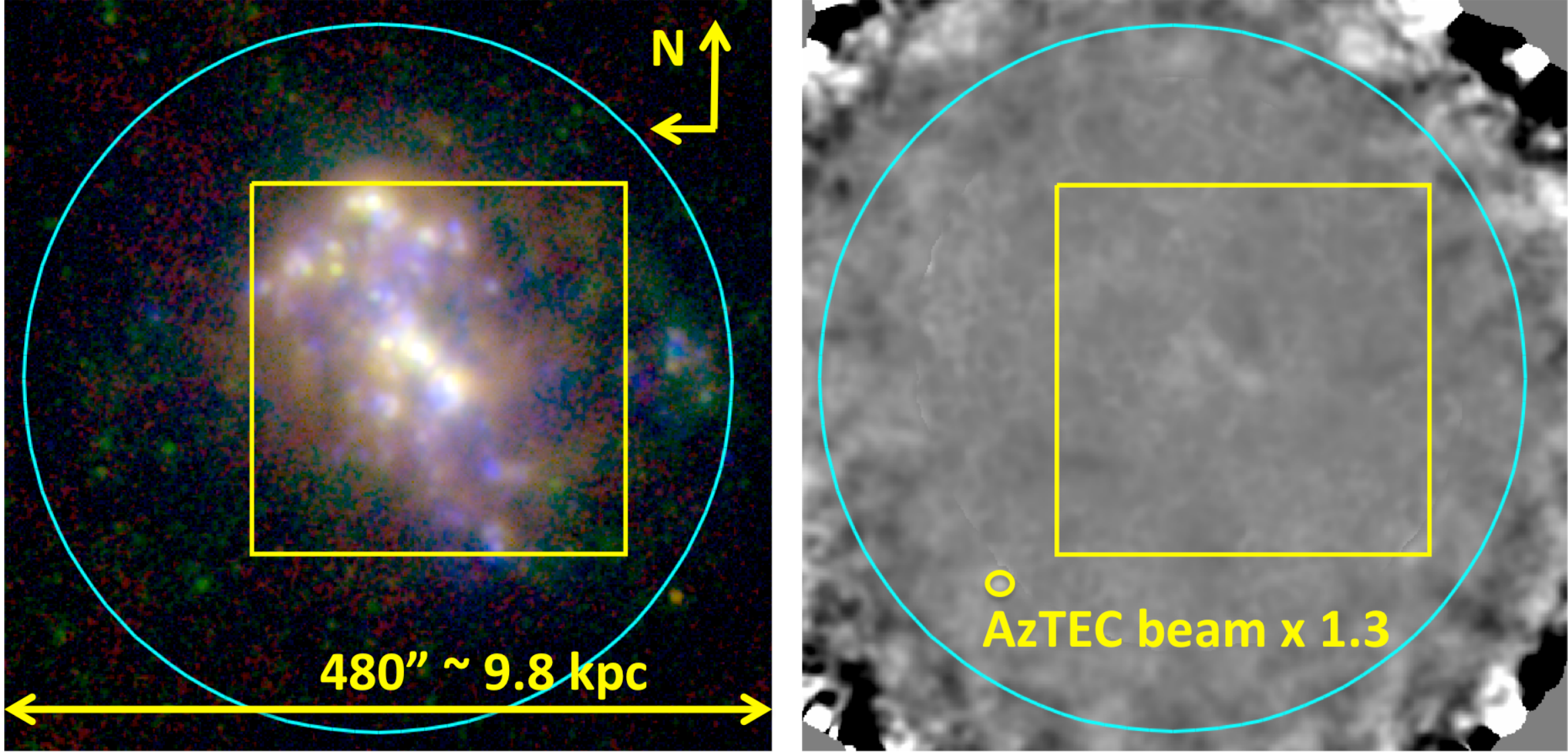}
\caption{A side--by--side comparison between a three--color composite (left) and the  AzTEC 1.1~mm emission (right) of NGC\,4449. The three colors in the left panel are: GALEX Far--ultraviolet (blue), SST/MIPS 24 $\mu$m (green), and HSO/PACS 70~$\mu$m (red) images, all at the native resolution of $\sim$5$^{\prime\prime}$--6.5$^{\prime\prime}$. The size of each panel is about 480$^{\prime\prime}$, corresponding to about 9.8~kpc at the distance of the galaxy. The cyan circle has a diameter of 450$^{\prime\prime}$, i.e., the size of the photometric aperture we use to measure the total flux from the galaxy. The yellow square marks the location and size of the Central Region analyzed in detail in this work. The dynamic  range of the AzTEC image shown in this figure is [$-$0.5,$+$0.5] MJy~sr$^{-1}$, and the size of the AzTEC beam is shown about 30\% larger than actual, for display purposes. North is up, East is left.} 
\label{fig1} 
\end{figure}

\clearpage
\begin{figure}[h]
\figurenum{2}
\includegraphics[scale=0.8]{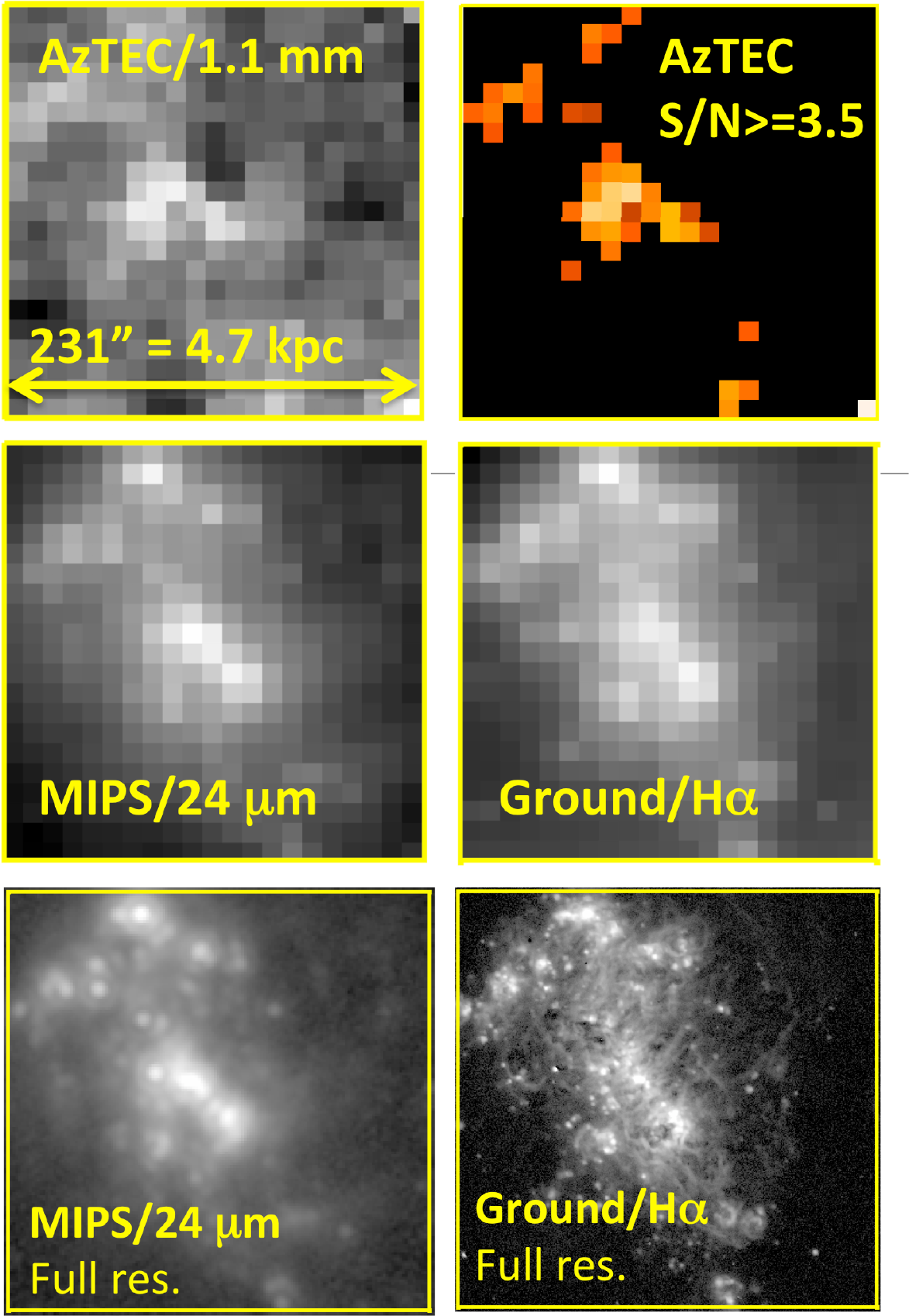}
\caption{Maps of the central region in NGC\, 4449 (the yellow rectangle in Figure~\ref{fig1}) at 1.1~mm, from AzTEC (top row), 24~$\mu$m, from SST/MIPS (center-- and bottom--left), and H$\alpha$, from ground--based (center-- and bottom--right). Top and center rows show the images after convolution to the HSO/PACS 160~$\mu$m PSF and resampling in spaxels of 11$^{\prime\prime}\times$11$^{\prime\prime}$. The images in the bottom row are shown at native PSF and resolution to help orient the reader. The AzTEC map is in linear scale, while the 24~$\mu$m and H$\alpha$ maps are shown with a logarithmic stretch. The dynamic range of the AzTEC image is [$-$0.5,$+$0.9] MJy~sr$^{-1}$. The low surface brightness areas of the 24~$\mu$m and of the H$\alpha$ images have been depressed to highlight the structure of the bright star-forming regions. The top--right panel, in heat scale, shows the location of the 11$^{\prime\prime}\times$11$^{\prime\prime}$ spaxels in the AzTEC image with signal--to--noise S/N$\ge$3.5. The central region, $\sim$4.7$\times$12.5~kpc$^2$ in de--projected size, is almost coincident with the area  mapped in CO(1-0) by \citet{Kohle1998, Kohle1999}. North is up, East is left. }
\label{fig2} 
\end{figure}

\clearpage
\begin{figure}[h]
\figurenum{3}
\plotone{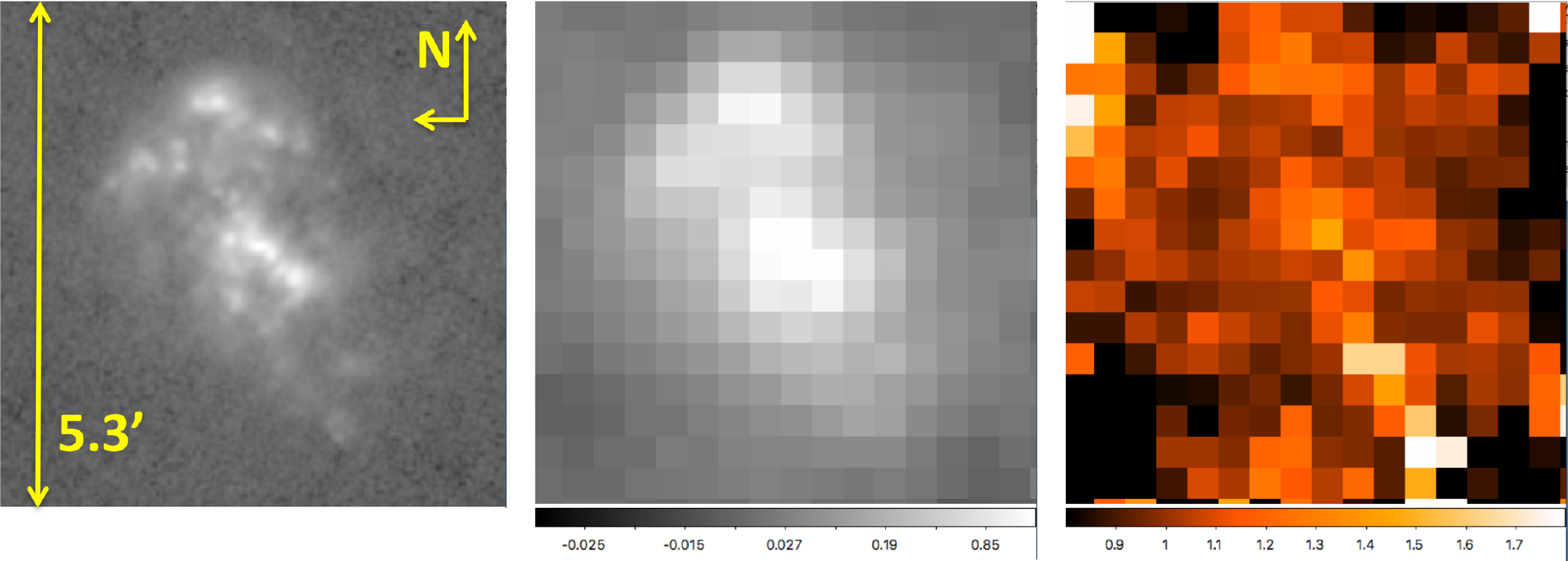}
\caption{A 5$^{\prime}$.3$\times$5$^{\prime}$.3 region centered on NGC\, 4449, showing: ({\bf Left}) the full resolution HSO/PACS70 image in grey scale; ({\bf Center}) the HSO/PACS70 image re-sampled to 20$^{\prime\prime}\times$20$^{\prime\prime}$ pixels, i.e., roughly the SST/MIPS70 PSF size, in grey scale; and ({\bf Right}) the PACS70/MIPS70 ratio image, also in 20$^{\prime\prime}\times$20$^{\prime\prime}$ pixels, in heat scale. The left and center images are shown to orient the reader as to the location of the galaxy relative to the ratio image. Scale bars are shown below the center image (in units of Jy) and the left image (adimensional). The PACS70/MIPS70 ratio is typically between 0.9 and 1.8, with a galaxy--wide ratio $\sim$1.3.} 
\label{fig3} 
\end{figure}

\clearpage
\begin{figure}[h]
\figurenum{4}
\plottwo{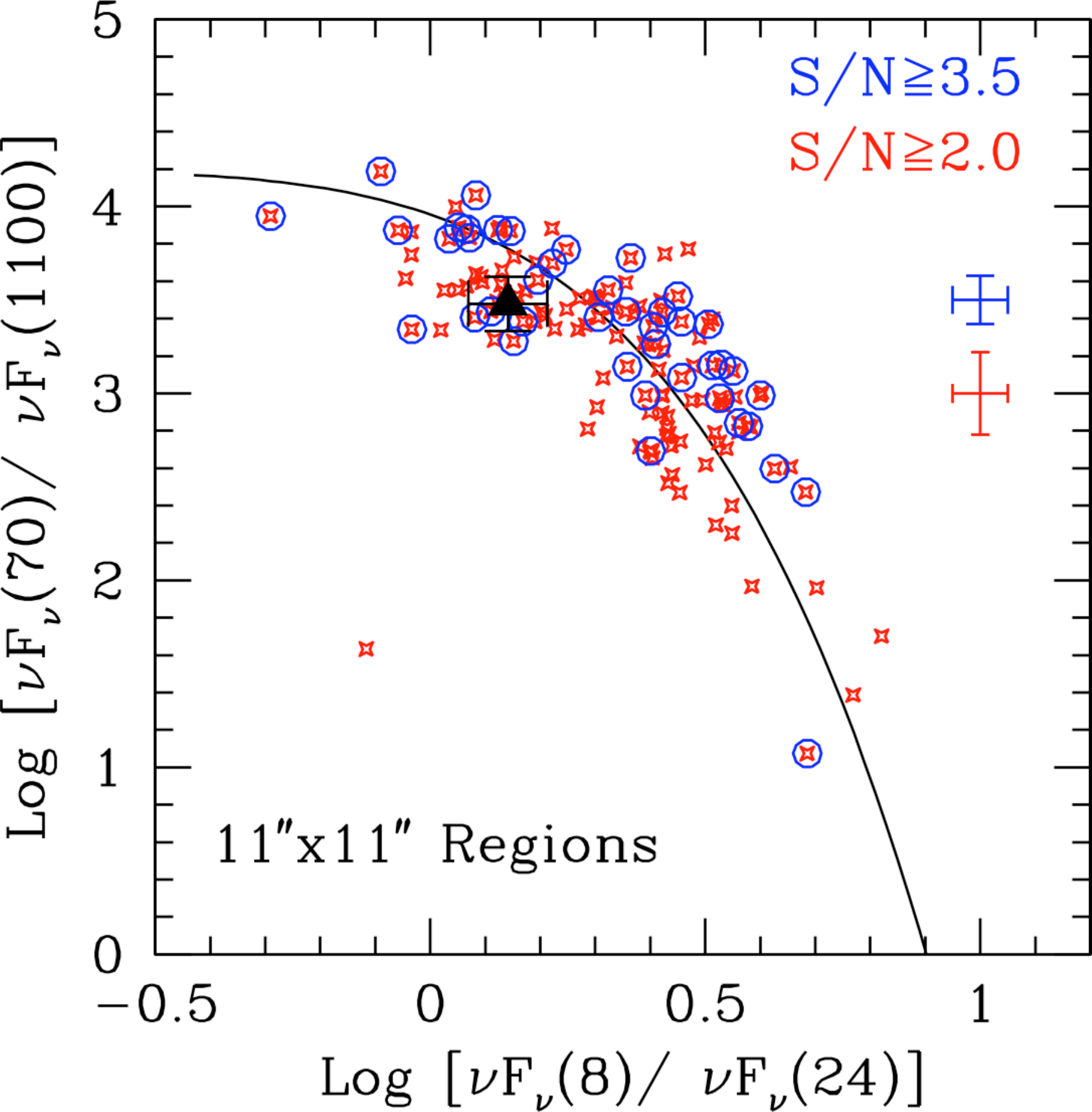}{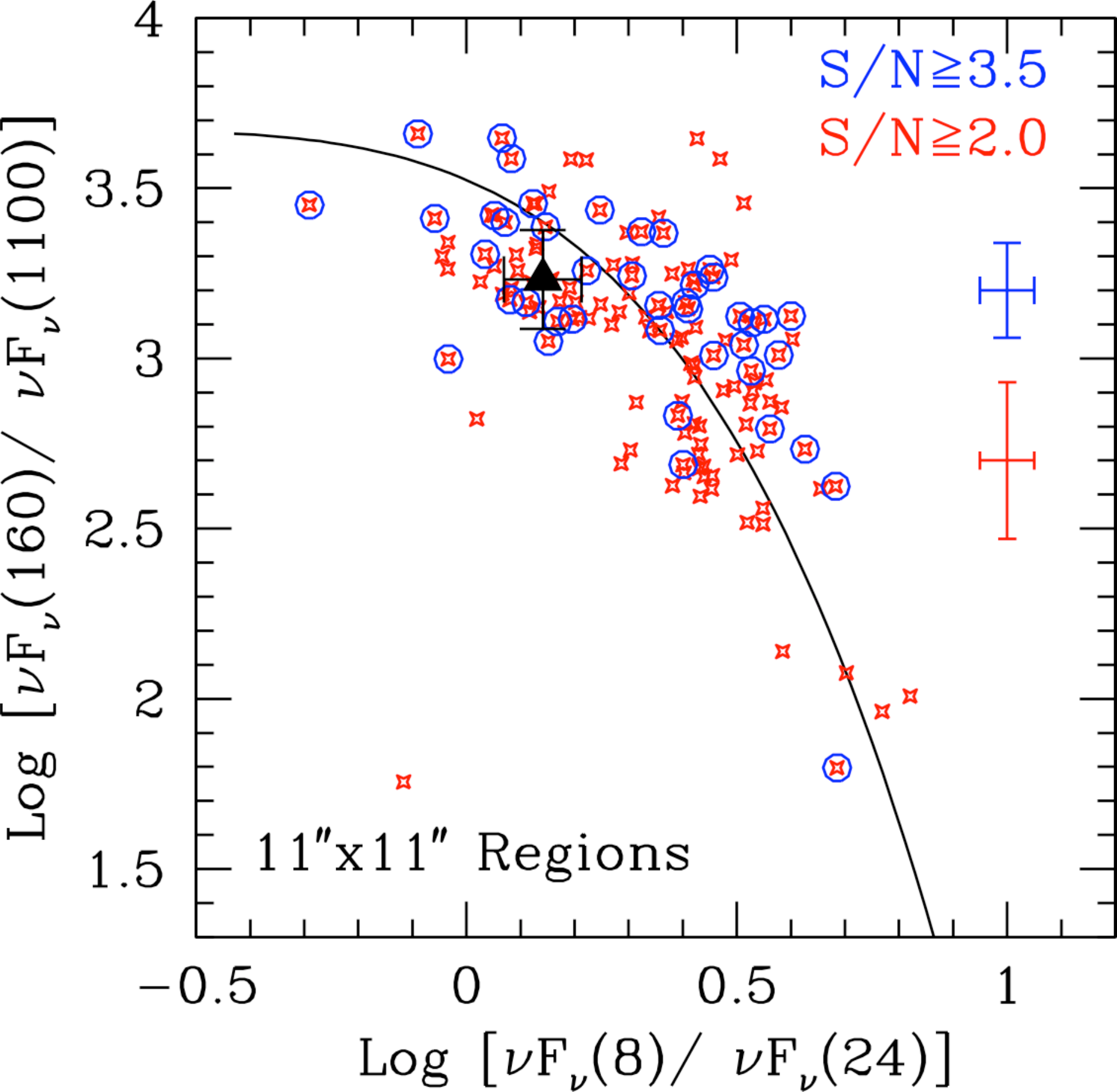} 
\plottwo{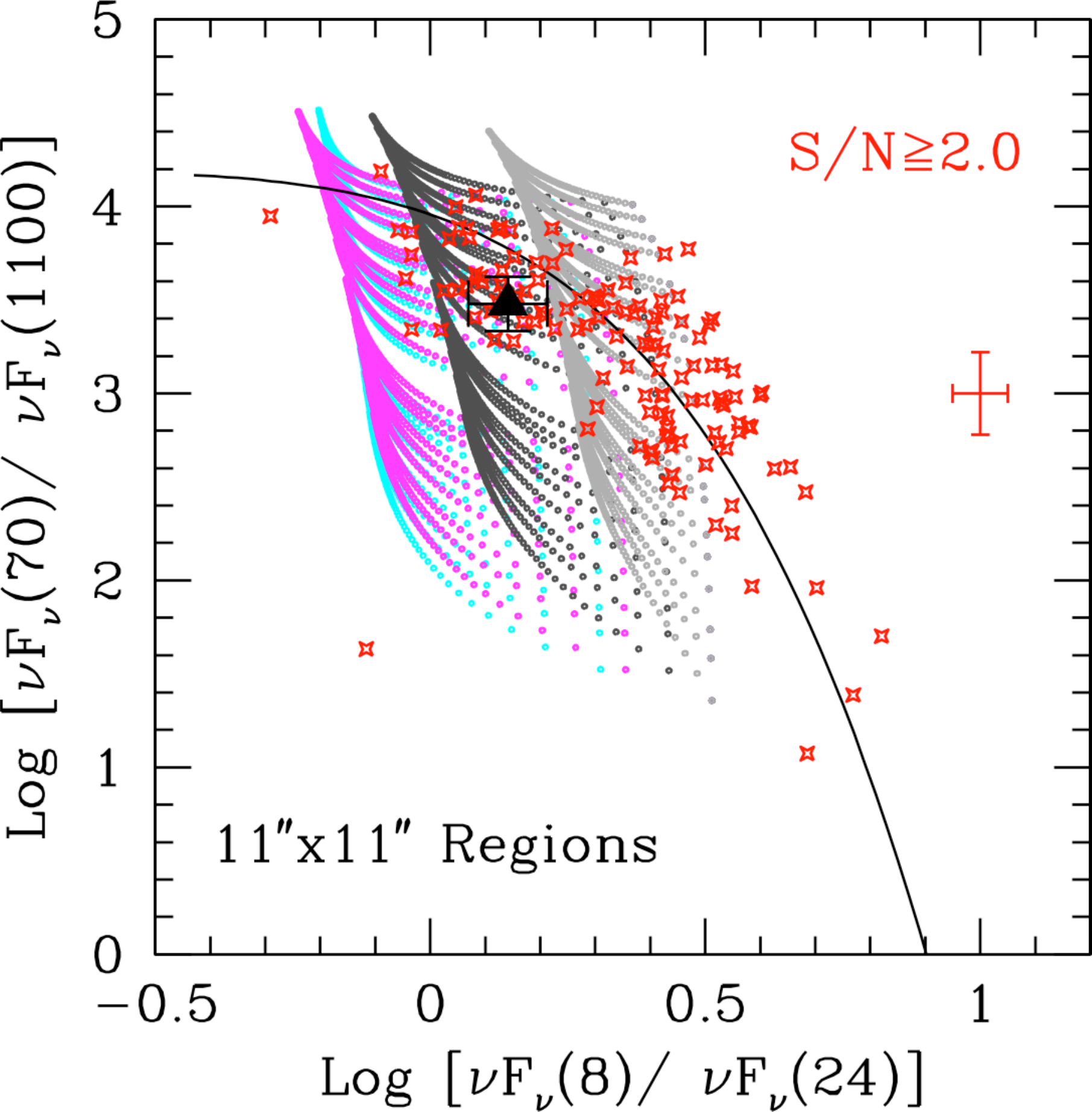}{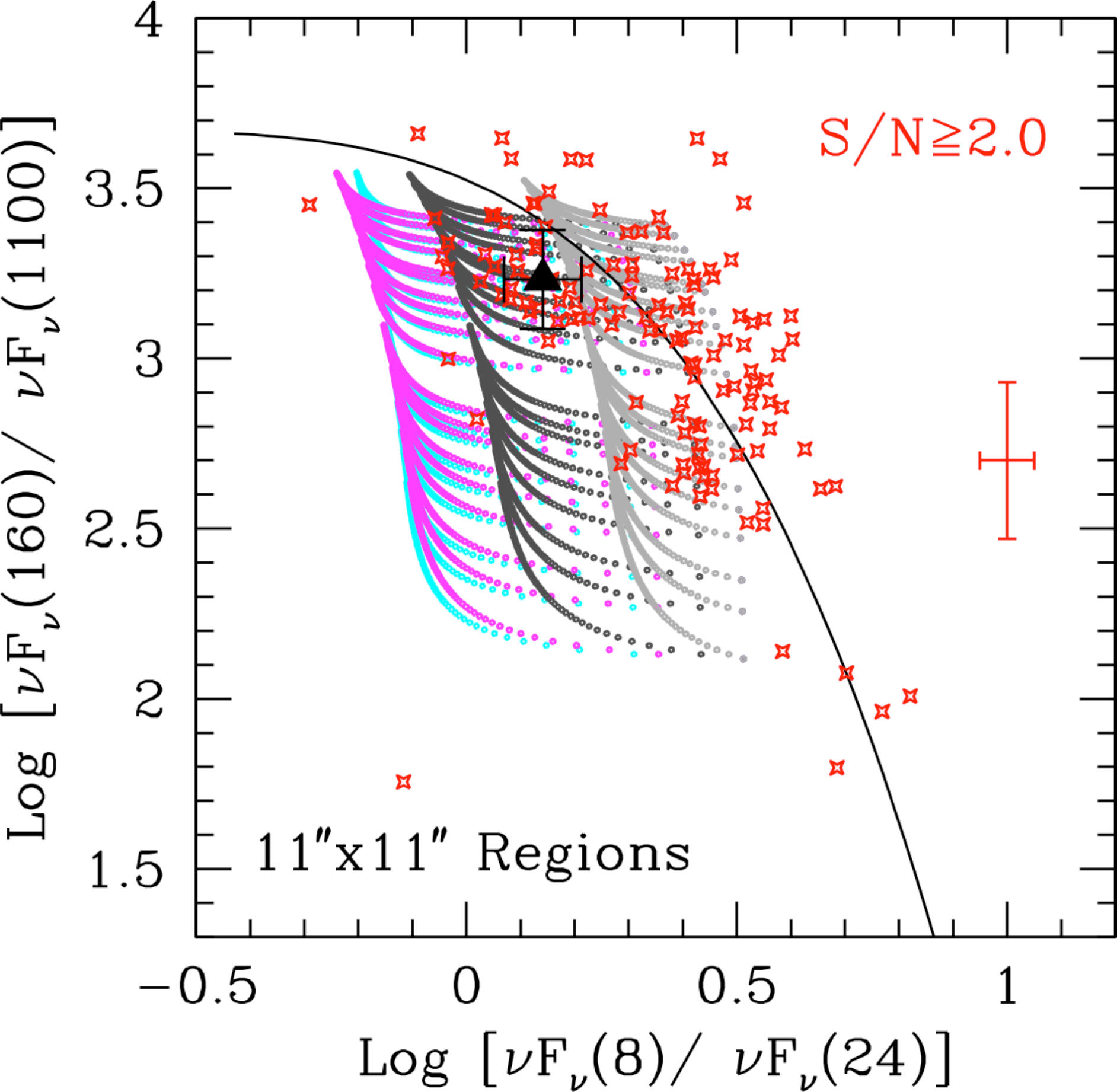} 
\caption{{\bf Top:} Infrared--mm color--color plots for the 11$^{\prime\prime}\times$11$^{\prime\prime}$ spaxels (corresponding to de--projected $\sim$223~pc $\times$ 595~pc) in the central region of NGC\,4449. Red asterisks indicate the 131 spaxels with S/N$_{1100~\mu m}\ge$2.0, and blue circles the 42 spaxels with S/N$_{1100~\mu m}\ge$3.5. Representative 1~$\sigma$ error bars are shown to the right in each panel. The data follow a well defined trend in both color--color plots, through which a best-fit curve is drawn (black line). The expression for each curve is given in the text. The large black triangle with error bars is the mean value of the entire central region of Figure~\ref{fig2}. {\bf Bottom:} The same plots as in the top panels, showing only the spaxels with S/N$_{1100~\mu m}\ge$2.0 (red asterisks) with their 1~$\sigma$ error bar. The locus occupied by the \citet{DraineLi2007} models for the range U$_{min}$=0.1,...,25 (bottom to top), U$_{\max}$=10$^3$ (light grey),10$^4$ (dark grey),10$^5$ (magenta), and 10$^6$ (cyan),  and $\gamma$=0.0,...,1.0 (right to left), at fixed q$_{PAH}$=3.2\% is shown by the colored tracks. The limit for log[$\nu F_{\nu}(8)/\nu F_{\nu}(24)$]$\le$0.5 is built into the \citet{DraineLi2007} models.}
\label{fig4} 
\end{figure}

\clearpage
\begin{figure}[h]
\figurenum{5}
\plotone{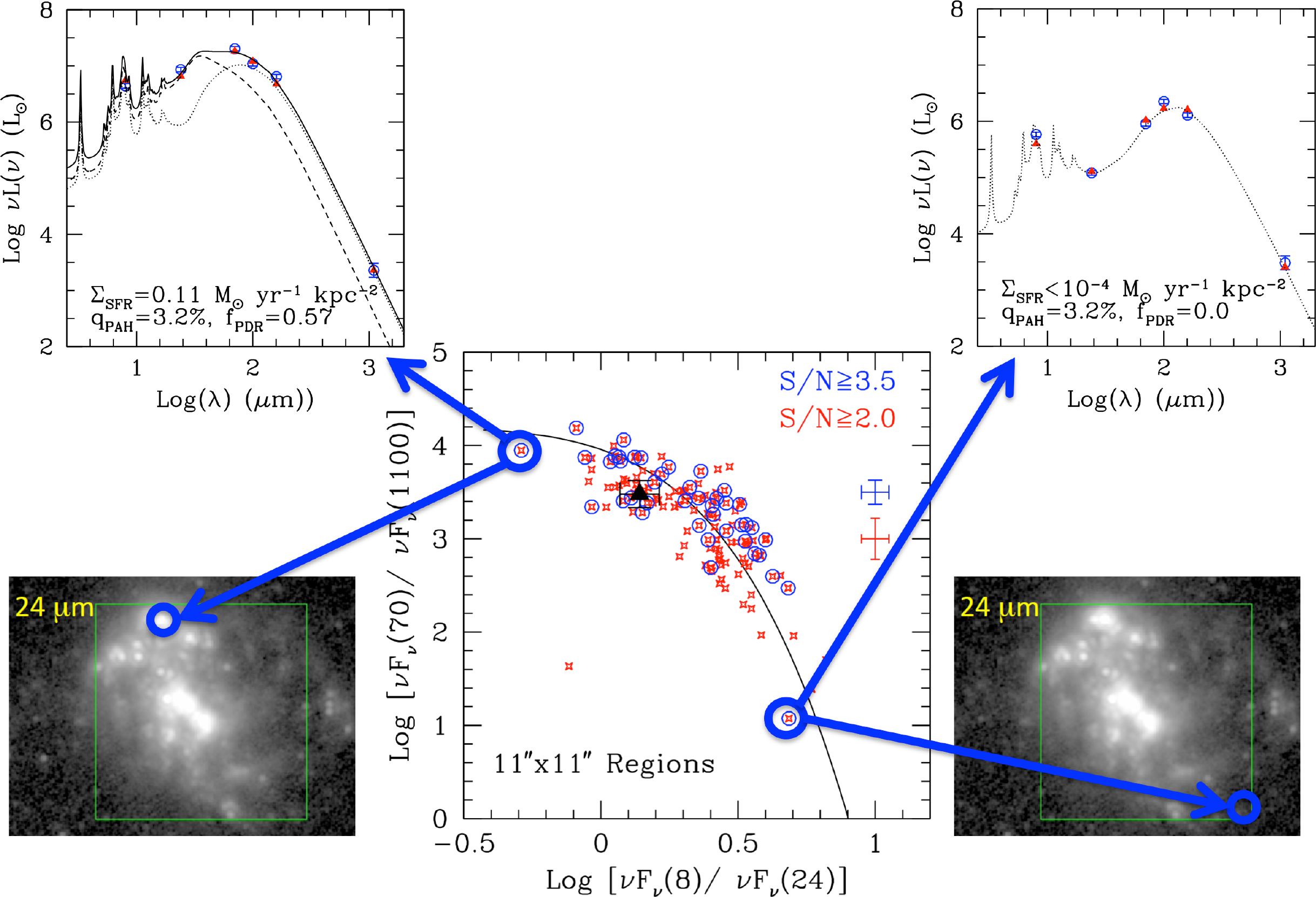} 
\caption{The same plot shown in the top--left--side panel of Figure~\ref{fig4} is now shown at the center of this Figure. On both the left-- and right--hand sides are: (top) the data and the best fit SED, and (bottom) the identification of the location within the central region of the S/N$_{1100 \mu m}\ge$3.5 spaxels with the most extreme color combinations in Figure~\ref{fig4}. The spaxel at the left of the infrared--mm color--color plot corresponds to a local peak of star formation, whose SED is well described by models that include 57\% PDR contribution. The spaxel at the right of the infrared--mm color--color plot corresponds to a region of very low  current star formation, and its SED is well described by a model that includes emission only from cold dust, i.e., f$_{PDR}$=0. The star formation rate surface density, $\Sigma_{SFR}$, is indicated within the SED plots of each spaxel.}
\label{fig5} 
\end{figure}

\clearpage
\begin{figure}[h]
\figurenum{6}
\plottwo{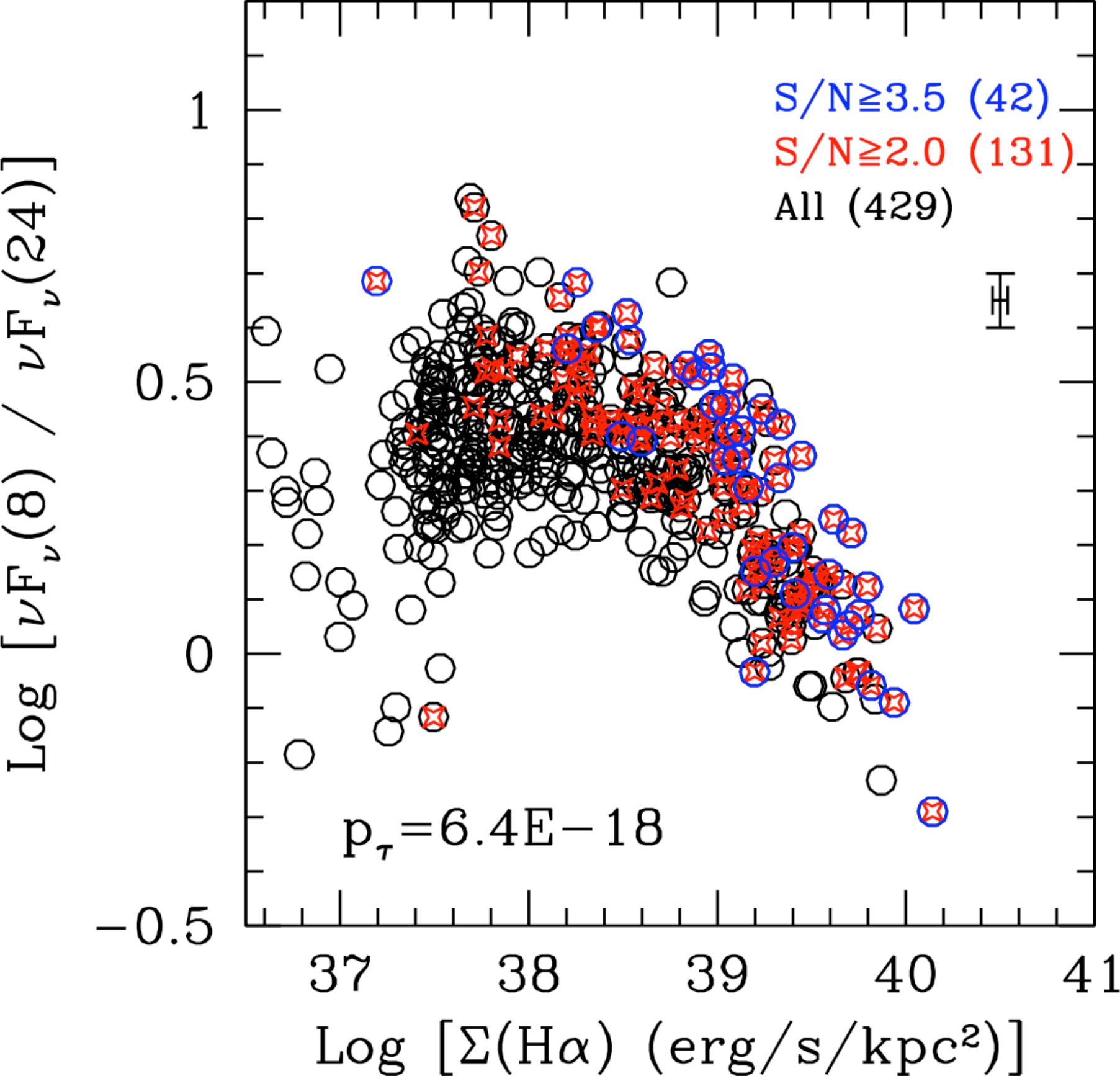}{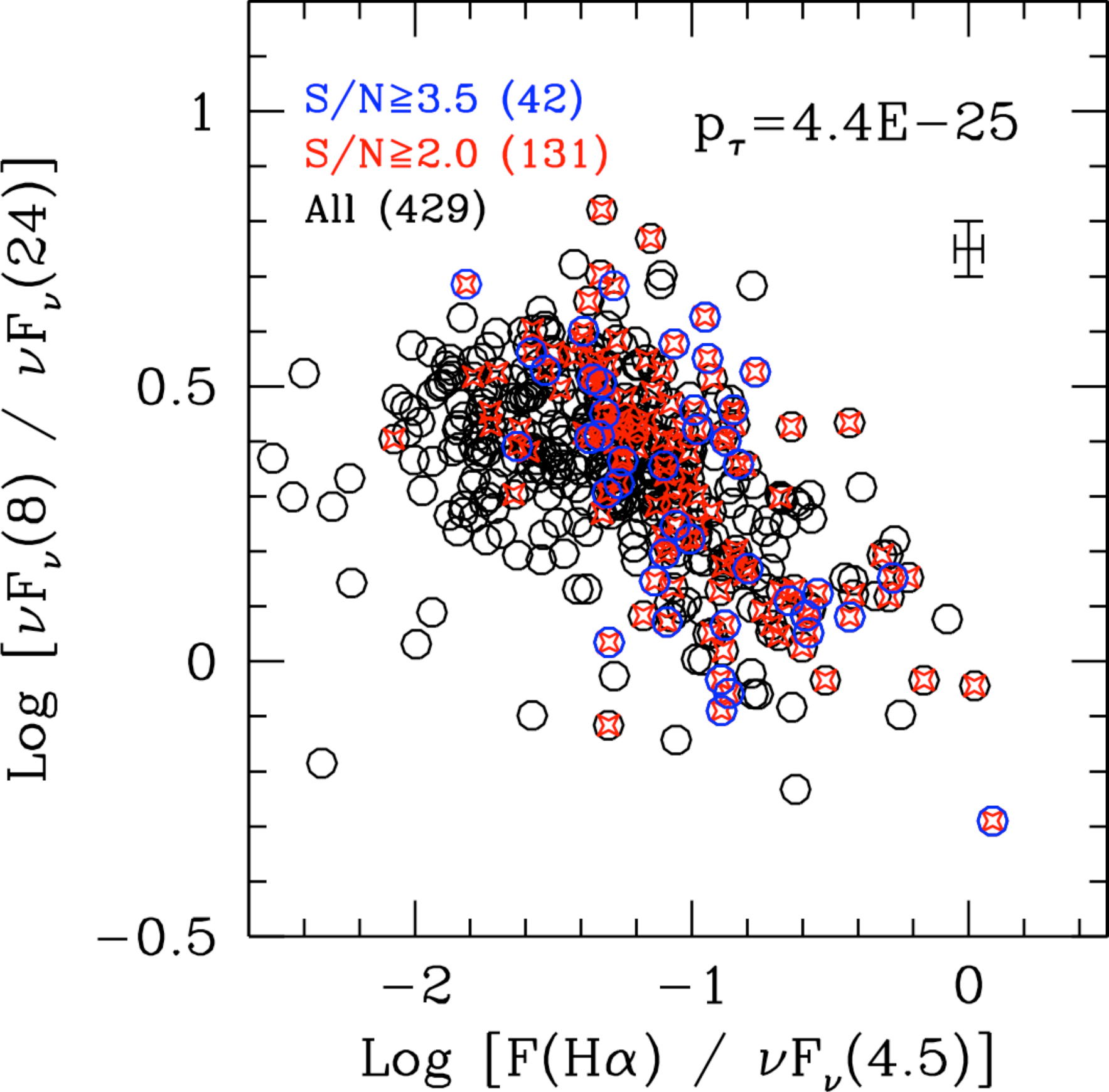} 
\caption{The ratio of 8/24 luminosity as a function of both the H$\alpha$ luminosity surface density (Left) and the H$\alpha$/4.5 luminosity ratio (Right), for all 11$^{\prime\prime}\times$11$^{\prime\prime}$ spaxels in the central region (black), those with S/N$_{1100 \mu m}\ge$2 (red) and those with S/N$_{1100 \mu m}\ge$3.5 (blue). The number of spaxels in each regime of S/N is listed in the Figure. The anti--correlations observed in both panels suggest that regions of high star formation rate (traced by H$\alpha$) relative to both the subtended area and the luminosity of the old stellar populations (traced by the 4.5~$\mu$m emission) have proportionally higher 24~$\mu$m luminosity relative to the 8~$\mu$m one. The non--parametric Kendall probability p$_{\tau}$, shown in the panels, is the probability that the two quantities along the axes are uncorrelated.}
\label{fig6} 
\end{figure}

\clearpage
\begin{figure}[h]
\figurenum{7}
\plottwo{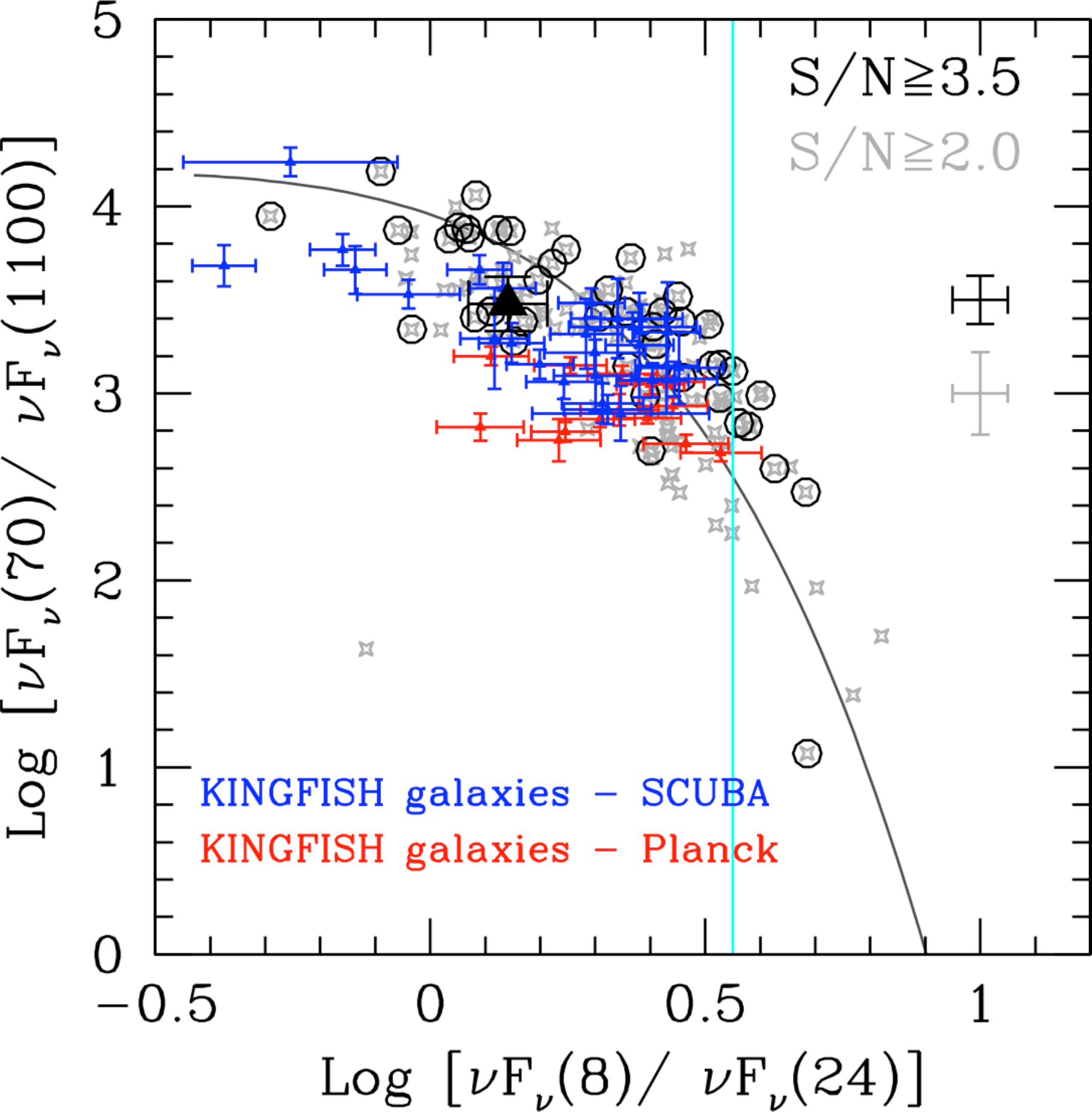}{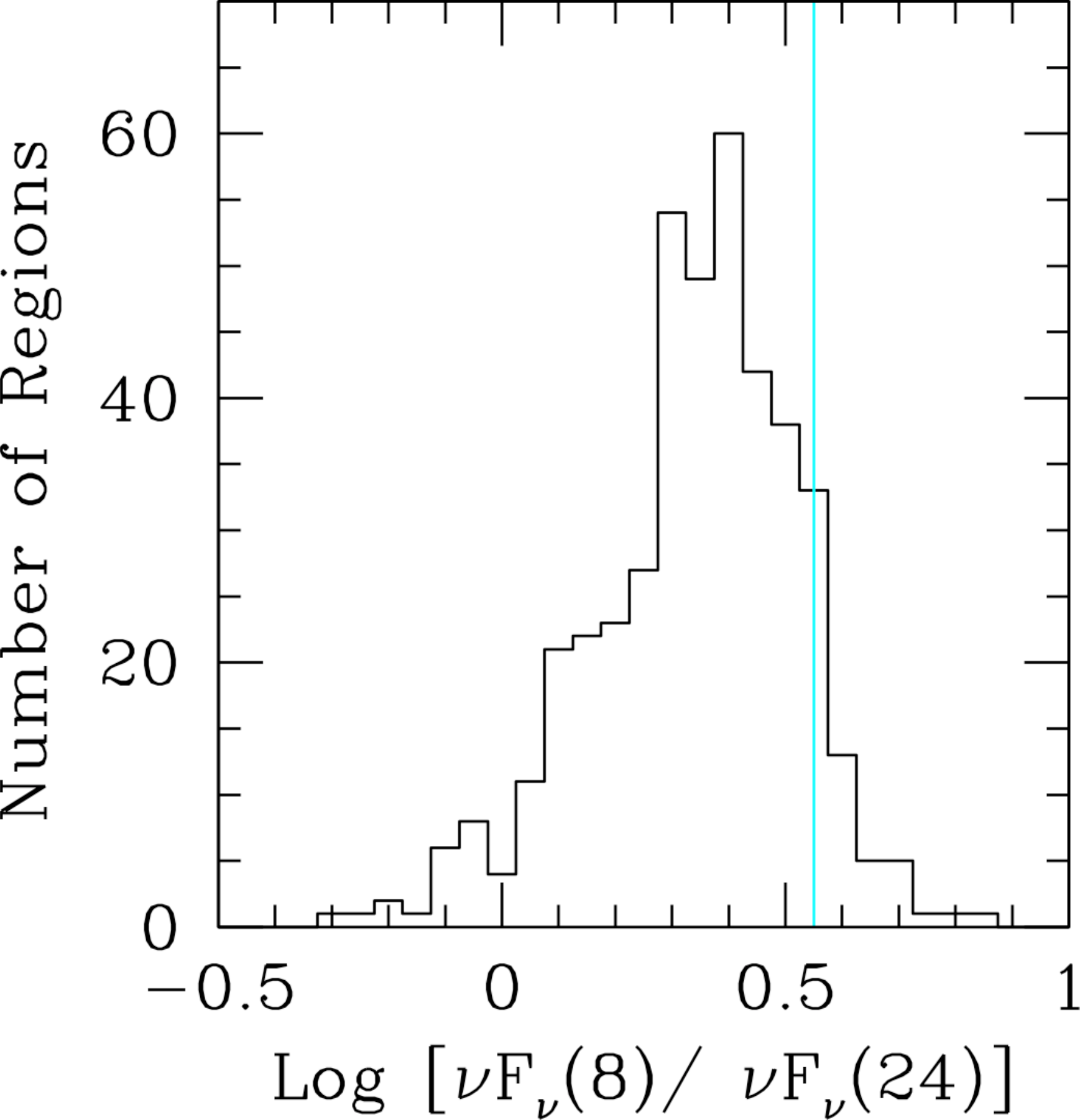} 
\caption{{\bf Left:} The same plot shown in the top--left panel of Figure~\ref{fig4} is now shown in grey--scale for the spaxels of NGC\,4449. The black triangle with error bars is the mean value of the central region of NGC\, 4449, as shown in Figure~\ref{fig4}, as well. The infrared--mm color--color integrated data of 41 nearby galaxies from \citet{Dale2017} are shown as blue and red symbols, together with 1~$\sigma$ error bars. The 1100~$\mu$m data of the galaxies are obtained from either the interpolation of the Planck data (red) or the extrapolation of the SCUBA 850~$\mu$m measurements (blue). We use the distribution of local galaxies to set an upper limit of Log[F(8)/F(24)] =0.55, indicated by the vertical cyan line, in order to include only spaxels to which the DL07 models can be applied. {\bf Right:} Histogram of the F(8)/F(24) color distribution of the spaxels in the central region of NGC\, 4449, irrespective of S/N$_{1100~\mu m}$. Most (91\%) of the spaxels are below the value Log[F(8)/F(24)]=0.55 (vertical cyan line), i.e., have values consistent with those of the nearby galaxies.}
\label{fig7} 
\end{figure}

\clearpage
\begin{figure}[h]
\figurenum{8}
\plottwo{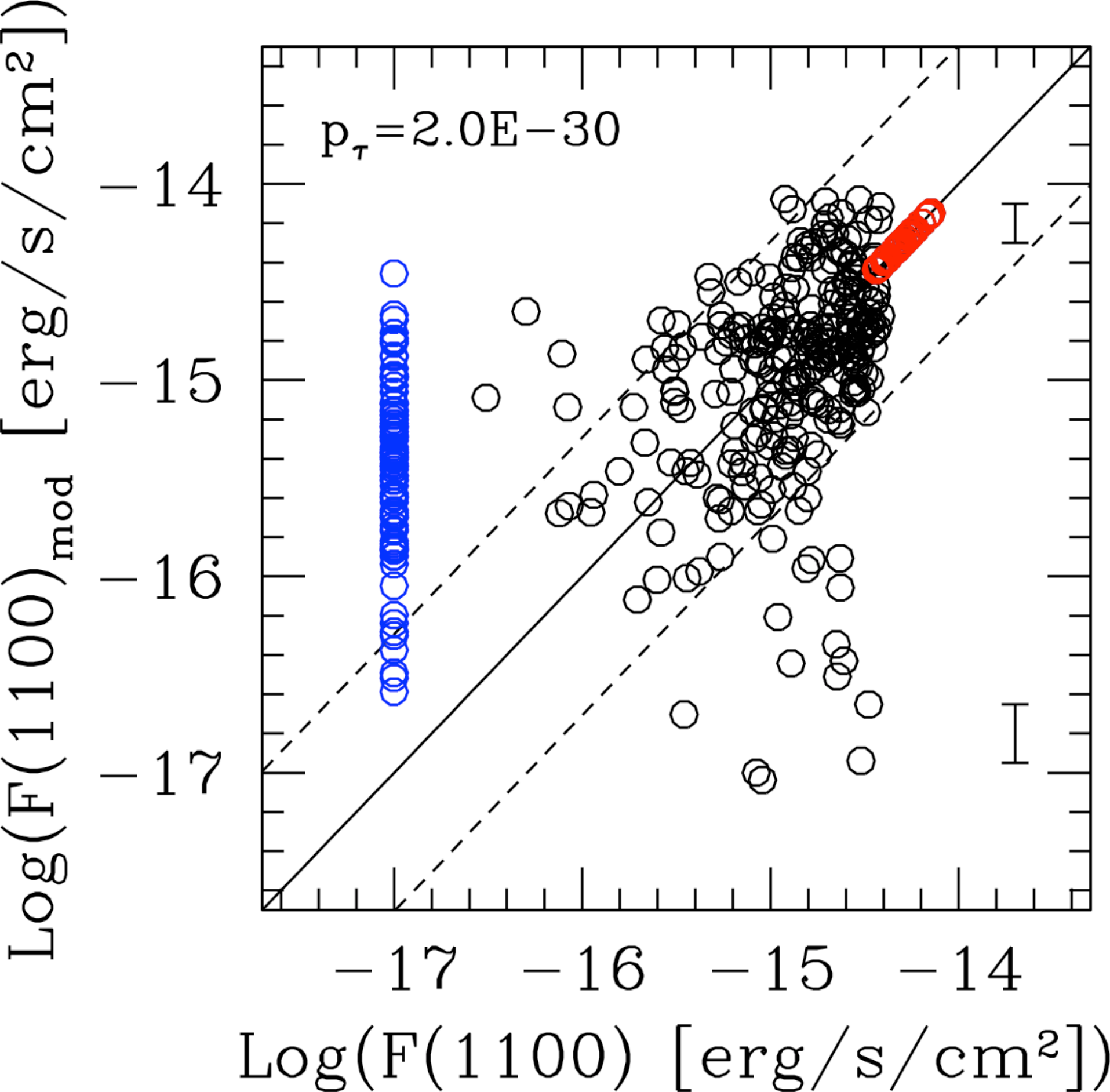}{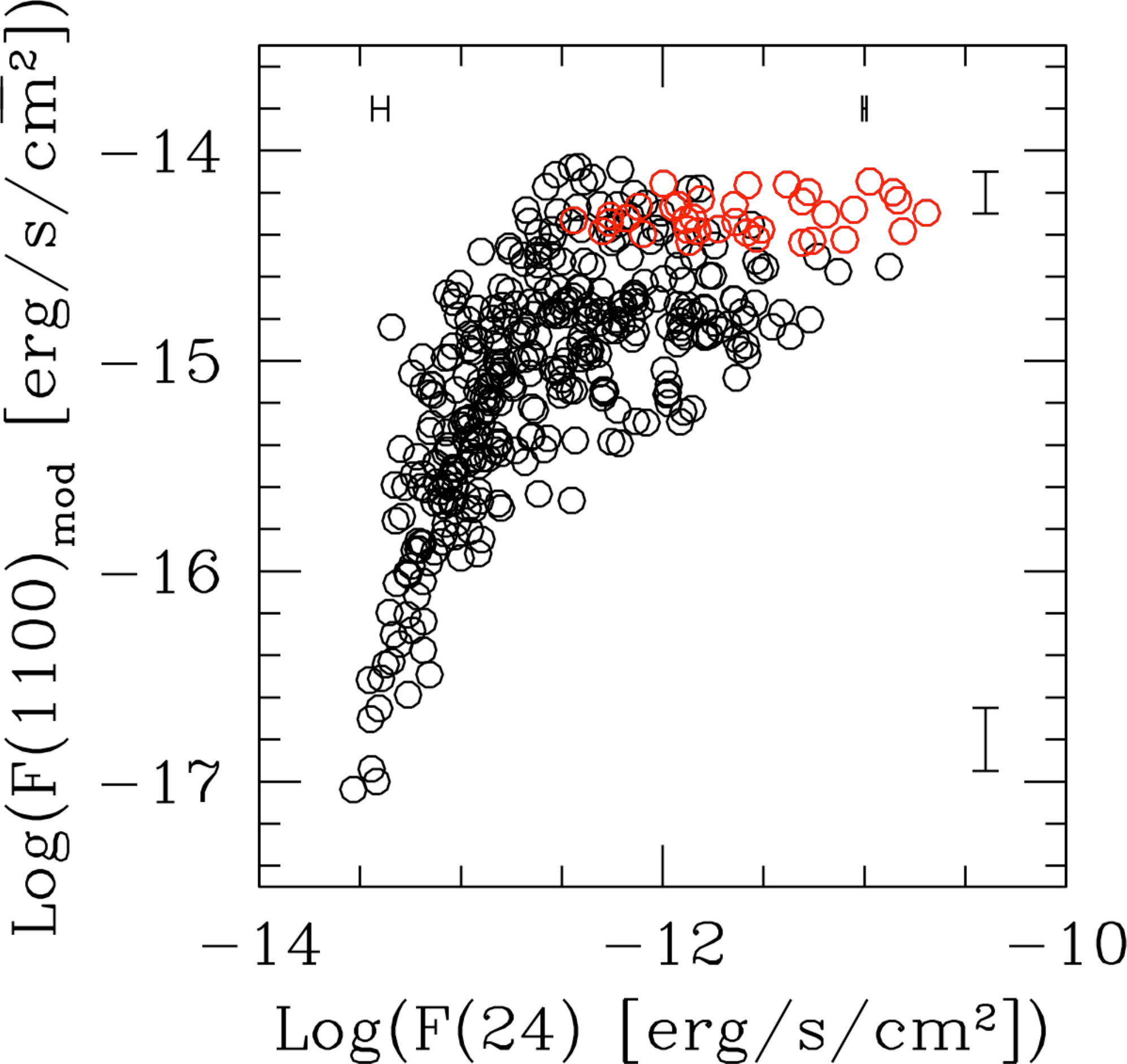} 
\caption{{\bf Left:} The measured flux at 1100~$\mu$m is compared with the flux, F$_{1100, mod}$,  obtained from the model of equation~2. Data with S/N$_{1100 \mu m}\ge$3.5 retain the original measurements in both axes (red circles), while the model is applied to all other data. The measurements are reported for all S/N$_{1100 \mu m}$ values (black circles), including negative/null values, which have been assigned an arbitrarily low flux for plotting purposes (blue circles). There is a general positive trend between the two sets of measurements, as highlighted by the 1--to--1 solid black line. The dash black lines mark the locus of factor 10 deviation from the 1--to--1 relation. The relation between the two quantities is highly significant; the Kendall $\tau$ probability of a random distribution is indicated by the parameter p$_{\tau}$ in the panel, and corresponds to a 14.5~$\sigma$ deviation from the null hypothesis. The Kendall $\tau$ test is applied only to the data with S/N$_{1100 \mu m}<$3.5 (i.e., excluding the red circles). This  lends further support to the relation linking mid--IR to far--IR/mm fluxes. Highly deviant points from the relation are discussed in the text.  {\bf Right:} The 1100~$\mu$m measurements from the model of equation~2 as a function of the flux at 24~$\mu$m. Typical 1~$\sigma$ error bars are shown for both F(24) (top of the panel) and F(1100) (right of the panel); for the latter we include the dispersion in the relation of equation~2 as inferred from Figure~\ref{fig4}. Red circles show the location of the 1100~$\mu$m measurements with S/N$>$3.5, to which no model is applied.}
\label{fig8} 
\end{figure}

\clearpage
\begin{figure}[h]
\figurenum{9}
\plottwo{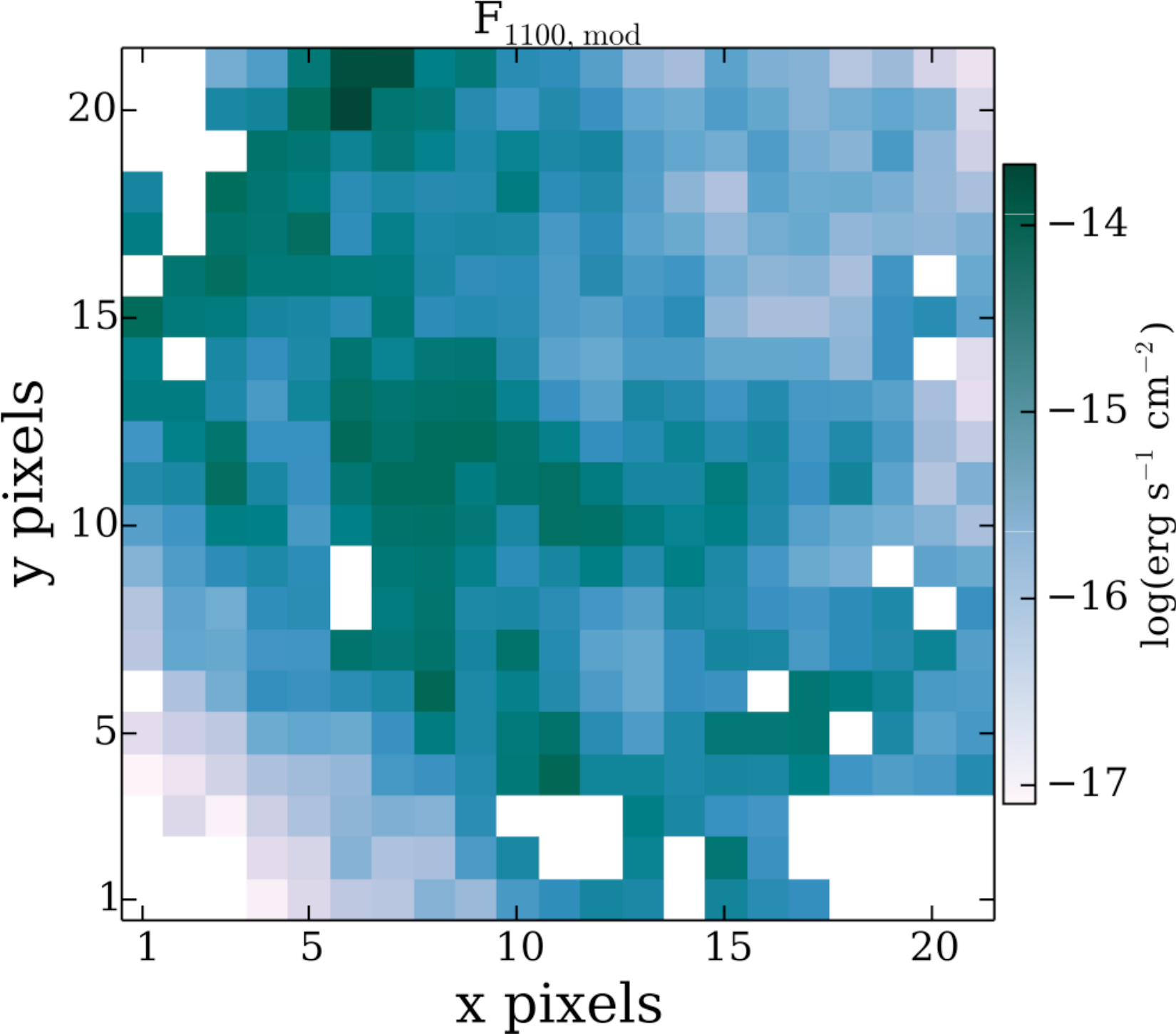}{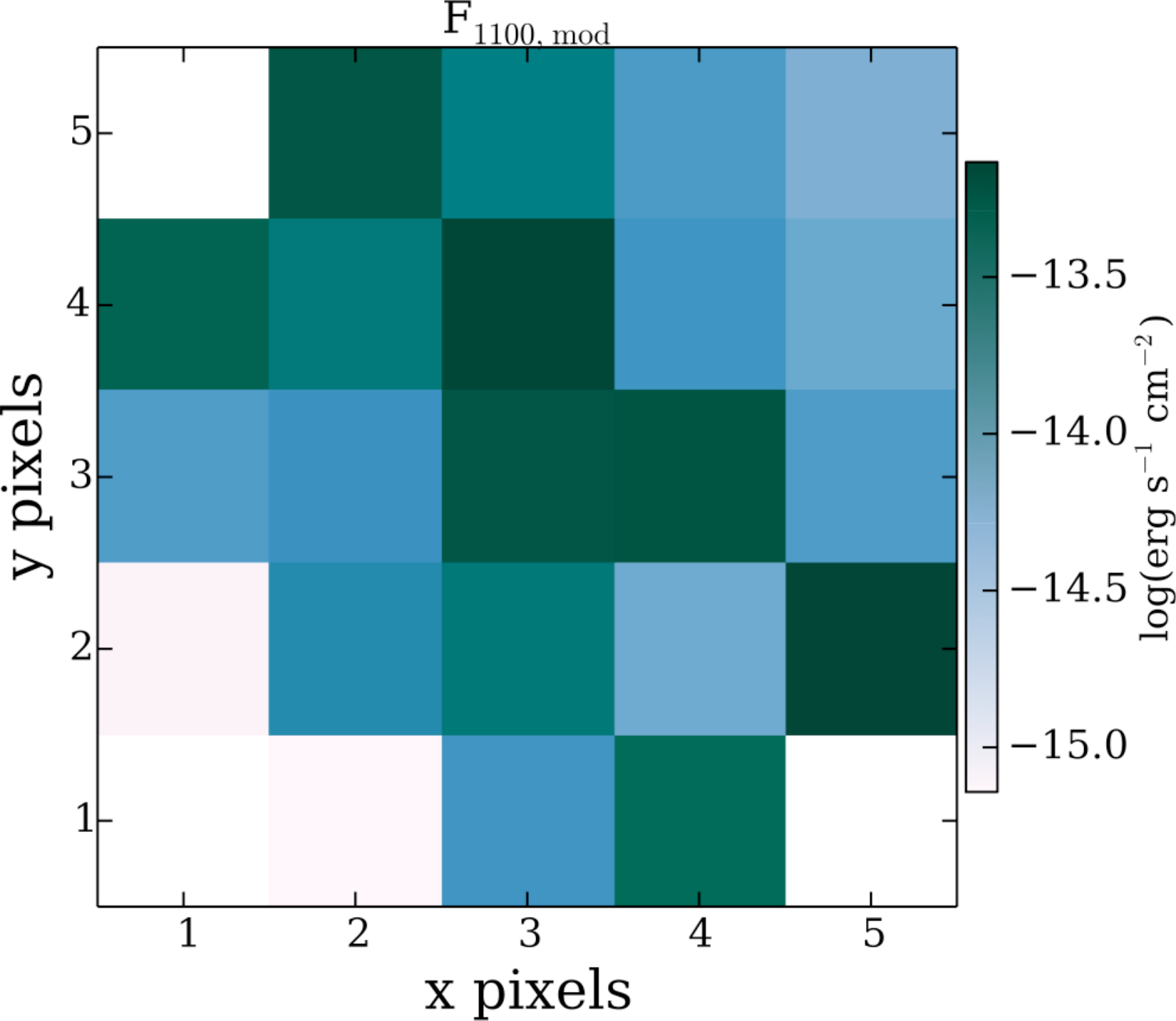} 
\caption{The maps at 1100~$\mu$m of the central region of NGC\, 4449 with 11$^{\prime\prime}\times$11$^{\prime\prime}$ resolution ({\bf left}) and 44$^{\prime\prime}\times$44$^{\prime\prime}$ resolution ({\bf right}). The 1100~$\mu$m fluxes are from the original measurements for spaxels with S/N$_{1100 \mu m}\ge$3.5, and from the model of equation~2 for the other spaxels; only spaxels with Log[F(8)/F(24)]$\le$0.55 are assigned valid F(1100) fluxes, for a total of 393 spaxels in the maps to the left. The units of the maps are shown by the scale to the right of each panel.}
\label{fig9} 
\end{figure}

\clearpage
\begin{figure}[h]
\figurenum{10}
\plottwo{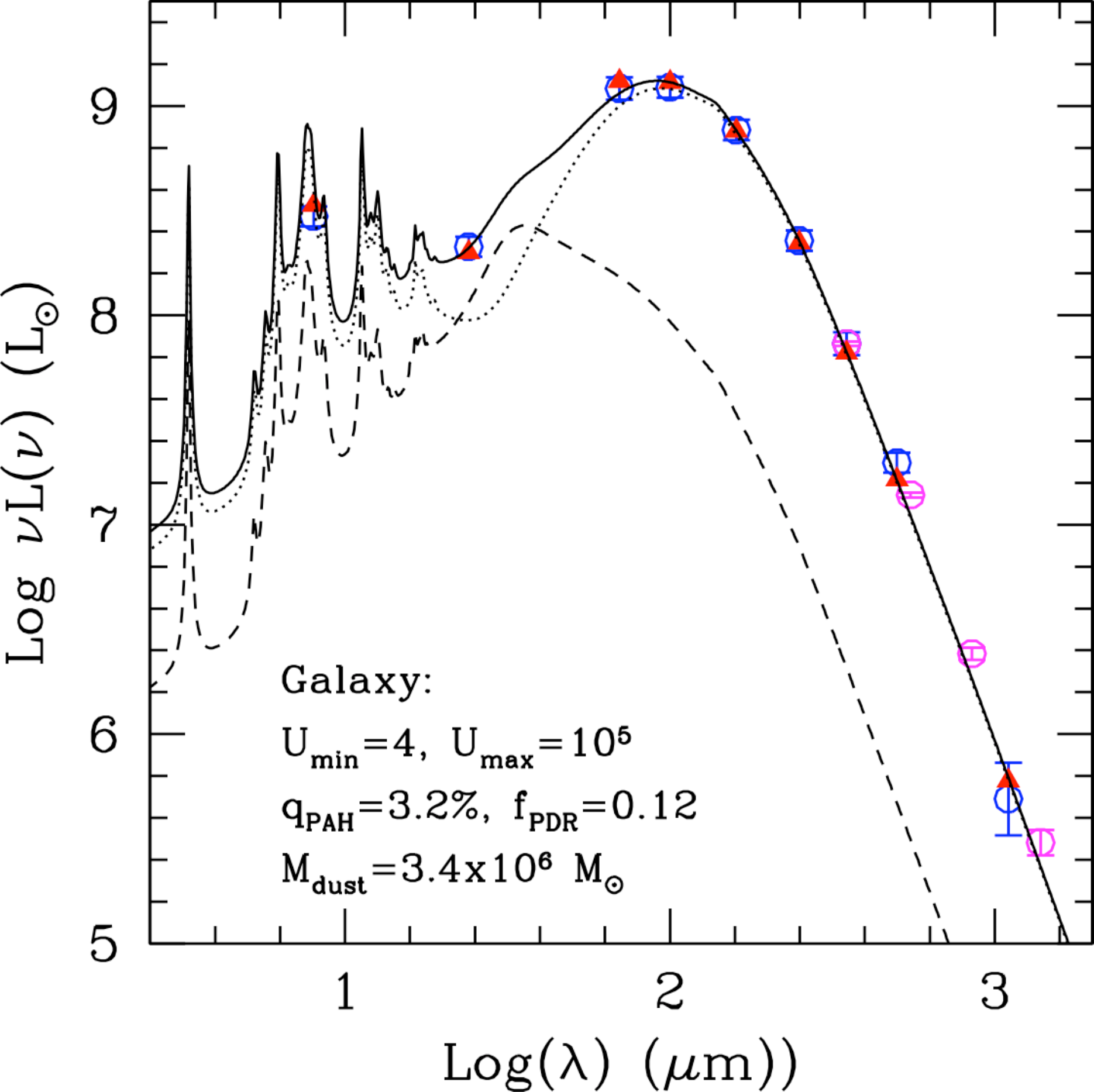}{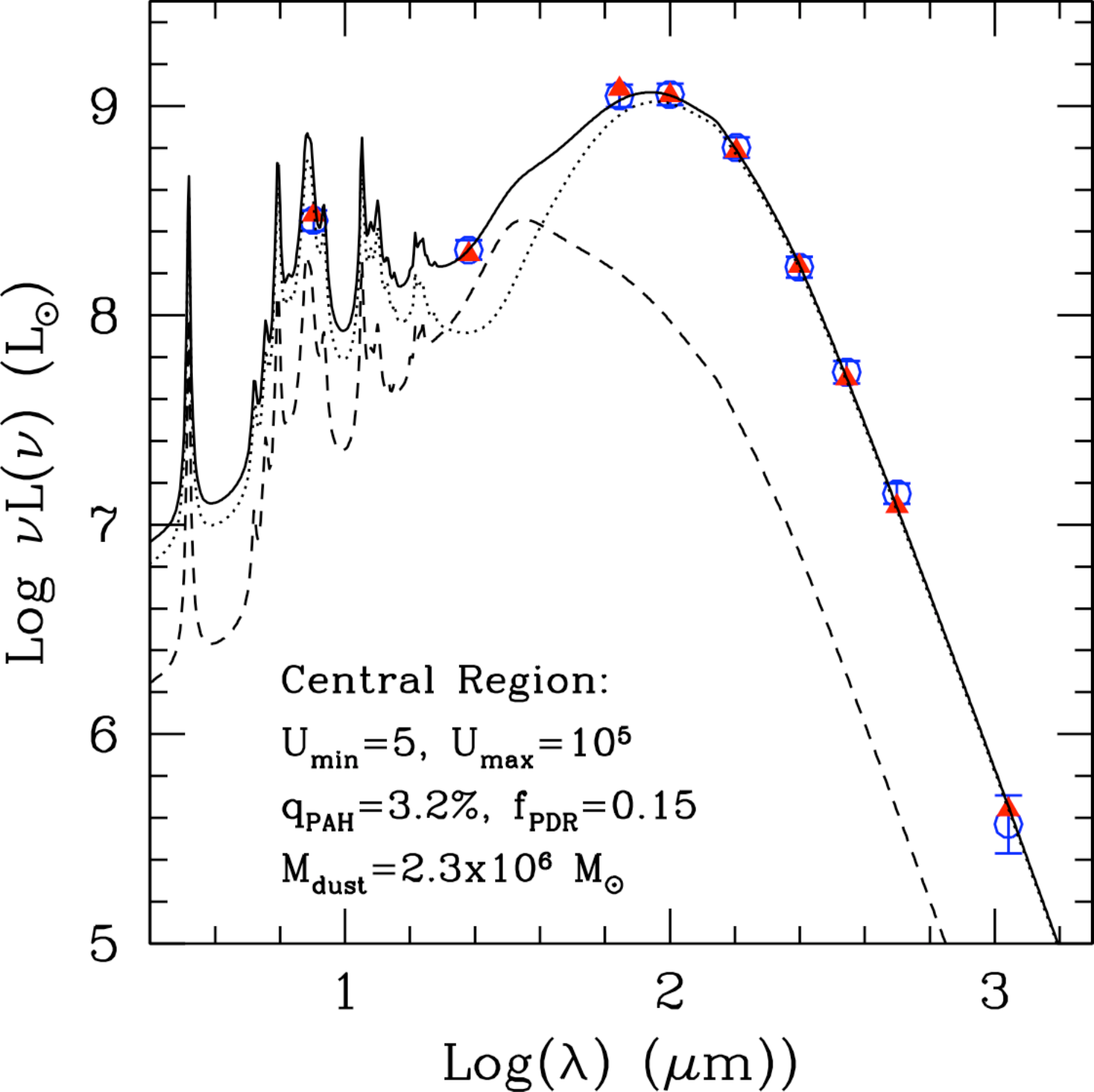}
\caption{The dust SED fits of the whole galaxy (left) and of the central region (right, $\sim$4.7$\times$4.7~kpc$^2$ de--projected to $\sim$4.7$\times$12.5~kpc$^2$). The black lines are the best fits through the data, the red triangles the resulting broad--band photometry from the best fit models, and the blue circles are the photometric data used in the fits with the 1~$\sigma$ error bars indicated. For the whole galaxy we also report the Planck photometric measurements with their uncertainties as magenta circles: these data are not used in the fits, but are used here to confirm the robustness of the model fits. The best fit SEDs are constructed from a dust mixture heated by two starlight intensity components:  the general interstellar medium (black dot line), described by the energy density parameter {\em U$_{min}$}, and regions with a power law 
distribution of intensities, $dM_{dust}/dU \propto U^{-2}$, between  {\em U$_{min}$}, and {\em U$_{max}$}  (black dash line). 
Recent results from the Planck Collaboration indicate that the dust masses from the \citet{DraineLi2007} models, and listed in the two panels, may need to be reduced by a factor as much as $\sim$1.5 (see section~6 for details).} 
\label{fig10} 
\end{figure}

\clearpage
\begin{figure}[h]
\figurenum{11}
\plotone{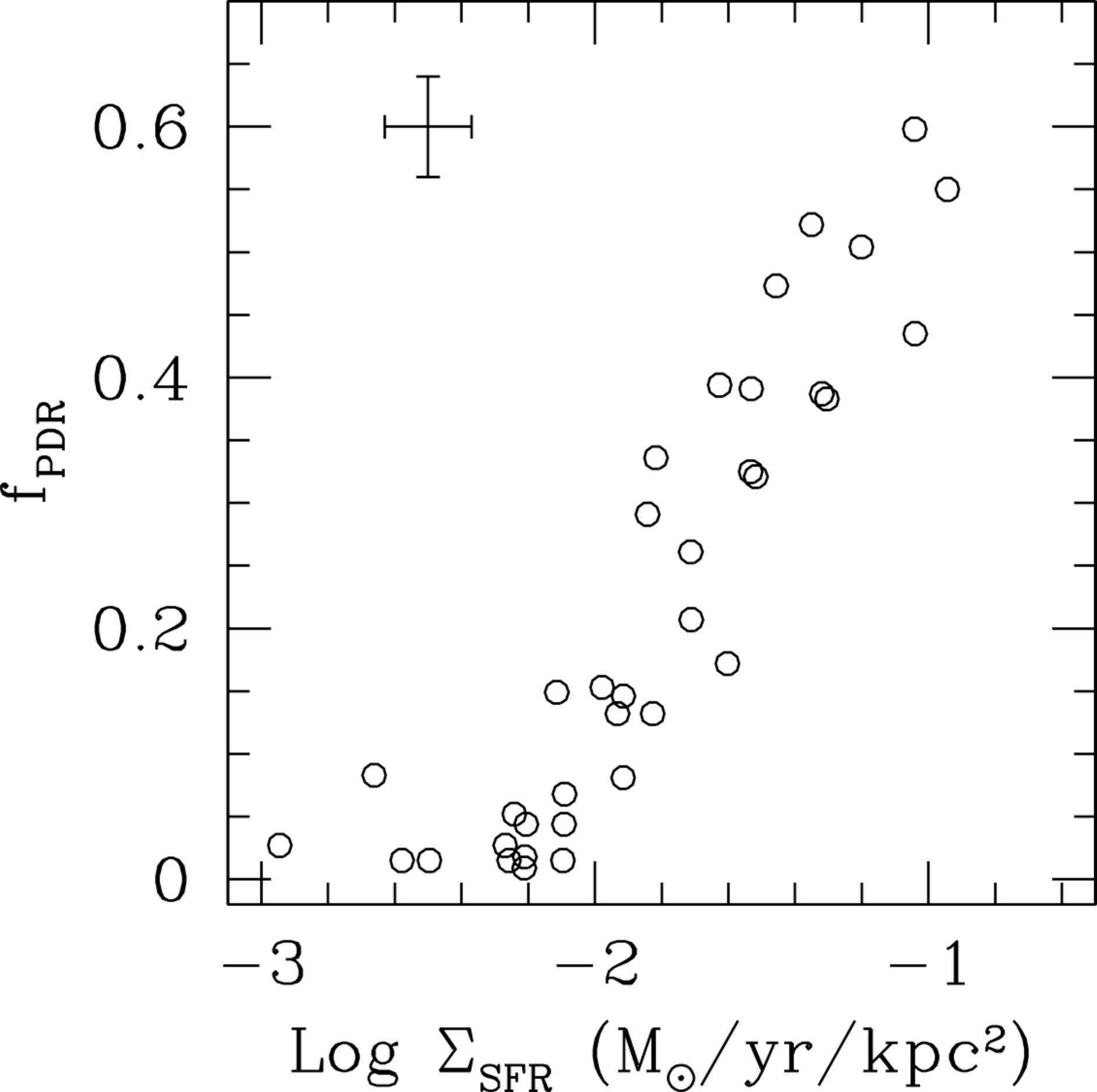} 
\caption{The range f$_{PDR}$ values, the fraction of the dust luminosity radiated from regions with U$>100$, as a function of the star formation rate surface density, $\Sigma_{SFR}$, for the 36 11$^{\prime\prime}\times$11$^{\prime\prime}$ spaxels in the central region with S/N$_{1100 \mu m}\ge$3.5 and Log[F(8)/F(24)] $\le$0.55. The positive correlation between the two quantities is expected, since stronger star forming regions will contribute a higher fraction of the dust luminosity.} 
\label{fig11} 
\end{figure}

\clearpage
\begin{figure}[h]
\figurenum{12}
\includegraphics[scale=1.0]{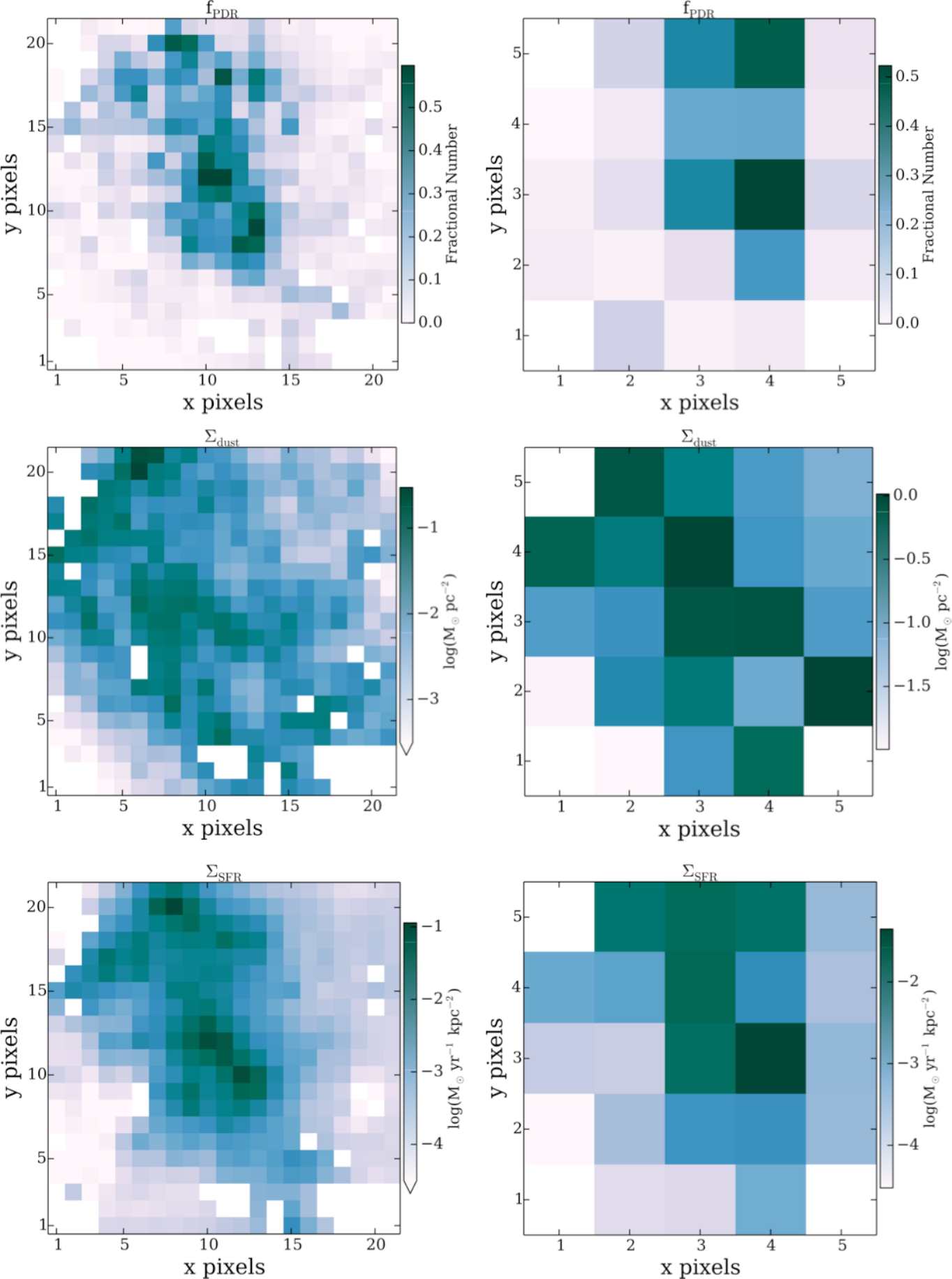} 
\caption{Maps of f$_{PDR}$, the fraction of dust luminosity emitted by star forming regions (top panels), $\Sigma_{dust}$, the dust surface density (central panels, in units of M$_{\odot}$~pc$^{-2}$), and $\Sigma_{SFR}$, the SFR surface density (bottom panels, in units of M$_{\odot}$~yr$^{-1}$~kpc$^{-2}$), both at 11$^{\prime\prime}\times$11$^{\prime\prime}$ (left column) and 44$^{\prime\prime}\times$44$^{\prime\prime}$ (right column) resolution. f$_{PDR}$ and  $\Sigma_{dust}$ are best--fitting parameters from the dust SED of each spaxel (section~7). $\Sigma_{SFR}$ is from equation~1. }
\label{fig12} 
\end{figure}

\clearpage
\begin{figure}[h]
\figurenum{13}
\plotone{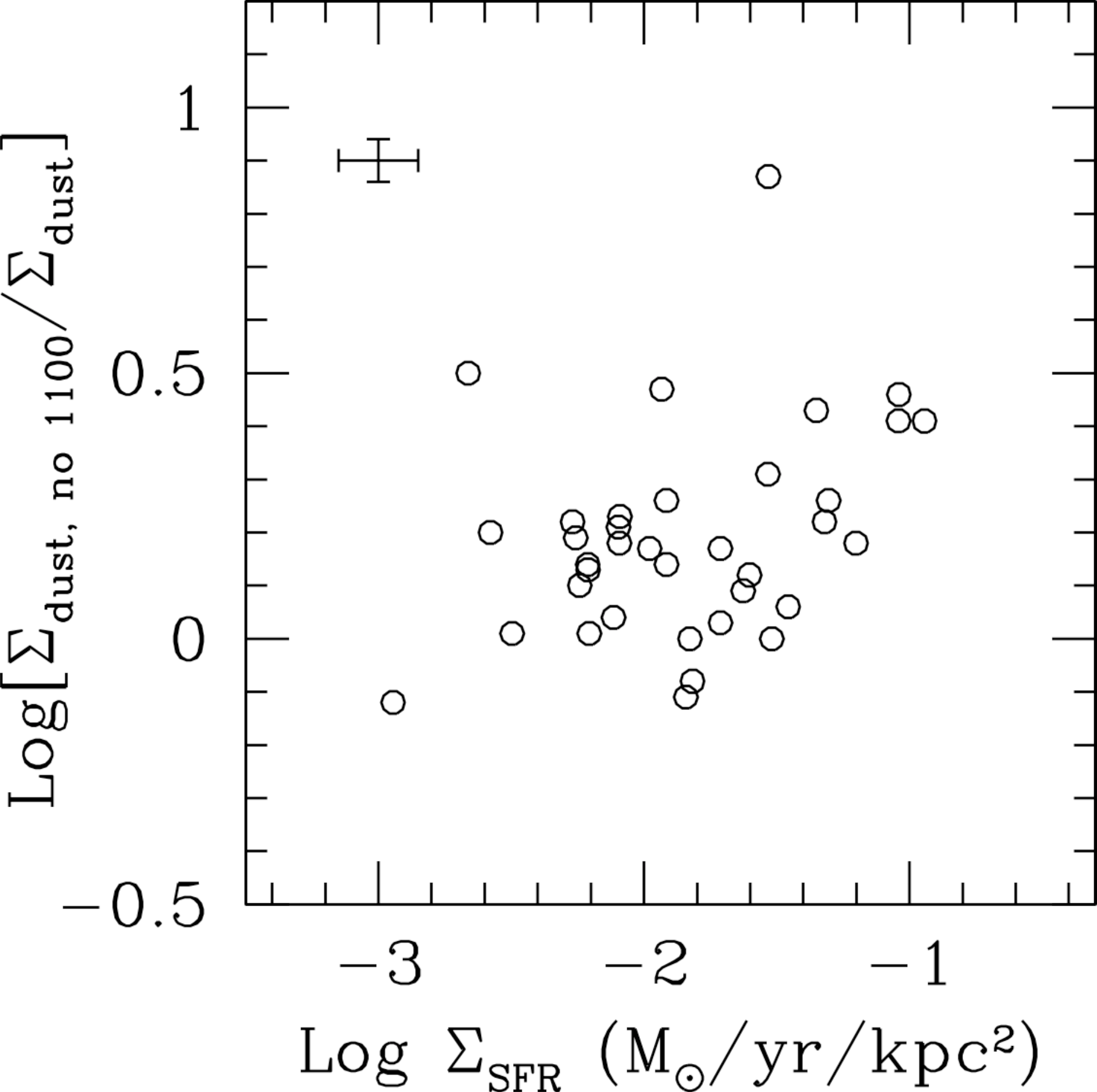} 
\caption{The ratio of the dust surface densities in the S/N$_{1100 \mu m}\ge$3.5 spaxels of the central region in NGC\, 4449, as determined from best fits of the IR SEDs including ($\Sigma_{dust}$) or excluding ($\Sigma_{dust, no\ 1100}$) the data points at 1100~$\mu$m. When excluding the 1100~$\mu$m data, the dust surface densities are systematically higher, with a median offset of about 60\%, and a slight trend for higher offsets in more strongly star--forming  regions, i.e., those with higher values of $\Sigma_{SFR}$.} 
\label{fig13} 
\end{figure}

\clearpage
\begin{figure}[h]
\figurenum{14}
\plotone{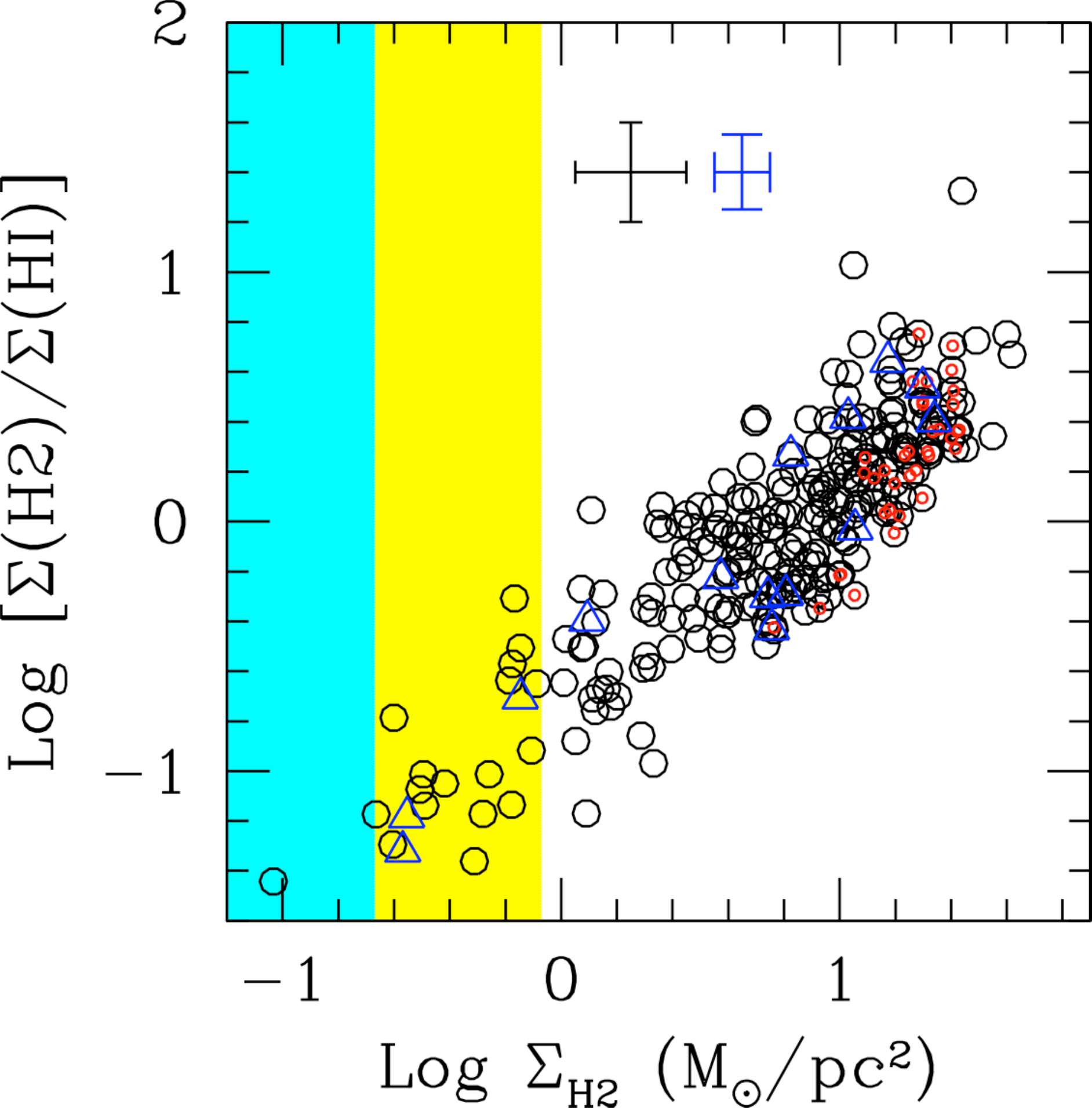} 
\caption{The H$_2$/HI surface density ratio as a function of the H$_2$ surface density, $\Sigma(H_2)$, for the 11$^{\prime\prime}\times$11$^{\prime\prime}$ (black circles) and 44$^{\prime\prime}\times$44$^{\prime\prime}$ (blue triangles) spaxels, respectively. Red circles are 11$^{\prime\prime}\times$11$^{\prime\prime}$ spaxels with S/N$_{1100 \mu m}\ge$3.5. The yellow region corresponds to the threshold defined in section~7 at 11$^{\prime\prime}\times$11$^{\prime\prime}$ resolution, while the cyan region is the same value at 44$^{\prime\prime}\times$44$^{\prime\prime}$, respectively. Typical error bars for the faint $\Sigma(H_2)$ are also shown. There are no significant differences in the trend and dynamic range of  the H$_2$/HI mass ratio, for  spaxels at the two resolutions. We request all our  H$_2$ values to be larger than the 1~$\sigma$ uncertainty (i.e., to be to the right of the color regions) to be considered valid measurements.  All 44$^{\prime\prime}\times$44$^{\prime\prime}$ spaxels are above the H$_2$ threshold (cyan region), and 95\% of the  11$^{\prime\prime}\times$11$^{\prime\prime}$ are (yellow region). } 
\label{fig14} 
\end{figure}

\clearpage
\begin{figure}[h]
\figurenum{15}
\plottwo{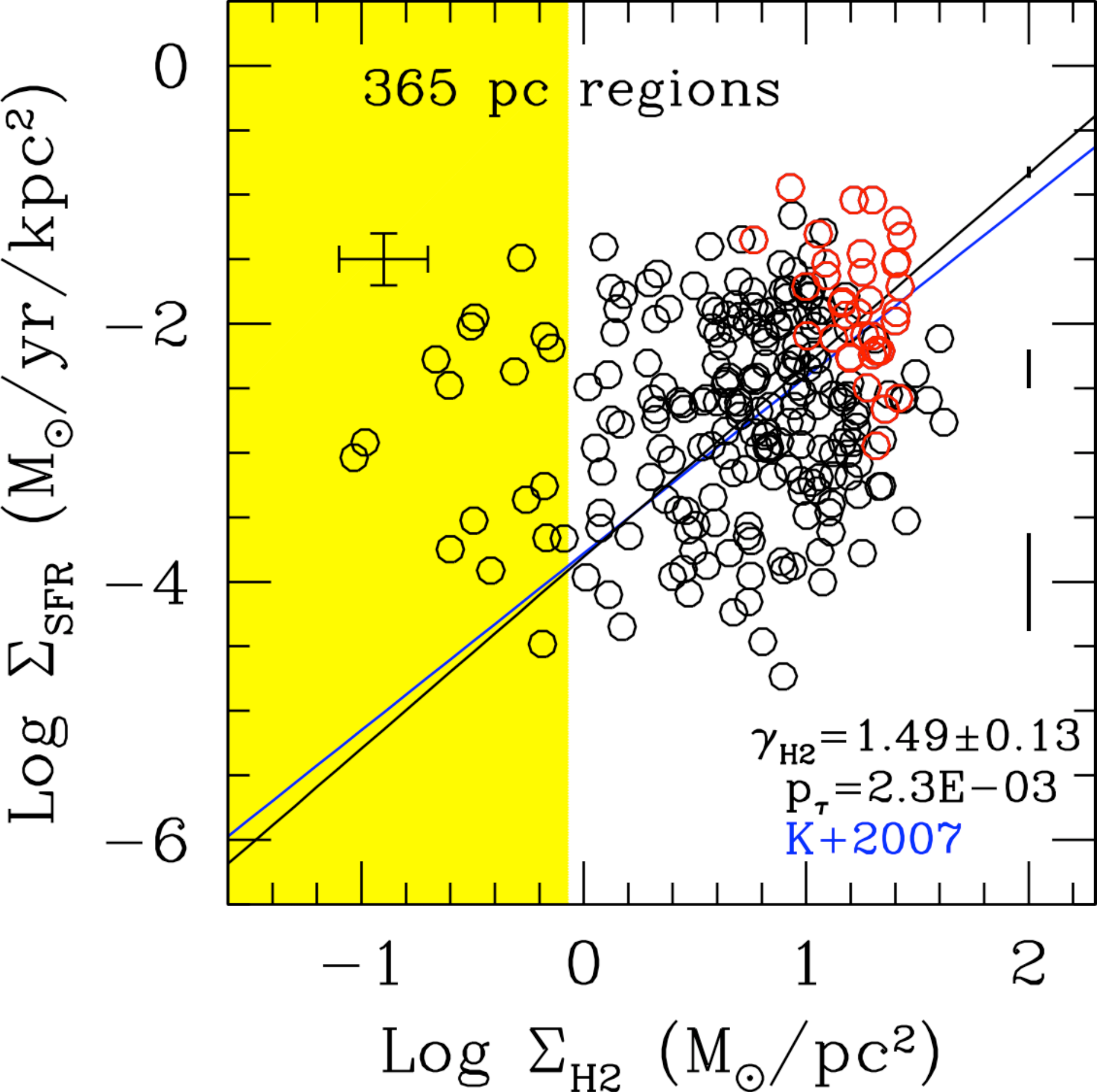}{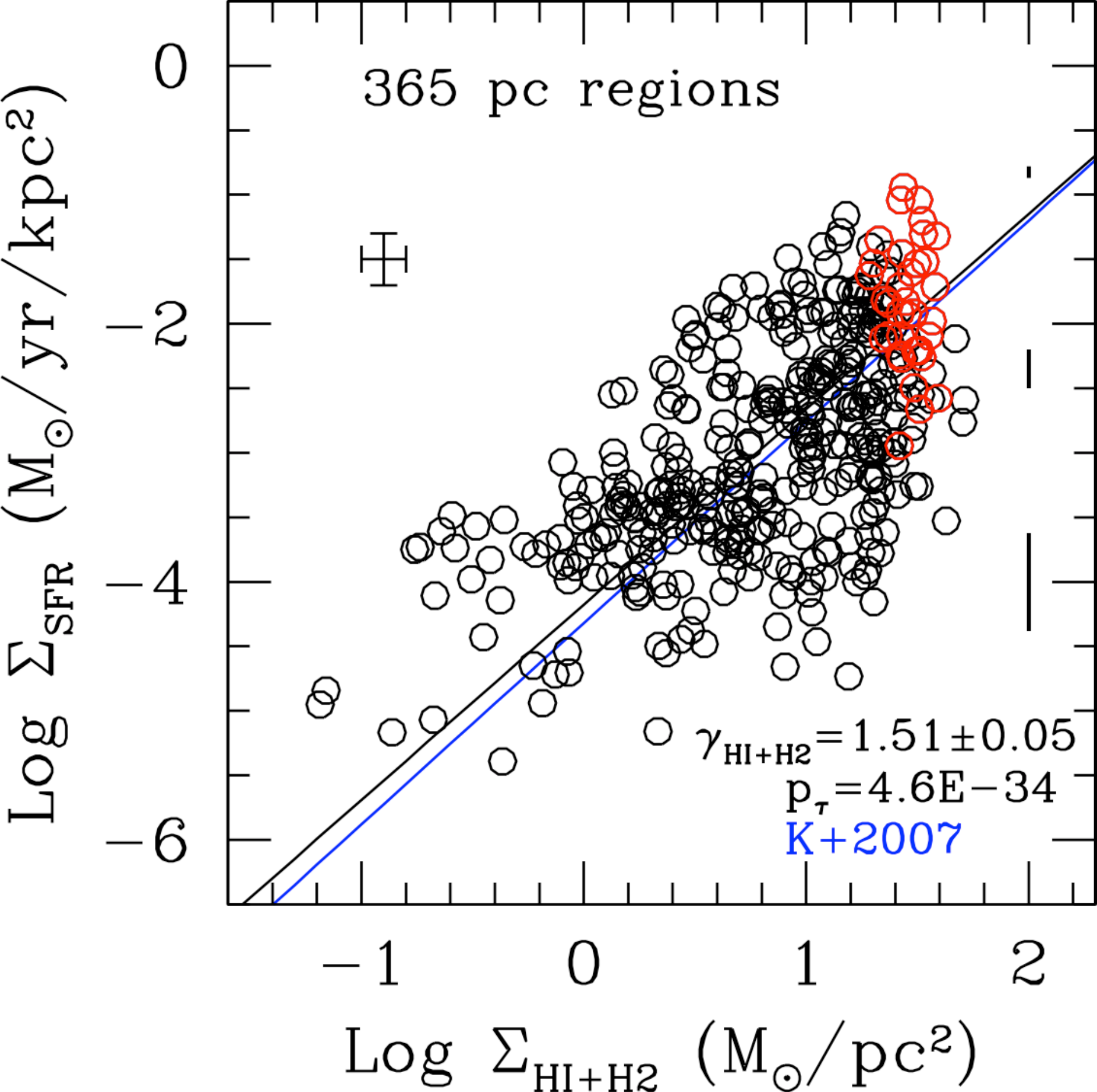} 
\plottwo{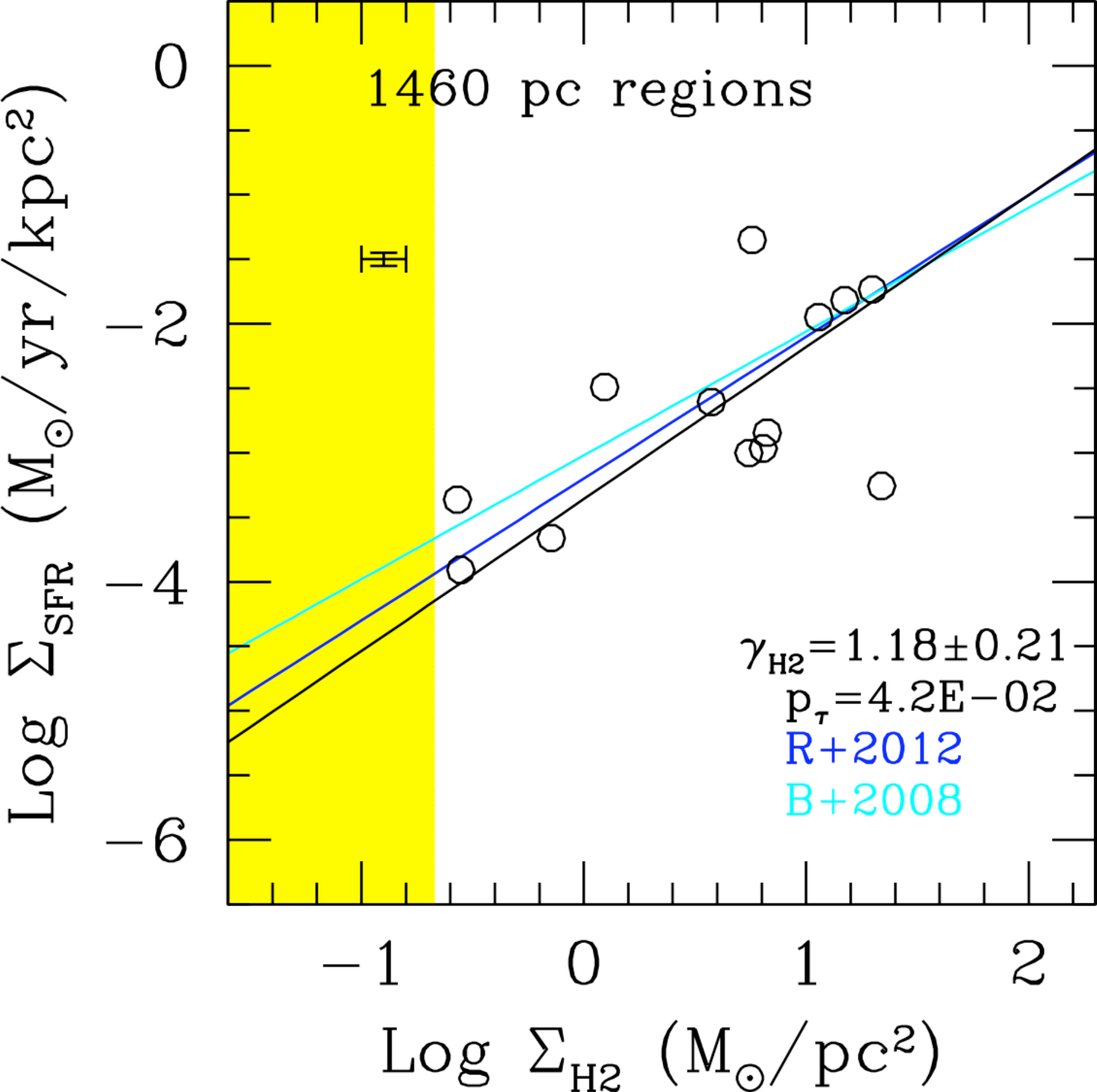}{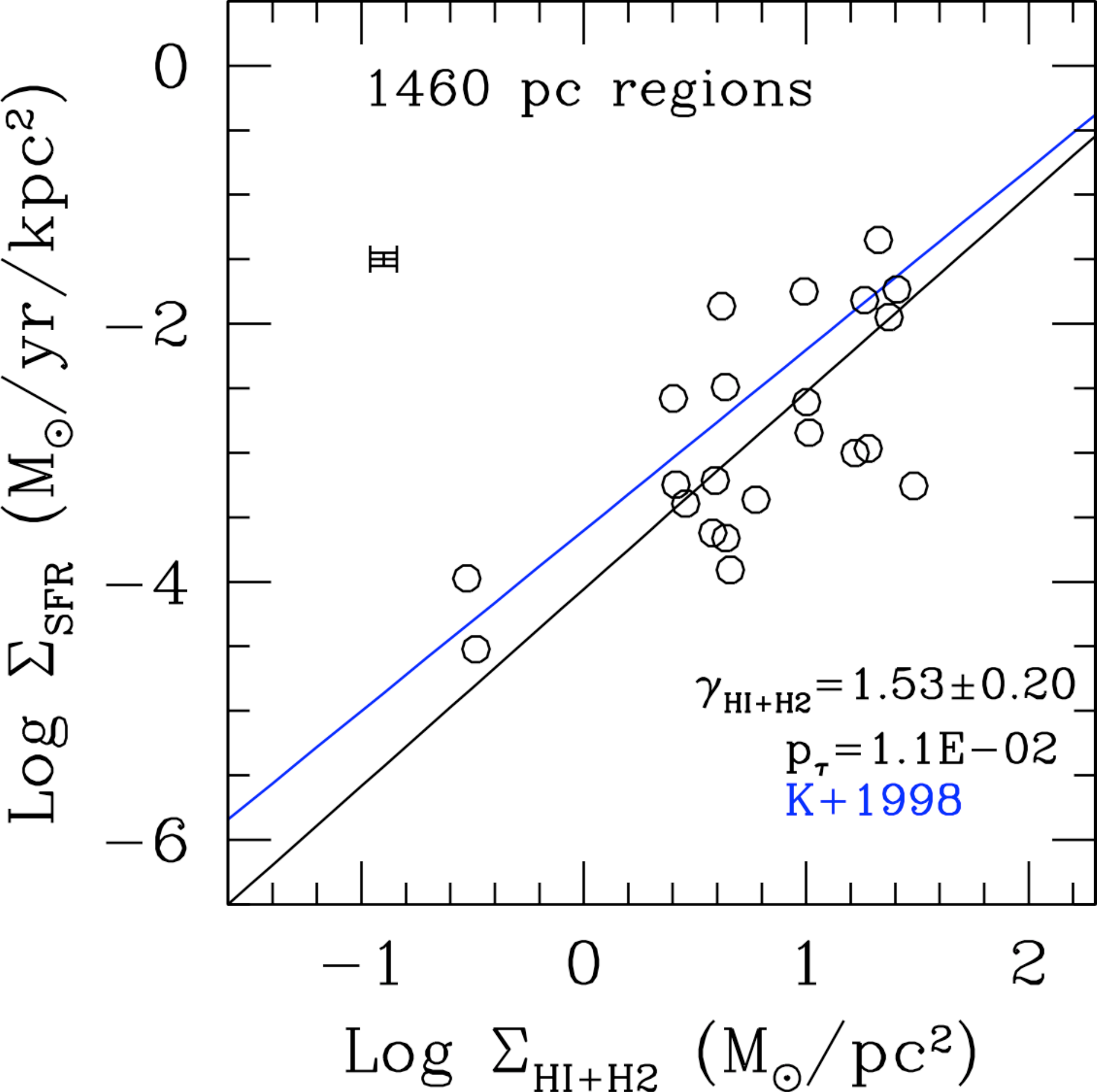} 
\caption{The scaling relations between the SFR  and  gas surface densities for two of the four spaxel sizes considered in this work: $\sim$360~pc {\bf (top panels)} and $\sim$1.5~kpc  {\bf (bottom panels)}. Both the relations for the molecular gas {\bf (left panels)} and the total, atomic$+$molecular, gas {\bf (right panels)} are shown. The data for each spaxel are shown as black circles; the red circles in the top panels are the spaxels with S/N$_{1100 \mu m}\ge$3.5. Representative 1~$\sigma$ uncertainties for the data are reported in each panel. The yellow--shaded region is the lower limit for the $\Sigma_{H2}$ data points (see Fig.~\ref{fig14} and text). The black vertical bars to the right of the top panels show the magnitude of the uncertainty due to stochastic sampling of the stellar IMF in the spaxels, which decreases for large  SFR surface densities \citep{Cervino2002}. Stochastic sampling is negligible for large galactic regions (bottom panels). The black lines show the OLS bi--sector linear best fits through the data, which are compared with best fits for other galaxies in the literature: \citet[][K+2007]{Kennicutt2007}, \citet[][R+2012]{Rahman2012}, \citet[][B+2008]{Bigiel2008}, and \citet[][K+1998]{Kennicutt1998}. The slopes of the best linear fits, $\gamma$, and the Kendall $\tau$ correlation probabilities, $p_{\tau}$, are listed in each panel. See text for details. }
\label{fig15} 
\end{figure}

\clearpage
\begin{figure}[h]
\figurenum{16}
\plottwo{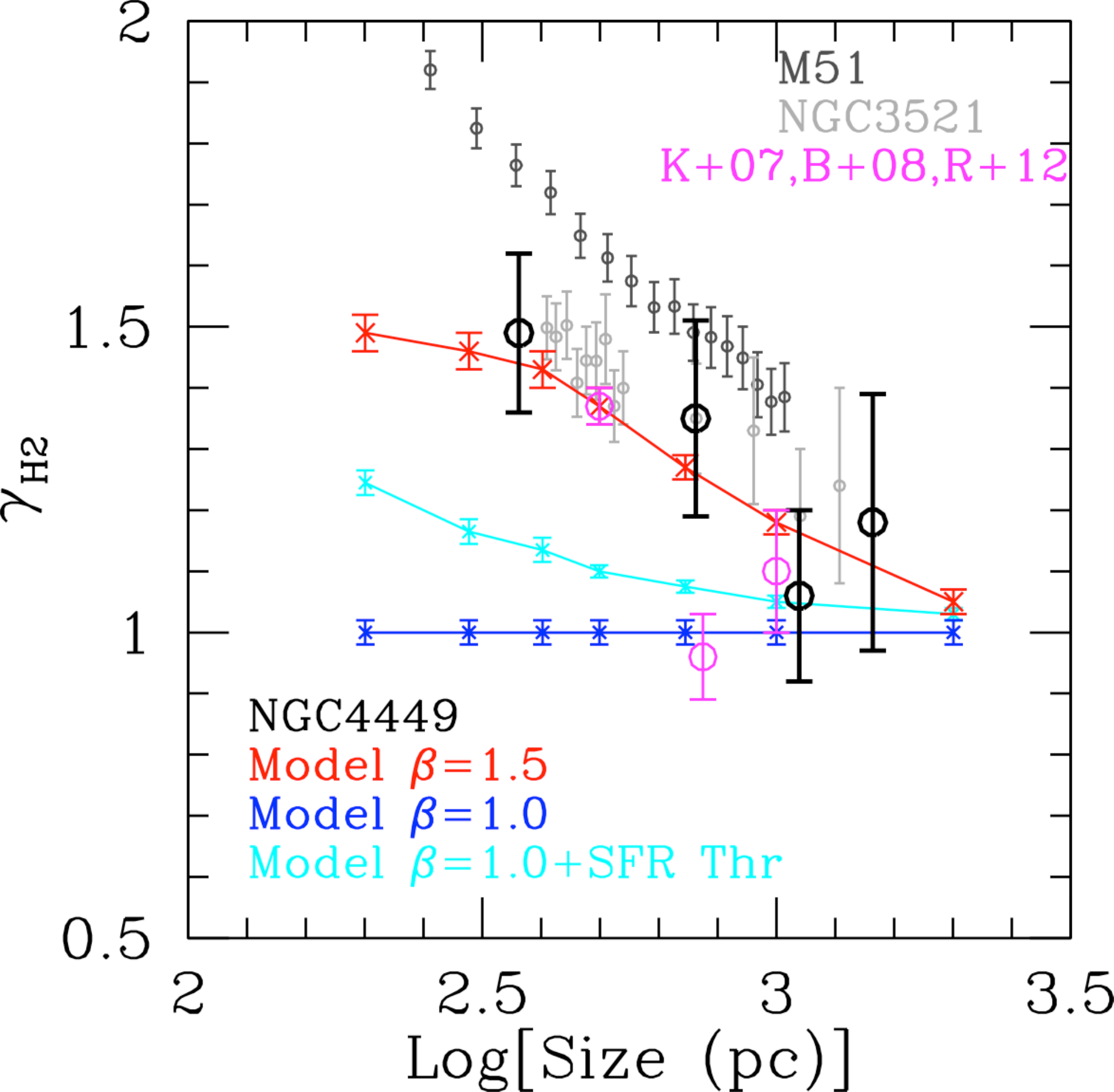}{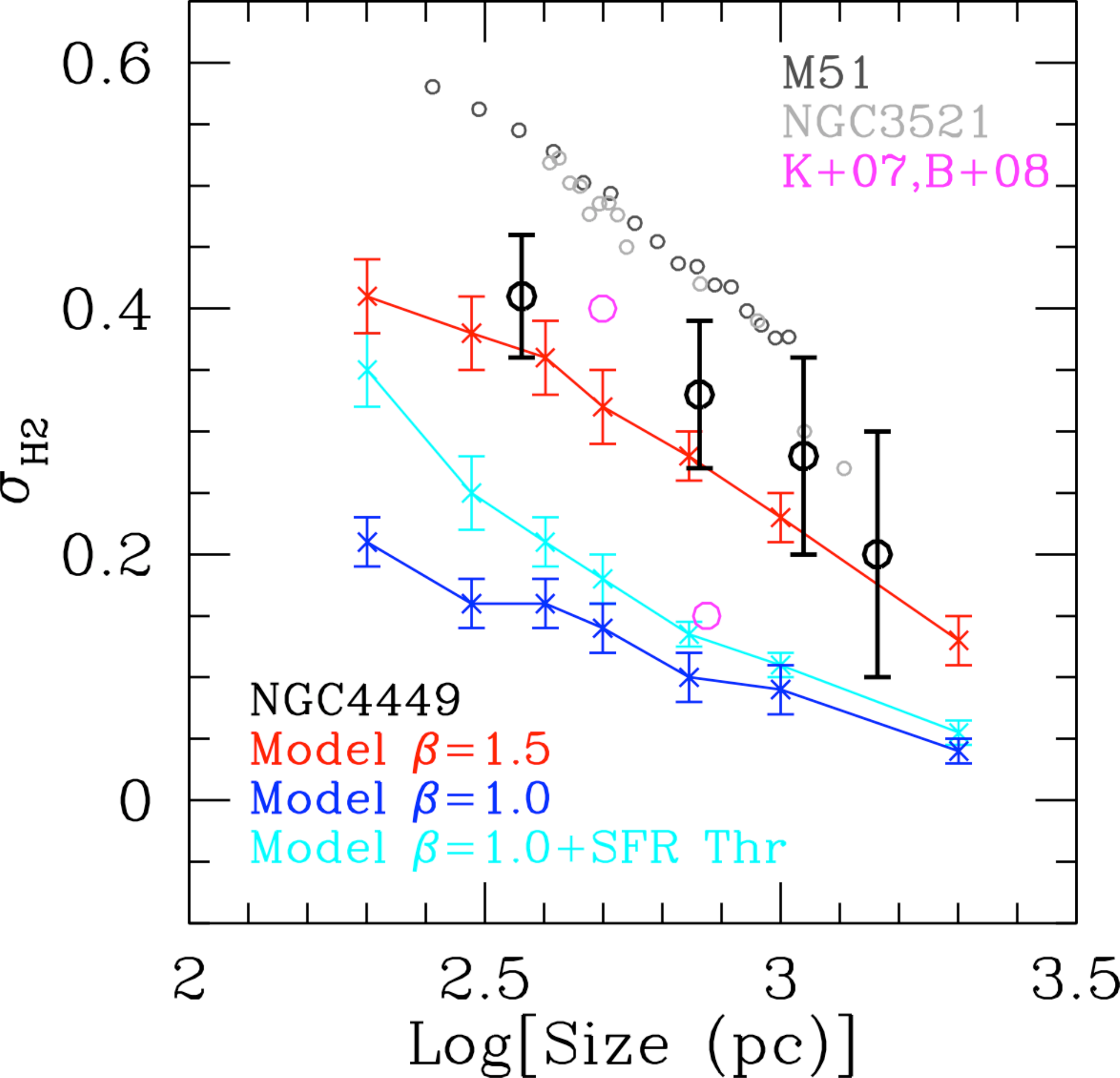} 
\caption{The slope, $\gamma_{H2}$, of the Log($\Sigma_{SFR}$)--Log($\Sigma_{H2}$) best fit linear relation {\bf (left panel)} and the scatter, $\sigma_{H2}$, of the data about the best--fit line  {\bf (right panel)}, as a function of the linear size of the galactic regions in NGC\,4449 used to perform the measurements. The data, with their 1~$\sigma$ uncertainties, are shown as black circles. Three simulations from \citet{Calzetti2012} are used for comparison, discriminated by the value of the exponent $\beta$ in the relation between SFR and molecular gas mass: SFR$\, \propto\, $M$_{H2}^{\beta}$. The case $\beta=1.5$ (red crosses and line), the case $\beta=1$ (blue crosses and line), and, finally,  the case $\beta=1$  with the lowest 18\% of the mass in clouds being devoid of star formation (cyan line).  Data from the literature are also shown on the panels: M51 and NGC3521 as dark and light grey circles, respectively, from \citet{Liu2011}; magenta circles show the location of the data from: \citet[][K+07, $\sim$500~pc]{Kennicutt2007}, \citet[][B+08, $\sim$750~pc]{Bigiel2008}, and \citet[][R+12, $\sim$1,000~pc]{Rahman2012}.}
\label{fig16} 
\end{figure}

\clearpage
\begin{figure}[h]
\figurenum{17}
\plotone{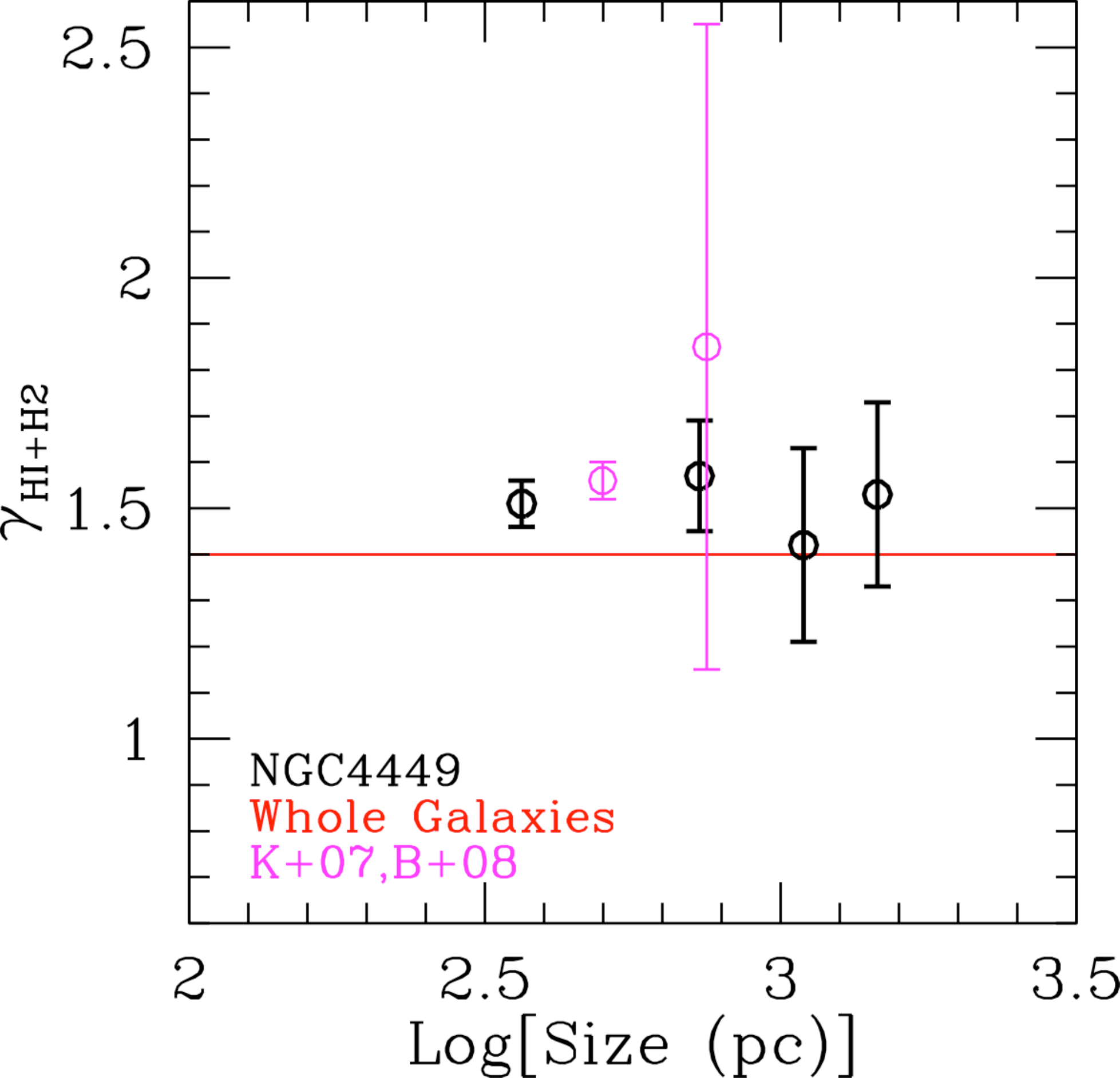} 
\caption{The slope $\gamma_{HI+H2}$, of the Log($\Sigma_{SFR}$)--Log($\Sigma_{HI+H2}$) best fit linear relation as a function of the size of the galactic regions in NGC\,4449. The relation observed for the total, atomic plus molecular, gas remains steep, around a slope of $\approx$1.5, for all region sizes, and agrees, within the uncertainties, with the slope of $\sim$1.4 measured for whole galaxies \citep[red horizontal line,][]{Kennicutt1998, KennicuttEvans2012}. Additional data from the literature are shown as magenta circles: \citet[][K+07, $\sim$500~pc]{Kennicutt2007}, and \citet[][B+08, $\sim$750~pc]{Bigiel2008}.}
\label{fig17} 
\end{figure}

\end{document}